\RequirePackage{fix-cm}
\documentclass[smallextended]{svjour3}       
\smartqed  
\usepackage{graphicx}
\usepackage{cite}
\usepackage[cmex10]{amsmath}
\usepackage{amssymb}
\usepackage{algorithmic}
\usepackage{graphics}
\usepackage{subfigure}
\usepackage{tabularx}
\usepackage{booktabs}
\usepackage{multirow}
\begin{document}

\title{High-quality Image Restoration from Partial Mixed Adaptive-Random Measurements
}


\author{Jun~Yang         \and
        Wei~E.I.~Sha     \and
        Hongyang~Chao    \and
        Zhu~Jin    
}


\institute{J. Yang \at
              School of Information Science and Technology, Sun Yat-sen University, Guangzhou 510006, China \\
              \email{juneryoung@gmail.com}           
           \and
           W. E.I. Sha (corresponding author)\at
              Department of Electrical and Electronic Engineering,the University of Hong Kong, Pokfulam Road, Hong Kong \\
             \email{wsha@eee.hku.hk; dr.weisha@gmail.com}
           \and
           H. Chao (corresponding author)\at
              School of Software, Sun Yat-sen University, Guangzhou 510006, China \\
              \email{isschhy@mail.sysu.edu.cn}
           \and
           Z. Jin \at
              School of Information Science and Technology, Sun Yat-sen University, Guangzhou 510006, China \\
              \email{zjin.sysu@gmail.com}
}

\date{Received: date / Accepted: date}

\maketitle

\begin{abstract}
A novel framework to construct an efficient sensing (measurement) matrix, called mixed adaptive-random (MAR) matrix, is introduced for directly acquiring a compressed image representation. The mixed sampling (sensing) procedure hybridizes adaptive edge measurements extracted from a low-resolution image with uniform random measurements predefined for the high-resolution image to be recovered. The mixed sensing matrix seamlessly captures important information of an image, and meanwhile approximately satisfies the restricted isometry property. To recover the high-resolution image from MAR measurements, the total variation algorithm based on the compressive sensing theory is employed for solving the Lagrangian regularization problem. Both peak signal-to-noise ratio and structural similarity results demonstrate the MAR sensing framework shows much better recovery performance than the completely random sensing one. The work is particularly helpful for high-performance and lost-cost data acquisition.
\keywords{Data acquisition \and Mixed adaptive-random sampling \and Total variation \and Compressive sensing}
\end{abstract}

\section{Introduction}

As a novel and revolutionary sensing (sampling) paradigm, compressive sensing (CS) theory has attracted much interest over the past few years. Now one can recover certain signals and images after directly acquiring far fewer samples or measurements in comparison with massive amounts of data collected in traditional data acquisitions \cite{romberg2007sensing,candes2006near,donoho2006compressed,lustig2007sparseMRI}. The sensing (measurement) matrix is essential to CS framework and must capture important information about the object of interest.

The basic principle of CS is that sparse signals can be recovered from very few measurements. A signal $ x = \{ x_n\}_{n=1}^N$ of length $N$ is said to be sparse in a basis space $\Psi$=$ \{\psi_n \}_{1\leqslant n \leqslant N}$ if transform coefficients $\langle x, \psi_n \rangle, 1\leq n \leq N $ are mostly zero; or nearly sparse in the space $\Psi$ if a dominant portion of these $N$ coefficients are either zero or very close to zero. The sparsity of $x$ in $\Psi$ is quantified by the number of significant (nonzero) coefficients $K$. The signal can be perfectly recovered from $ M = O(K\log(N/K))$ observations with a high probability.

Given $ M $ measurements $ y = \Phi x $, with $ \Phi $ producing the random projections, standard CS recovers $ x $  from $ y $ by using the following constrained optimization problem:
\begin{equation}\label{eqCSframe}
\min_x \|\Psi^T x\|_p \quad s.t. \quad y = \Phi x
\end{equation}
where $p$ is usually set to be 1 or 0, guaranteeing the sparse solution of the vector $ \Psi^T x $. $\| \ast \|_1 $ is $ \ell_1 $ norm, i.e. the summation of the absolute value of all the elements in a vector. While $\| \ast \|_0$ is $ \ell_0 $ norm, counting the nonzero entries of a vector. The $ \ell_1 $  minimization problem of (1) can be solved by a linear programming \cite{TaoT2005}. Other recovery algorithms have been recently proposed also, including gradient projection sparse reconstruction \cite{Wright2007}, matching pursuit \cite{tropp2007signal}, and iterative thresholding methods \cite{DeMolC2004}.

The recovery performance of CS depends significantly on the measurements and its sensing strategies. In the literature \cite{baraniuk2008simple} Baraniuk et al. prove that the sensing matrix satisfying the Johnson-Lindenstrauss lemma also holds true for the restricted isometry property (RIP) in compressive sensing. Furthermore, a sparse random projection \cite{li2006very}, which is almost as accurate as the conventional random projection, is proposed to reduce computational cost in the measurement process. Moreover, a structurally random matrix (SRM) \cite{do2012fast} is also constructed for fast and efficient CS, where the incoherence between SRM and sparsifying transforms is comparable to that between completely random sensing matrix and the transforms. However, all the measurement processes above are nonadaptive, i.e., the sensing matrix is predefined or fixed. On one hand, the predefined random sampling obeys the incoherence condition (with the RIP). Thus it is a fascinating character of CS as well, where all the measurements are equally important. On the other hand, a completely random sampling pattern capturing information of object aimlessly becomes inefficient for reconstructing high-resolution data. For example, the completely uniform random sampling \cite{lustig2007sparseMRI,candes2006robust} in k-space for magnetic resonance imaging application performs worse than the nonuniform random sampling, where low frequency components are densely sampled.

In CS, each measurement is a projection from the whole signal, which consumes a large number of computer resources. In this paper, we sense the signal in a different and simple way, which partially sample or ``select" the signal elements in spatial domain.

Could one acquire more important information of object in spatial (time) domain with fewer measurements? Could one improve the completely random sampling in spatial domain? In this work, we involve the object-dependent and ``most important" edge information of object, which can be extracted from a low-cost sampling procedure with much lower sampling rate, into the sensing matrix. The novel mixed adaptive-random (MAR) sensing hybridizes adaptive edge measurements obtained from a low-resolution image with uniform random measurements predefined for the high-resolution image to be reconstructed. The adaptive edge measurements can be regarded as a way of measurement learning. This is the basic idea of the paper and also our contribution. To the best of our knowledge, it is the first time we have introduced the mixed sampling concept into the compressed partial sampling framework. We will employ the MAR sensing matrix to directly acquire a compressed image representation and meanwhile adopt total-variation (TV) regularizer to recover the high-resolution image. Numerical examples demonstrate significant advantages of the MAR sensing framework over the completely random sensing one.

\section{Mixed Sampling Protocol}
\begin{figure}[h]
\scriptsize
\centering
\includegraphics[width=0.6\textwidth]{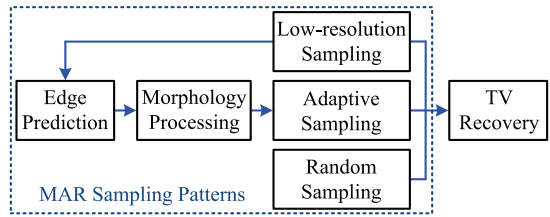}  
\caption{The schematic diagram for the mixed adaptive-random sampling protocol.}
\label{fig_csframwork}
\end{figure}
Fig.~\ref{fig_csframwork} shows the schematic diagram for the MAR sampling protocol. Here we assume the image $f(x,y)$ as a function in 2D Hilbert space $L(R)\times L(R)$. The MAR sensing matrix can be constructed by the following procedures:

\textbf{Step 1}, sampling a low-resolution image $f_l$ with an extremely low cost to predict the edge information of the high-resolution image $f$ to be recovered. Regarding practical hardware implementation, the low-resolution image requires much fewer photosensitive elements (such as $128\times128$) instead of millions (such as $1024\times1024$) required in convectional data acquisition for the high-resolution image. Mathematically, we have
\begin{equation}\label{eq1}
\Gamma(f)\approx\Gamma(f_p)=\Gamma(I(f_l))
\end{equation}
where $\Gamma(f)=1$ for the edge pixels of $f$, otherwise $\Gamma(f)=0$. The interpolation operator $I$ maps the low-resolution image $f_l$ to the predicted high-resolution one $f_p$. The $\Gamma$ denotes the edge detection operator that can be implemented with the Sobel edge detector \cite{parker1997algorithms,canny1986edgedection} and binary thresholding. As a result, real edges of the high-resolution image $\Gamma(f)$ can be approximated by the predicted edges $\Gamma(f_p)$.

\textbf{Step 2}, due to a possible inaccuracy of the predicted edges, morphological operations can be used to generate an adaptive sampling pattern around the edges of $f$.
\begin{equation}\label{eq2}
S_a=M^p(\Gamma(f_p))
\end{equation}
where $S_a$ is the adaptive sampling pattern and $M^p$ is the binary morphological operator on the edges of the predicted image $f_p$. The morphological operator involves dilation $M_d^p$ and closing $M_c^p$ (dilation followed by erosion), which can be found at MATLAB image processing toolbox. Additionally, $M_n^p$ suggests no morphological operation is executed. After understanding the role of edges in computer vision and image processing, we suppose that the image pixels that located at edges or near the edges are more important than those located at smooth regions. Consequently, involve the adaptive sampling pattern into the sensing procedure is highly reasonable.

\textbf{Step 3}, generating the random sampling pattern $S_r$ with a 2D uniform distribution $U(0,1)\times U(0,1)$ and with a binary thresholding for controlling the sampling ratio. The completely random sampling, which acquires pixels at edges and smooth regions uniformly, captures the image profile information and guarantees the RIP and incoherence condition.

\textbf{Step 4}, we mix the random and adaptive sampling patterns via a union operation to get the new MAR sampling pattern (sensing matrix with $0/1$ elements).
\begin{equation}\label{eq3}
S_m=S_a\bigcup S_r\bigcup S_l
\end{equation}
where $S_l$ is the low-resolution sampling pattern corresponding to $f_l$, $S_a$ is the adaptive sampling pattern defined in step 2, and $S_r$ is the random sampling pattern defined in step 3. In other words, we reuse (do not resample) the pixels of $f_l$ obtained at the Step 1 for saving the measurements.

To physically acquire the pixels corresponding to the MAR sensing matrix $S_m$, we may use integrated circuits to control reset transistors (or switches) in complementary metal-oxide-semiconductor (CMOS) camera. As a result, only a portion of photodetectors and amplifiers (with respect to $S_l$ and $S_m\backslash S_l$) are turned on. Compared to traditional image acquisitions, the MRA sensing saves electrical power and increases lifetime of image sensors. Most importantly, the MRA sensing can be generalized to other data acquisitions where the most important information of object is adaptively extracted and learned via a low-cost sampling.

For convenience, the sensing ratio of the MAR sensing matrix $\eta_1$ is defined as the number of nonzero elements of $S_m$ over the dimension of $S_m$ (i.e. image size of $f$). The adaptive sampling ratio $\eta_2$ is defined as the number of nonzero elements of $S_m\backslash S_r$, which is the complement of $S_r$ in $S_m$, over that of $S_m$.
\begin{equation}\label{eq4}
\eta_1=\frac{\sum_{i,j}S(i,j)}{Dim(S_m)},\quad\eta_2=1-\frac{\sum_{i,j}S_r(i,j)}{\sum_{i,j}S_m(i,j)}
\end{equation}
The sensing ratio $\eta_1$ could be considerably smaller and thus measurement cost can be reduced. In addition, the adaptive sampling ratio $\eta_2$ cannot be too large to satisfy the RIP and incoherence condition.

At the end of this section, we would like to discuss the RIP of our sensing matrix. A linear measurement (sensing) operator $\mathcal{A}: \mathbb{C}^{N_1\times N_2}\rightarrow \mathbb{C}^m$ has the RIP of order $s$ and level $\delta\in(0,1)$, if

\begin{equation}\label{eqRIP}
(1-\delta)\|X\|_2^2 \leq \|\mathcal{A}(X)\|_2^2 \leq (1+\delta)\|X\|_2^2~~~ \text{for all $s$-sparse}~ X \in \mathbb{C}^{N_1\times N_2}
\end{equation}
where $\|X\|_2$ denotes the Frobenius norm of the image X. The smallest such $\delta$ that \eqref{eqRIP} holds is denoted by $\delta_s$ and called the RIP constant.

In \cite{needell2013stable}, Needell and Ward adopted the bivariate Haar transform to state and prove the following theorem

\newtheorem{thm}{Theorem}
\begin{thm}
Consider $n,m,s\in \mathbb{N}$, and let $N=2^n$. There is an absolute constant $C>0$ such that if $\mathcal{A}:\mathbb{C}^{N\times N}\rightarrow \mathbb{C}^m$ is such that, composed with the inverse bivariate Haar transform, $\mathcal{AH}^{-1}:\mathbb{C}^{N\times N}\rightarrow \mathbb{C}^m$ has the restricted isometry property of order $Cs\log^3(N)$ and level $\delta<1/3$, then the following holds for any $X\in\mathbb{C}^{N\times N}$. If noisy measurements $y = \mathcal{A}(X) + \xi$ are observed with noise level $\|\xi\|_2\leq\varepsilon$, then
\begin{equation*}
\hat{X}= \mathop {\arg\min }\limits_Z \|Z\|_{TV}~~such~that~~ \|\mathcal{A}(Z)-y\|_2 \leq \varepsilon
\end{equation*}
satisfies
\begin{equation}
\|X-\hat{X}\|_2 \lesssim {\frac{{\| \nabla X - (\nabla X)_s\|_1}}{{\sqrt{s} }}+ \varepsilon}
\end{equation}
\end{thm}
From the theorem, when $\mathcal{A}$ is an identity operator and $\mathcal{H}$ is the orthogonal bivariate Haar transform, the measurements $\mathcal{A}$ will satisfy the RIP. Considering the proposed sensing strategy, which can be regarded as an approximation of the identity operator, could satisfy the RIP. We will further discuss the RIP condition at the Section of Numerical Results.

\section{Recovery Algorithm}

After using the MAR sensing matrix to directly acquire a compressed image representation, the recovery algorithm is essential to reconstruct a high-quality image with a high resolution. The greedy pursuit algorithm \cite{tropp2007signal,dai2009subspace,donoho2012sparse} offers a ${\ell_0}$ minimization for sparse reconstruction. Linear programming \cite{donoho2006compressed} and other convex optimization algorithms \cite{hale2008fixed,van2008probing,candes2006robust,dai2009physics} have been proposed to solve the ${\ell_1}$-minimization also. The TV regularizer was introduced by Rudin, Osher and Fatemi in \cite{rudin1992nonlinear} and became popular in recent years \cite{dai2009physics,rudin1992nonlinear,alliney1994an,osher2003image}.

Needell and Ward show that there are choices of underdetermined linear measurements (constructed from RIP matrices) for which the TV minimization program is guaranteed to recover images stably and robustly up to the best $s$-term approximation of their gradient \cite{needell2013stable}. Fixing integers $m$, $N$ and $s$ such that $m\geq C_1 s \log(N^2/s)$, the reconstruction error satisfies
\begin{equation}\label{eq5}
\| X-\hat{X} \|_2 \leq C_2 \log(N^2/s)\left( {\frac{{\| \nabla X - (\nabla X)_s\|_1}}{{\sqrt{s} }}+ \varepsilon} \right)
\end{equation}
where $X$ and $\hat{X}$ are original and reconstructed images, $ \varepsilon$ is the noise level of the measurement, $\nabla X$ is the gradient of $X$, and $C_1$ and $C_2$ are universal constants. The above inequality relies on the compressibility of the bivariate Haar wavelet transform (More details can be found at \cite{needell2013stable}). The fundamental reason why the MRA sensing could achieve better performance is that the sensing matrix contains the edge information, which captures the $s$ most important nonzero $\nabla X$.

For reconstructing the high-resolution image $f$ from the measurements (compressed image representation) $g$, a Lagrangian regularization problem should be solved, i.e.
\begin{equation}\label{eq6}
\begin{aligned}
&\mathop {\min }\limits_f \left\{{\int {\left( {g - S_mf} \right)} ^2}dxdy + \alpha \int {\sqrt {{\left(\frac{df}{dx}\right)^2} + {\left(\frac{df}{dy}\right)^2}} dxdy} \right.\\
&\left.+ \beta \int {\sqrt {{{\left( {Tf} \right)}^2}} dxdy}\right\}
\end{aligned}
\end{equation}
where $S_m$ is the MAR sensing operator, $\alpha$ and $\beta$ are Lagrangian multipliers, and $dx$ and $dy$ are the differential operators. The second term is the TV regularizer; and the third-term relates to ${\ell_1}$-minimization with a sparsifying transform operator $T$. According to the variational principle, we have
\begin{equation}\label{eq7}
\begin{aligned}
&\frac{\delta O(f)}{\delta f}=2{S_m^{*}}\left( {g - S_mf} \right)- \alpha \frac{d}{{dx}}\left( {\frac{{d{f}/dx}}{{\sqrt {{{(d{f}/dx)}^2} + {{(d{f}/dy)}^2}} }}} \right) \\
&- \alpha \frac{d}{{dy}}\left( {\frac{{d{f}/dy}}{{\sqrt {{{(d{f}/dx)}^2} + {{(d{f}/dy)}^2}} }}} \right)
+\beta {T^{*}}\left(\frac{{T{f}}}{{\sqrt {{{\left( {T{f}} \right)}^2}} }}\right)
\end{aligned}
\end{equation}
where $O(f)$ is the objective functional given in \eqref{eq6}, $\delta$ is the variational operator, and $S_m^{*}$ and $T^{*}$ are adjoint operators of $S_m$ and $T$, respectively. In this work, we did not focus on the recovery algorithm and set $\beta$ to zero for fast and simple reconstruction. With the help of nonlinear conjugate gradient method \cite{rudin1992nonlinear,hager2006survey} and \eqref{eq7}, the Lagrangian regularization problem \eqref{eq6} can be solved.

\section{Numerical Results}

In this section, numerical performances of the proposed MAR sensing matrix will be evaluated. Without loss of generality, we assume $Dim(S_m)=Dim(f)=256\times256$. The sensing ratio $\eta_1$ and adaptive sampling ratio $\eta_2$ defined in \eqref{eq4} can be tuned by the threshold values of the binary thresholding procedures of Steps 1 and 3 at Section 2. We will demonstrate that incorporation of edge information to the sensing procedure can pronouncedly improve the recovery performance. In the beginning, the MAR sensing performance for different edge extraction methods are investigated. Then, we compare recovery results by the MAR sensing matrix to those by the completely random sensing matrix. Finally, we will discuss the influence of $\eta_2$ on the recovery performance and compare our method with the standard CS and other random partial sampling methods.

The low-resolution image $f_l$ is numerically generated by downsampling the original high-resolution image $f$ by a factor of $4$, i.e. $Dim(f_l)=64\times 64$. Using the bicubic interpolation method \cite{keys1981cubic}, we can get the predicted image $f_p$ (Step 1 of Section 2). The edges of $f$ and $f_p$ can be extracted by the Sobel method (Step 1 of Section 2). For simple notations, $M_n$, $M_d$ and $M_c$ correspond to the edges of $f$ with null morphological operation, dilation and closing. Similarly, $M_n^p$, $M_d^p$ and $M_c^p$ correspond to the edges of $f_p$ (Step 2 of Section 2). Moreover, we use abbreviations of $S_r$ and $S_m$ to denote sensing methods using the completely random matrix and MAR matrix, respectively (Step 4 of Section 2).

\begin{figure}[!t]
\scriptsize
\centering
\subfigure[Phantom]{\label{fig:edgePSNR_a}\includegraphics[width=0.49\textwidth]{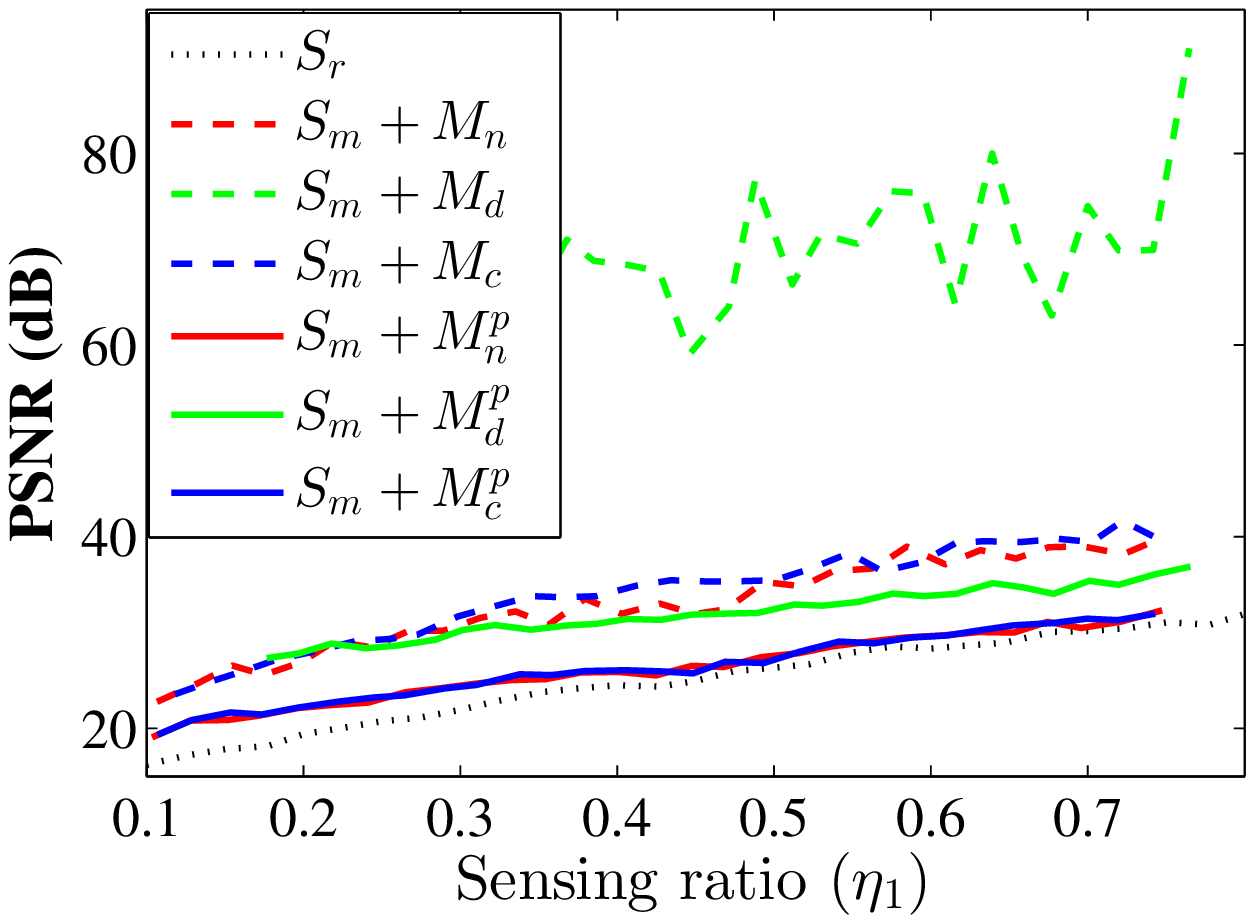}}
\subfigure[Fruits]{\label{fig:edgePSNR_b}\includegraphics[width=0.49\textwidth]{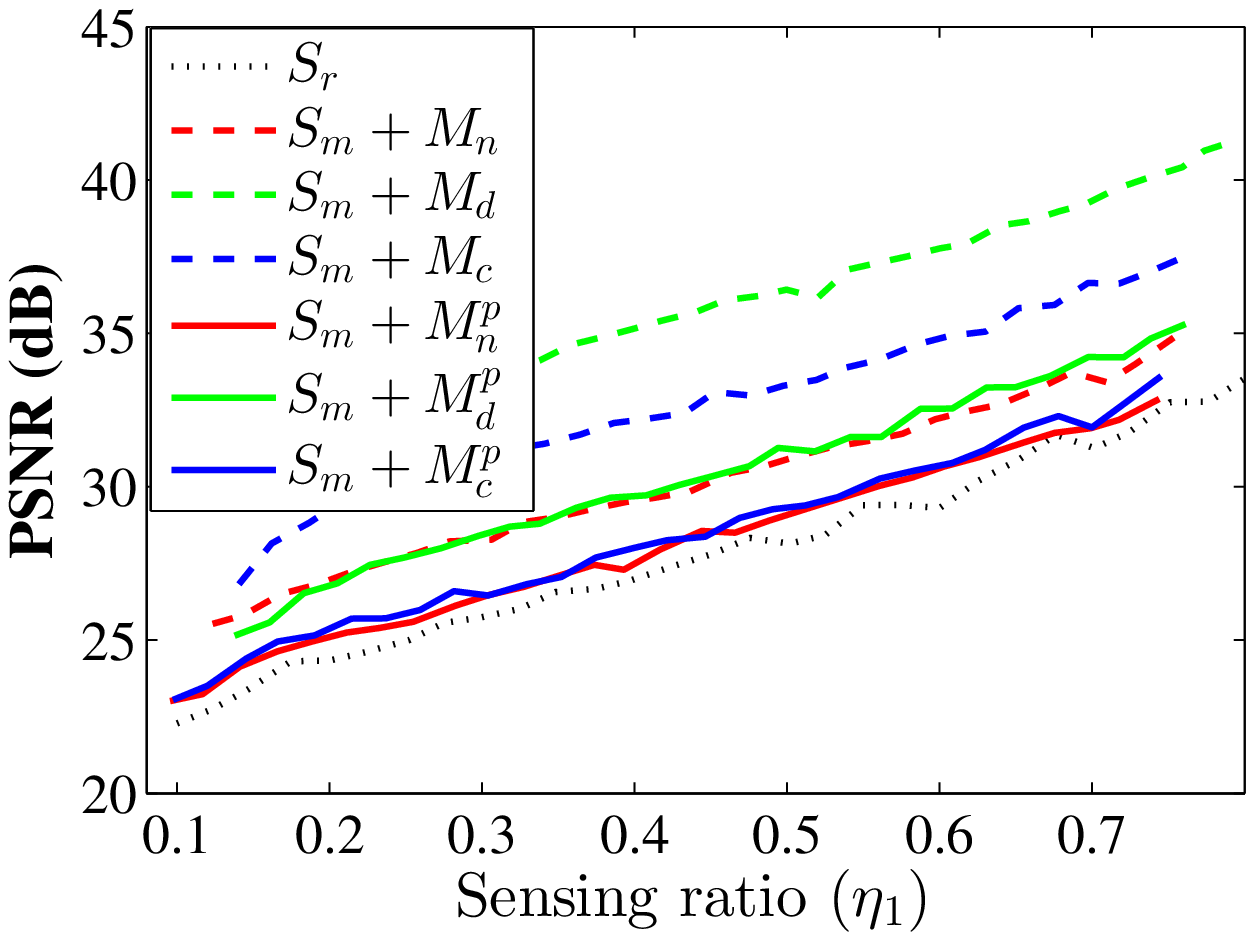}}  \\ 
\subfigure[Lena]{\label{fig:edgePSNR_c}\includegraphics[width=0.49\textwidth]{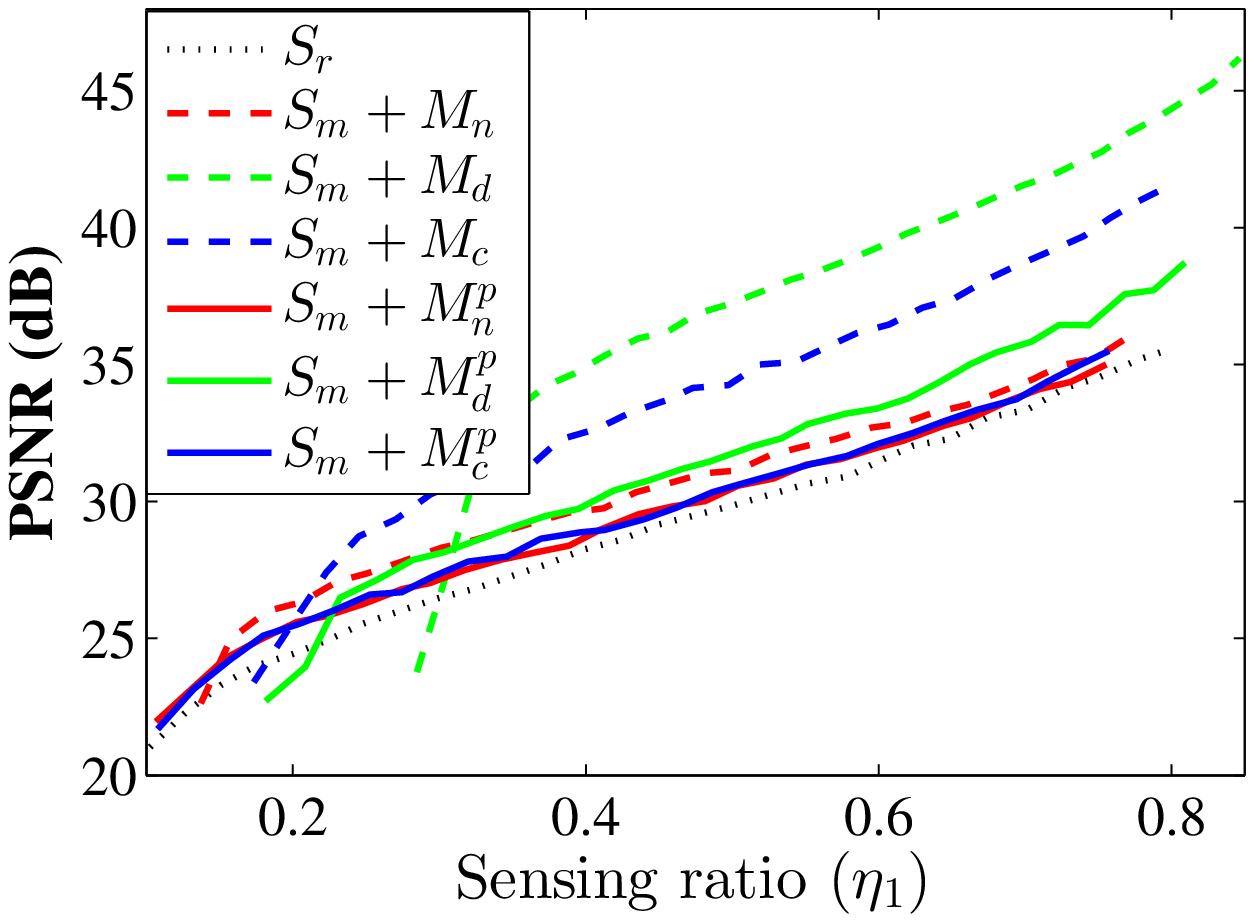}}
\subfigure[Boat]{\label{fig:edgePSNR_d}\includegraphics[width=0.49\textwidth]{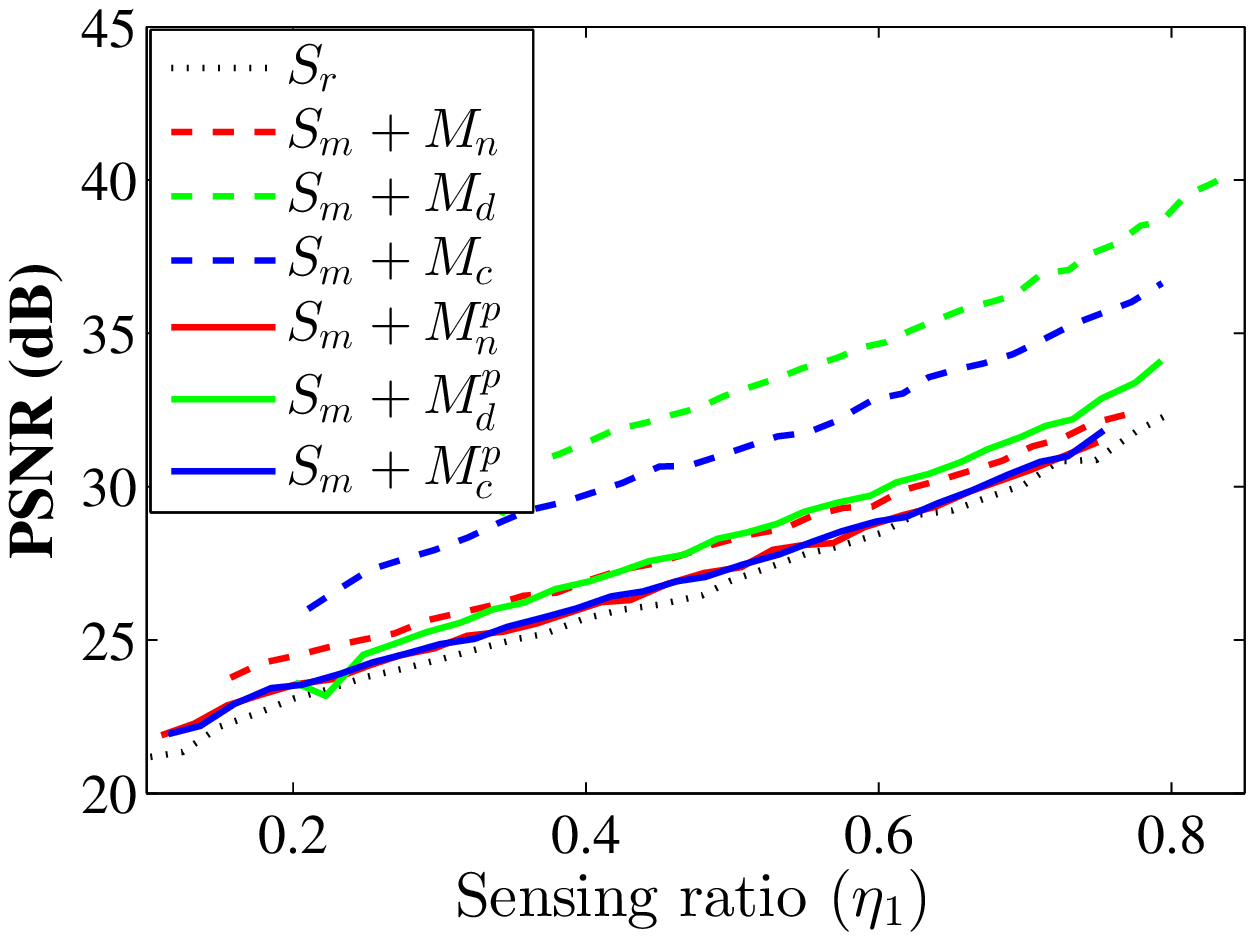}}  \\ 
\caption{PSNR performance as a function of the sensing ratio $\eta_1$ for the completely random sensing $S_r$ and MAR sensing $S_m$.}
\label{fig_edgePSNR}
\end{figure}

\begin{figure}[htbp]
\scriptsize
\centering
\subfigure[]{\label{fig:edgeImg_a}\includegraphics[width=0.24\textwidth]{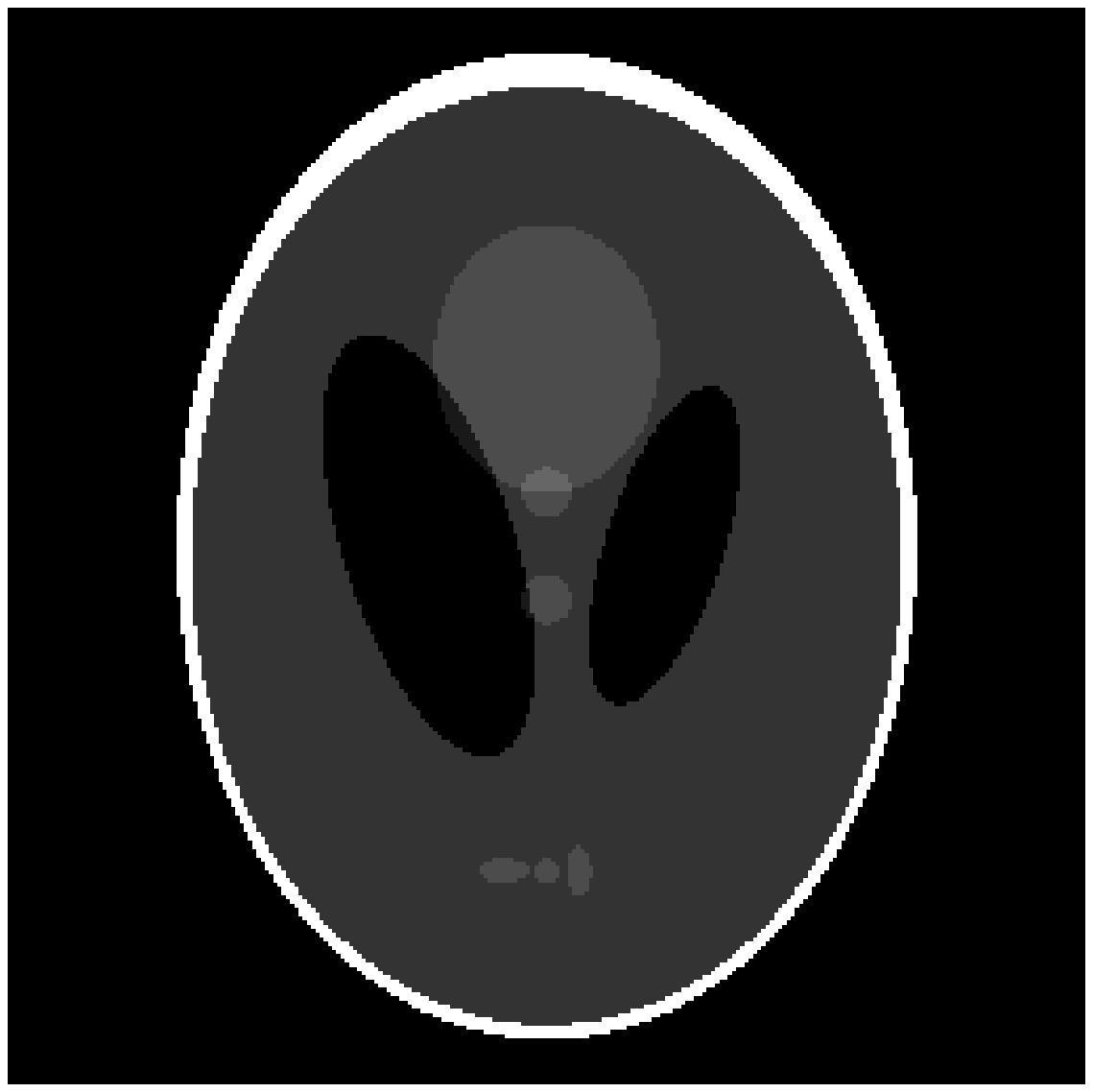}}
\subfigure[]{\label{fig:edgeImg_c}\includegraphics[width=0.24\textwidth]{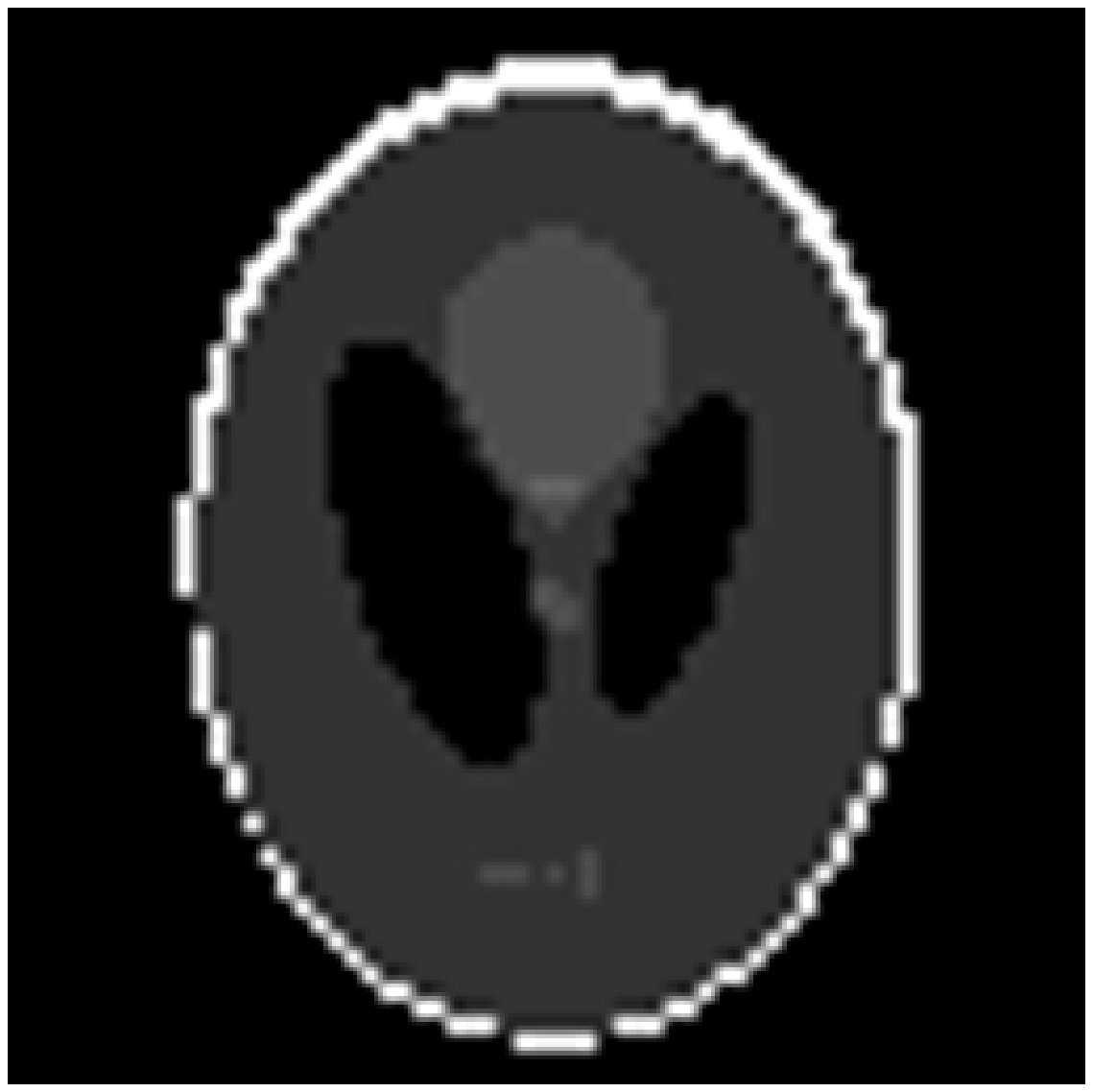}}
\subfigure[]{\label{fig:edgeImg_b}\includegraphics[width=0.24\textwidth]{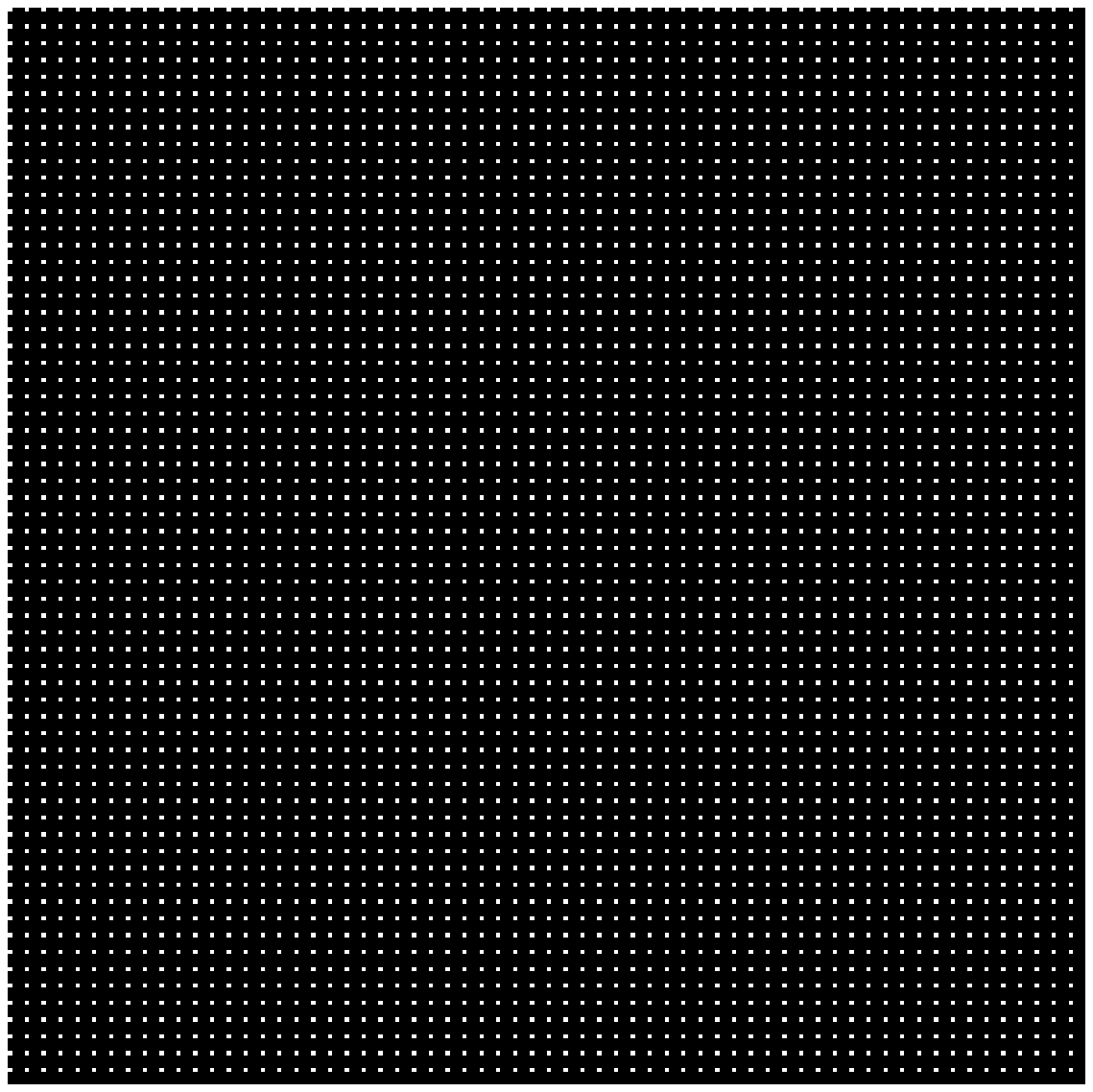}}
\subfigure[]{\label{fig:edgeImg_d}\includegraphics[width=0.24\textwidth]{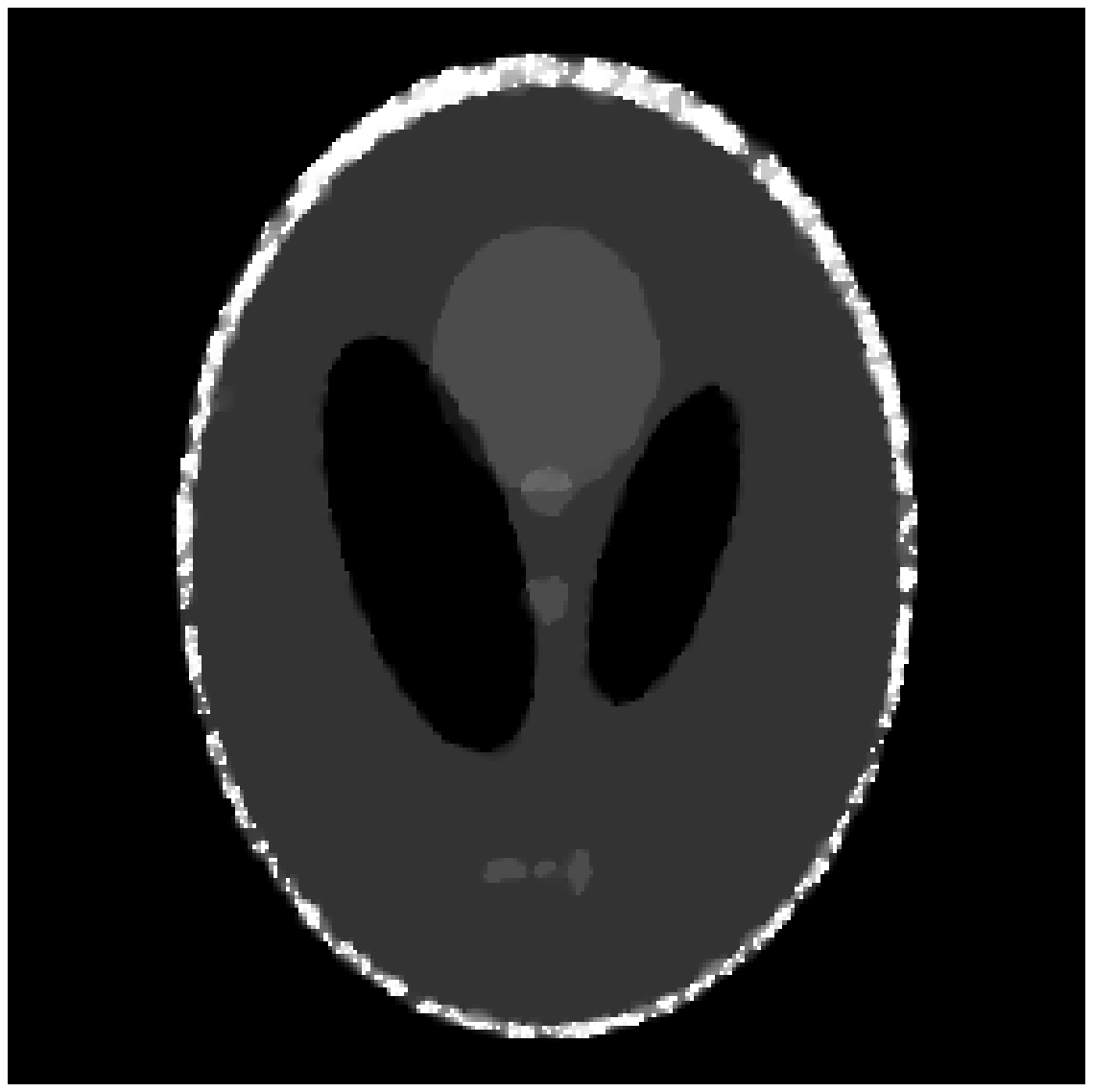}}       \\ 

\subfigure[]{\label{fig:edgeImg_e}\includegraphics[width=0.24\textwidth]{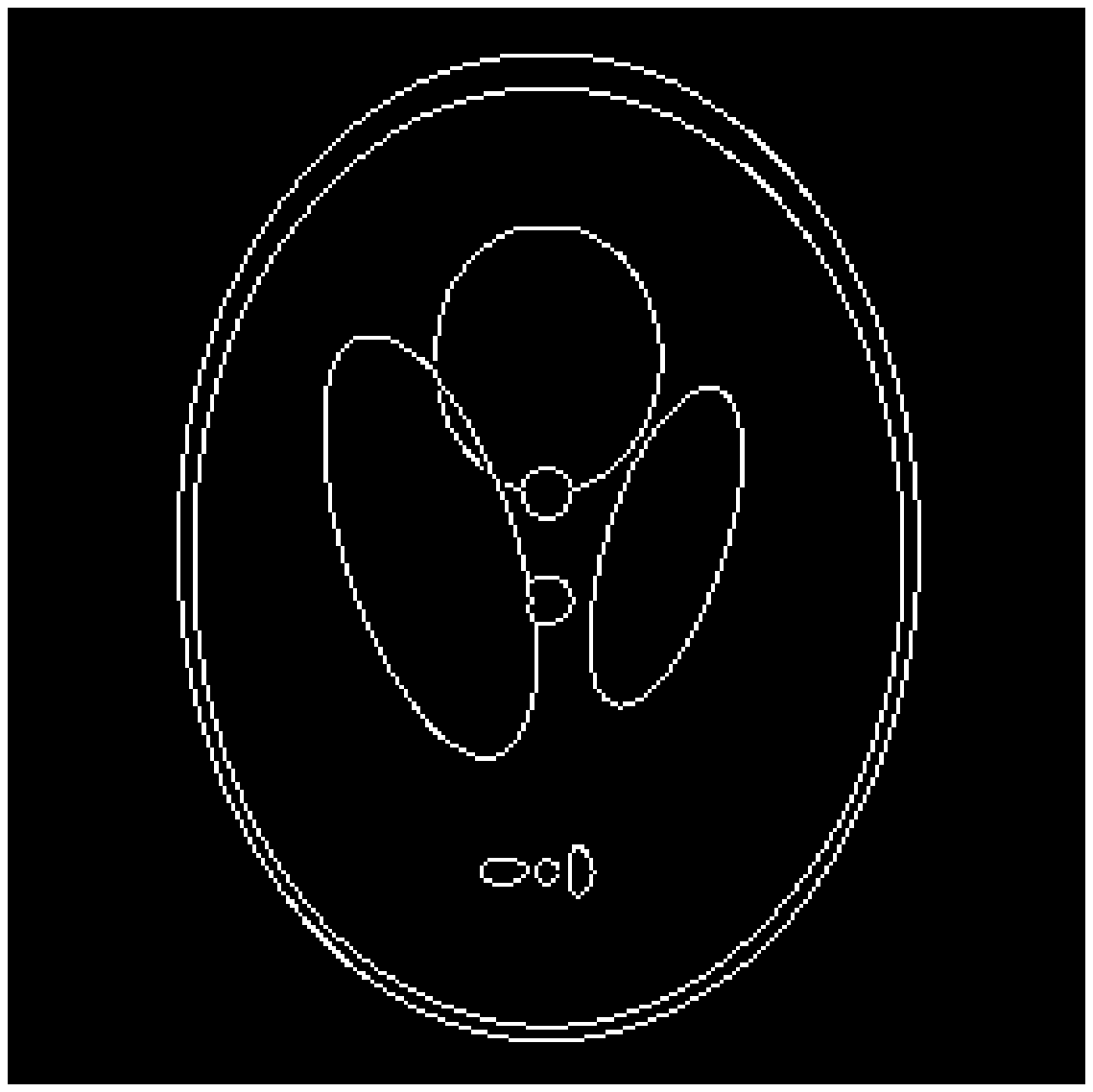}}
\subfigure[]{\label{fig:edgeImg_f}\includegraphics[width=0.24\textwidth]{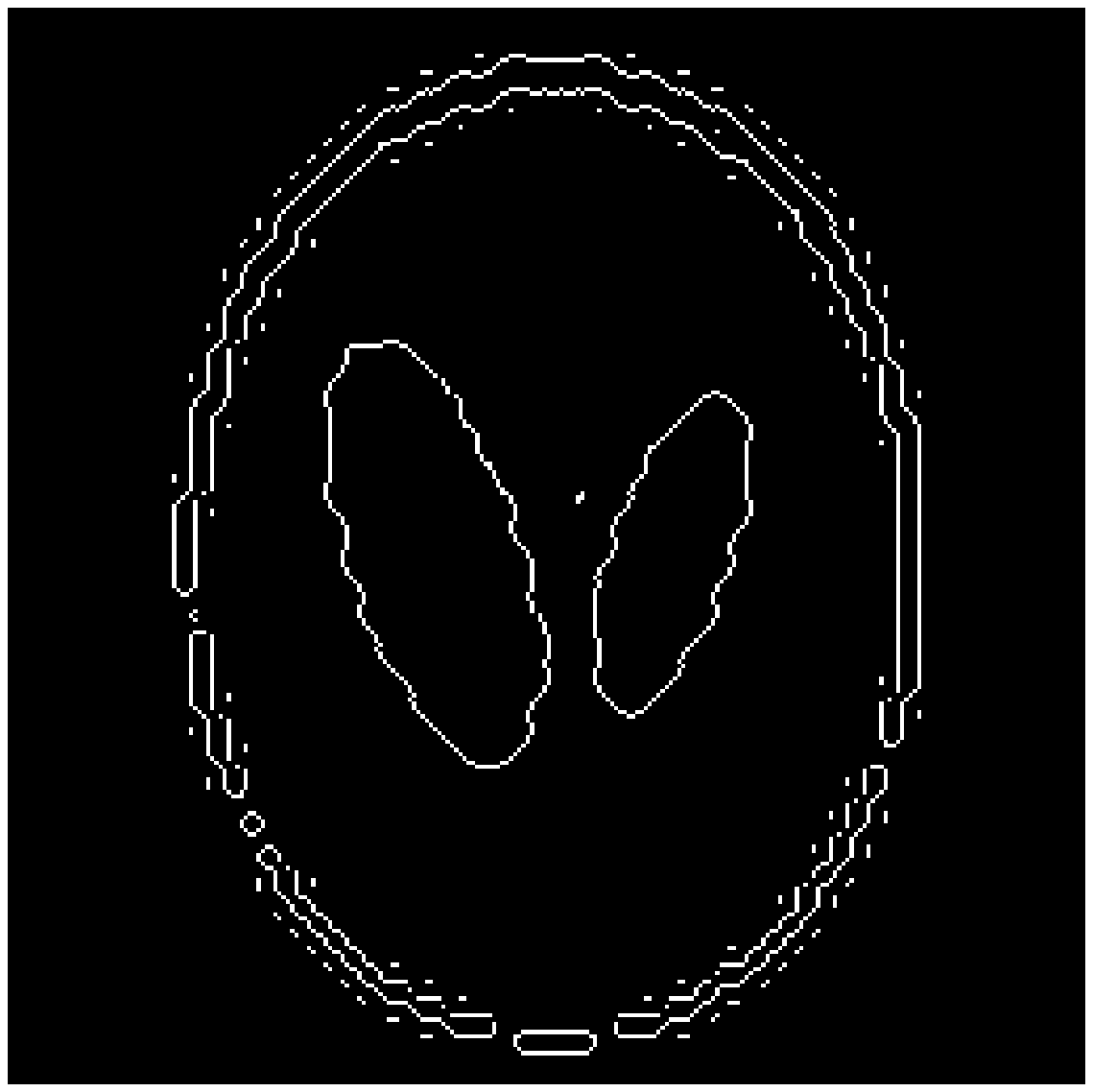}}
\subfigure[]{\label{fig:edgeImg_g}\includegraphics[width=0.24\textwidth]{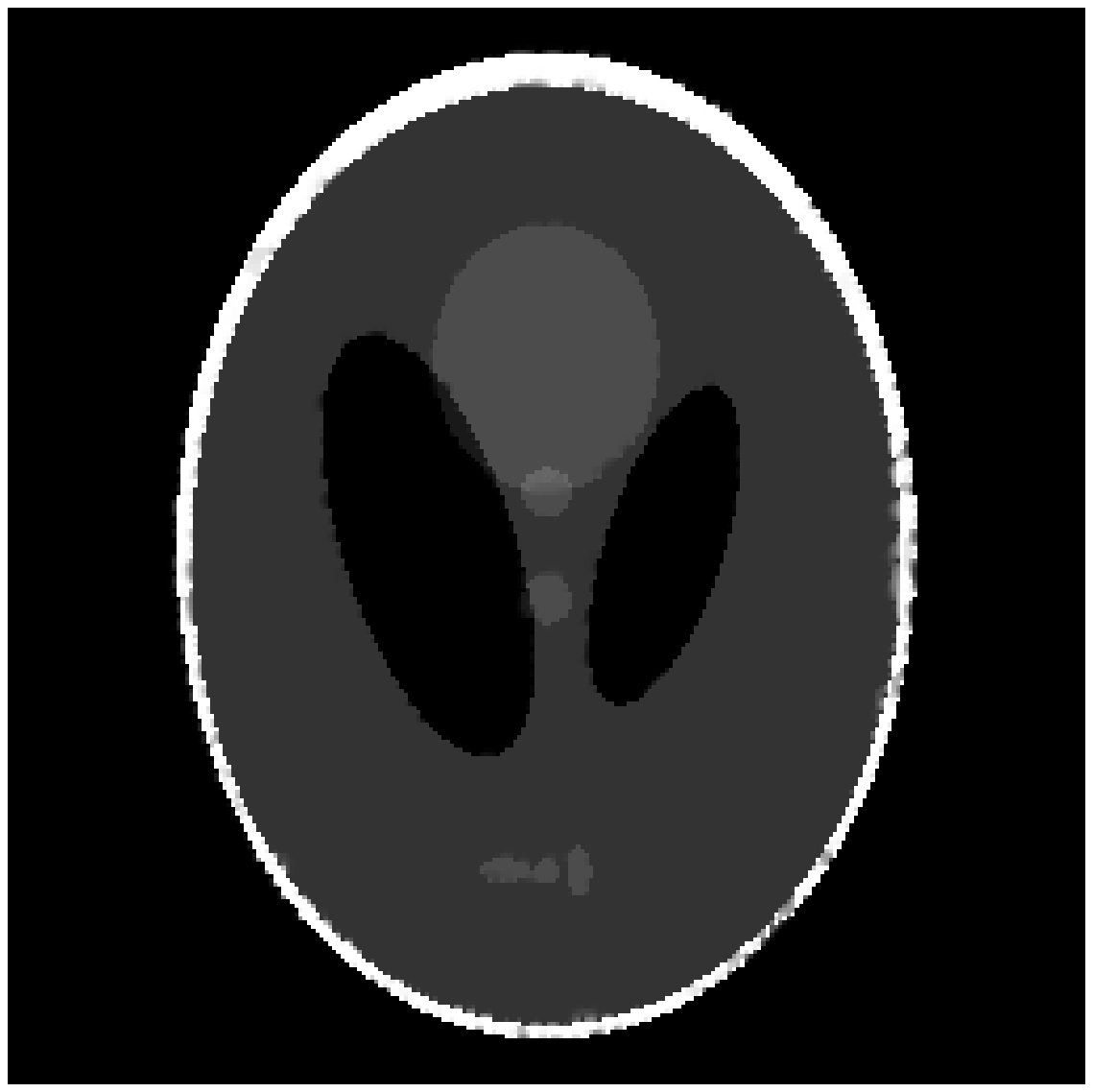}}
\subfigure[]{\label{fig:edgeImg_h}\includegraphics[width=0.24\textwidth]{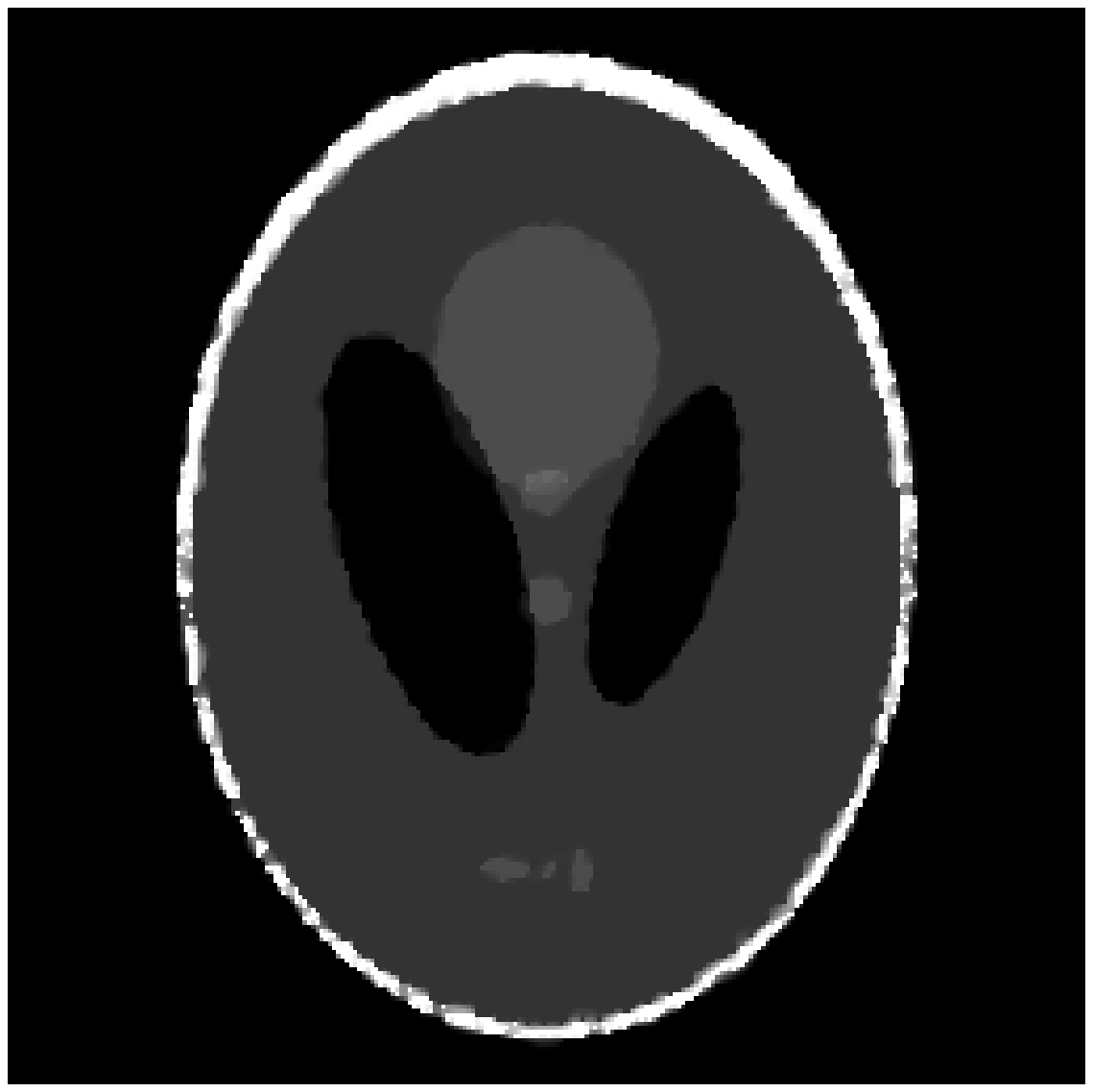}}     \\ 

\subfigure[]{\label{fig:edgeImg_i}\includegraphics[width=0.24\textwidth]{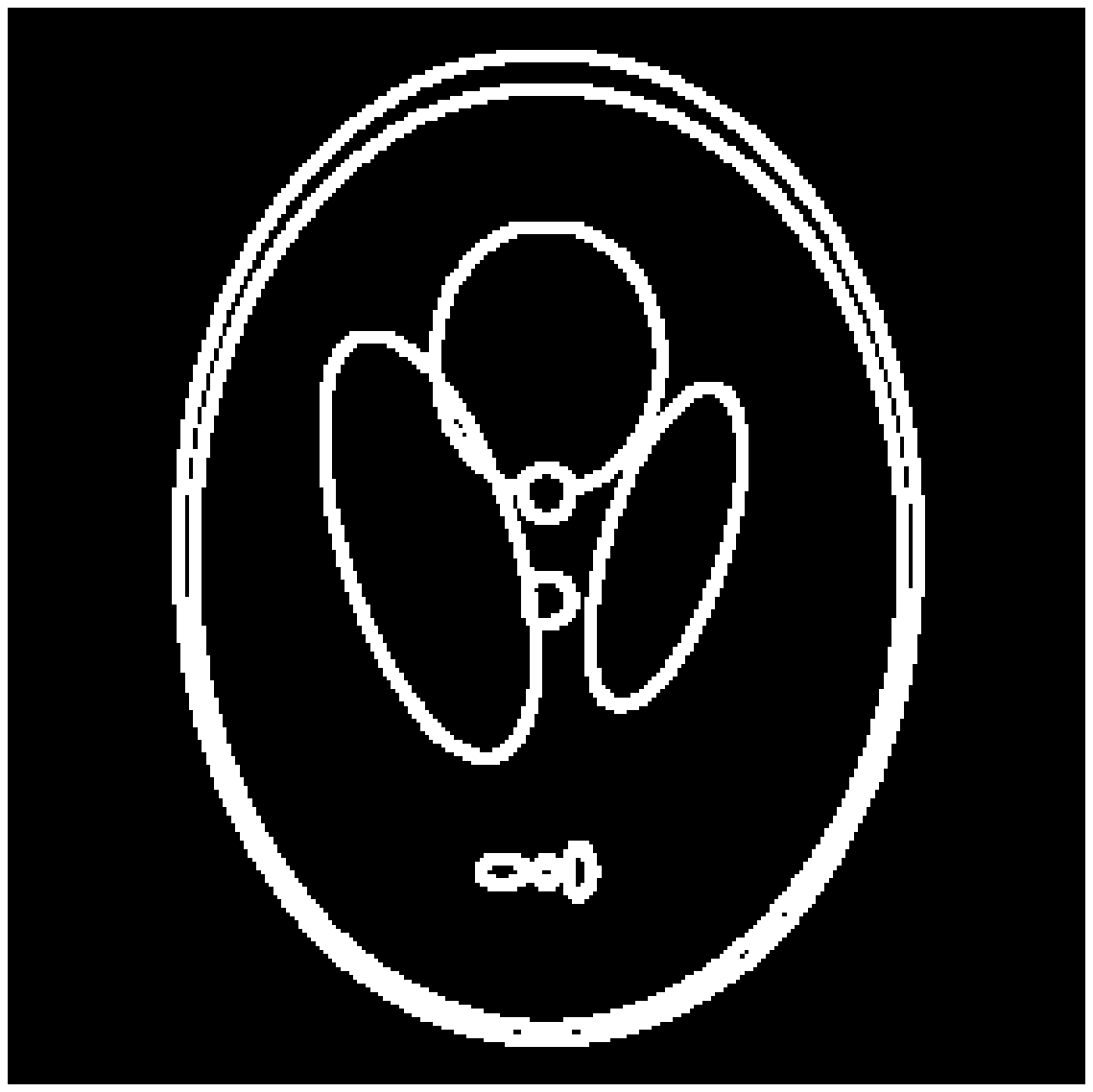}}
\subfigure[]{\label{fig:edgeImg_j}\includegraphics[width=0.24\textwidth]{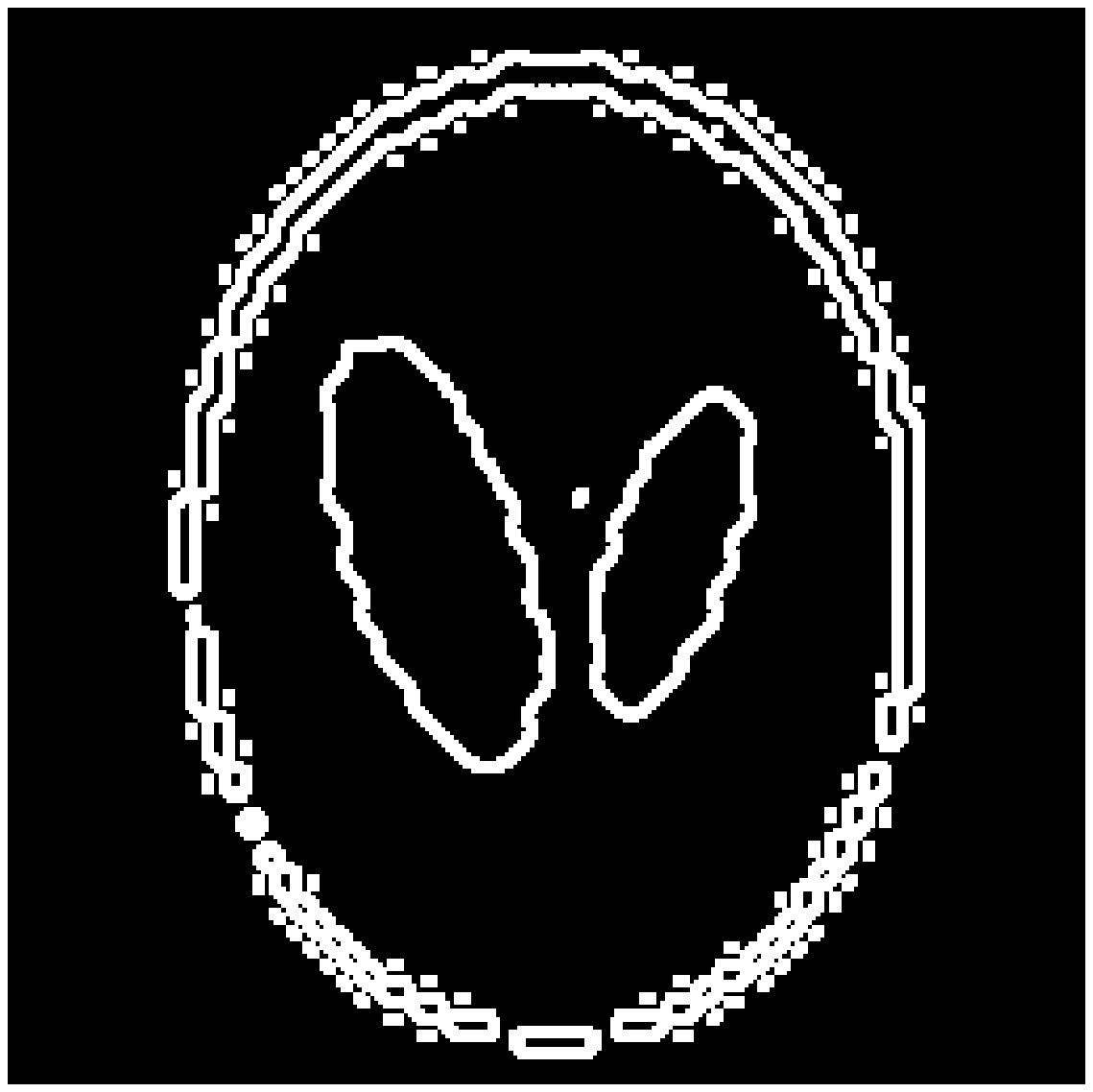}}
\subfigure[]{\label{fig:edgeImg_k}\includegraphics[width=0.24\textwidth]{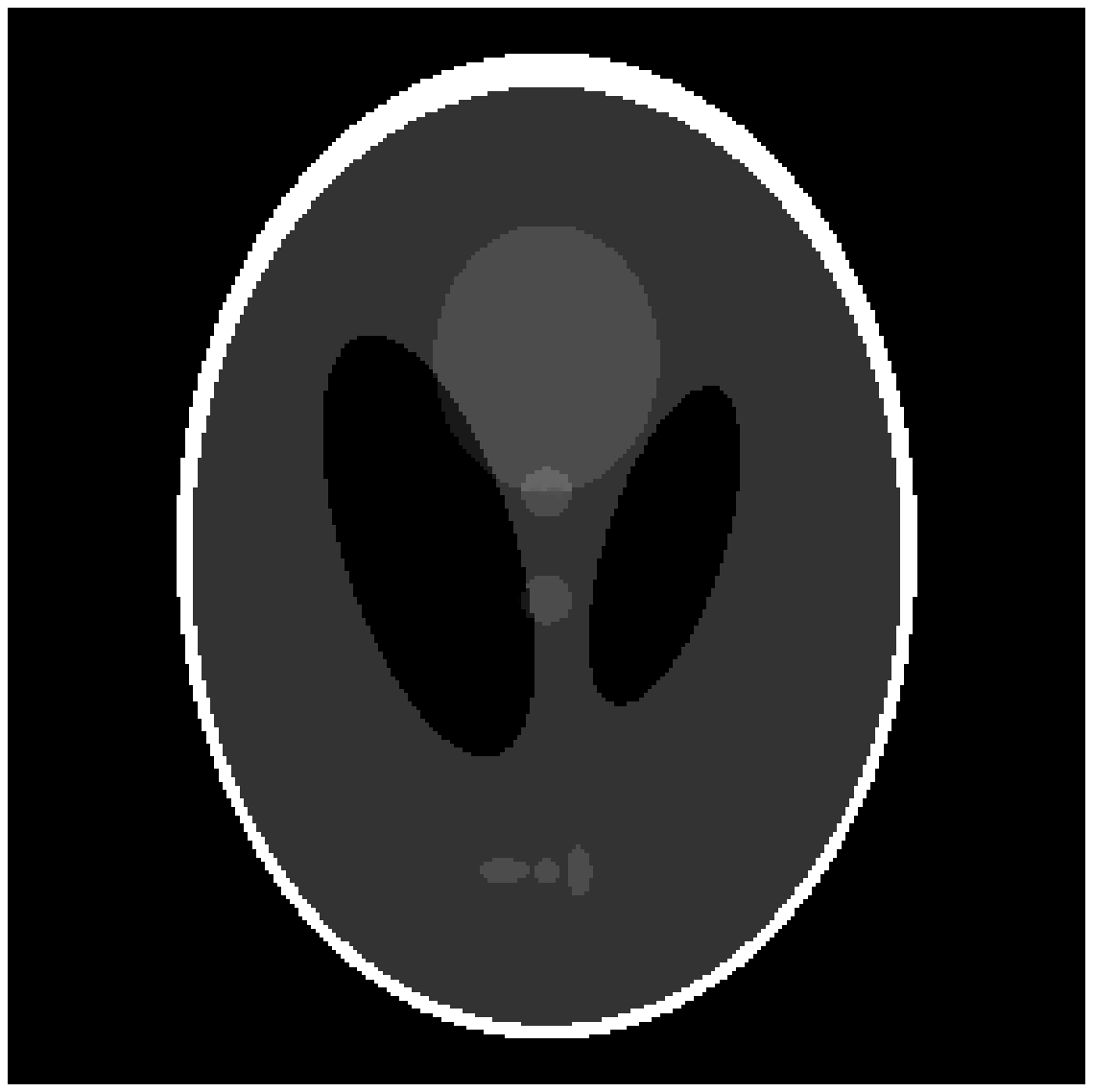}}
\subfigure[]{\label{fig:edgeImg_l}\includegraphics[width=0.24\textwidth]{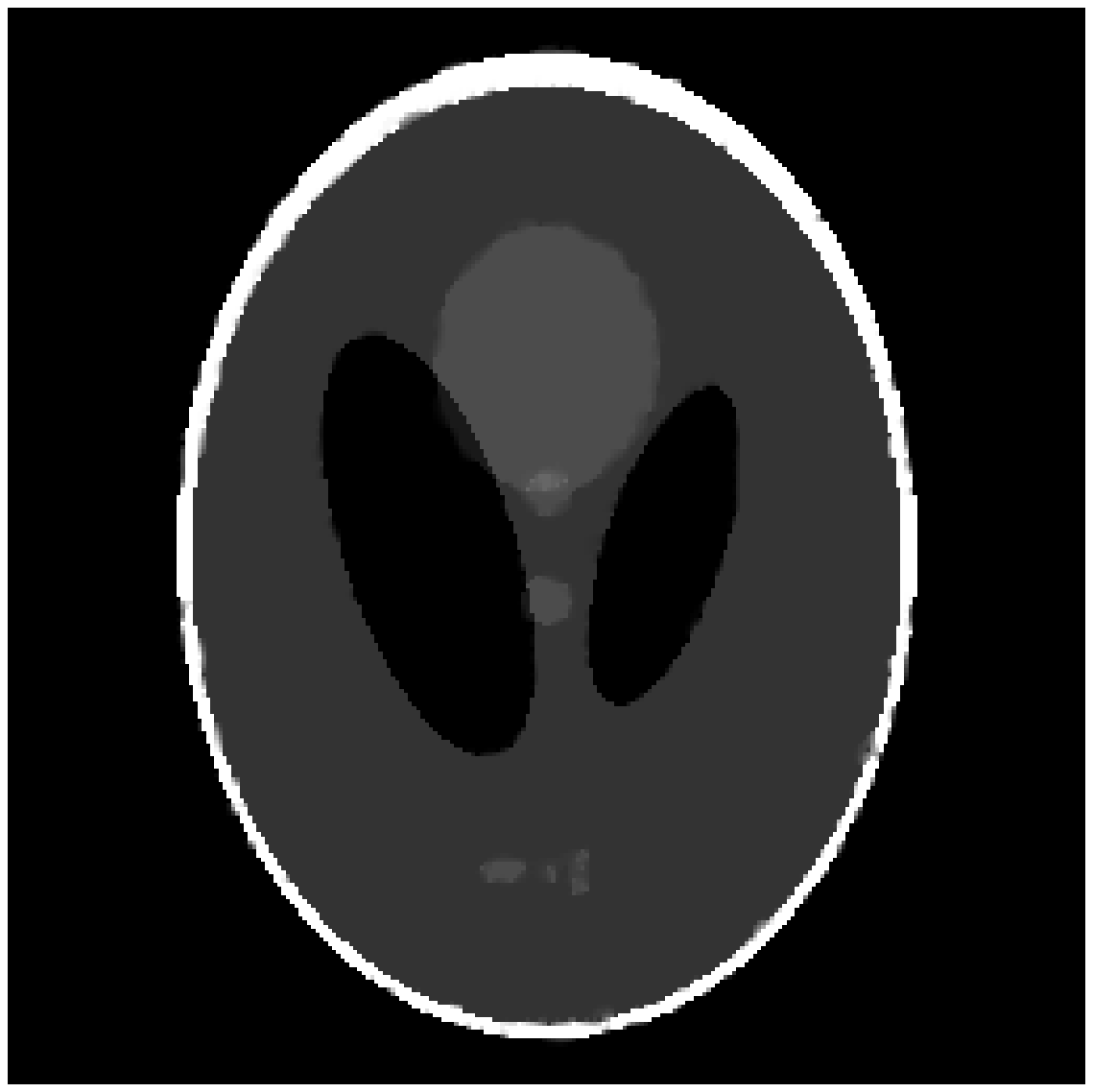}} \\ 

\subfigure[]{\label{fig:edgeImg_m}\includegraphics[width=0.24\textwidth]{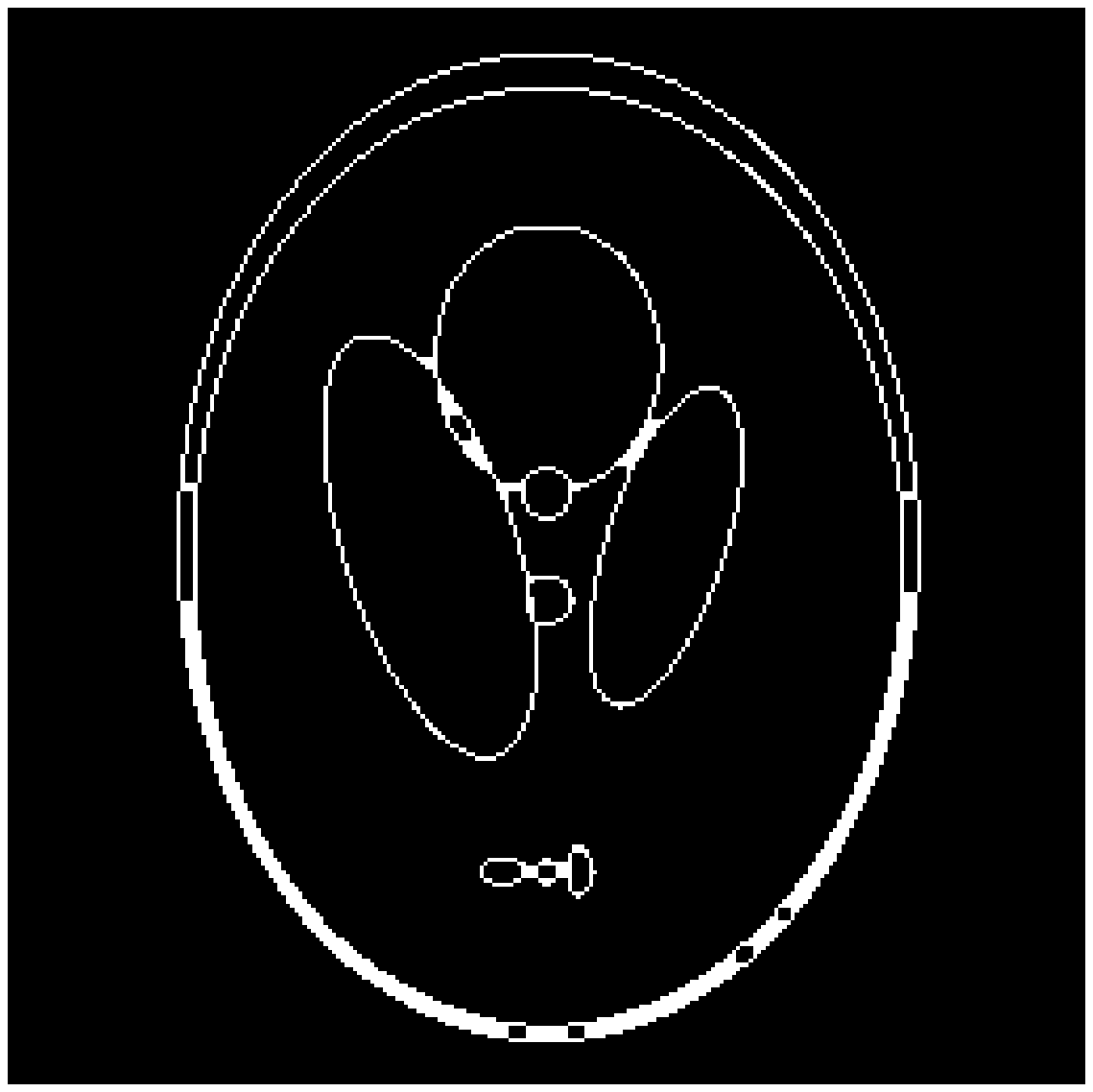}}
\subfigure[]{\label{fig:edgeImg_n}\includegraphics[width=0.24\textwidth]{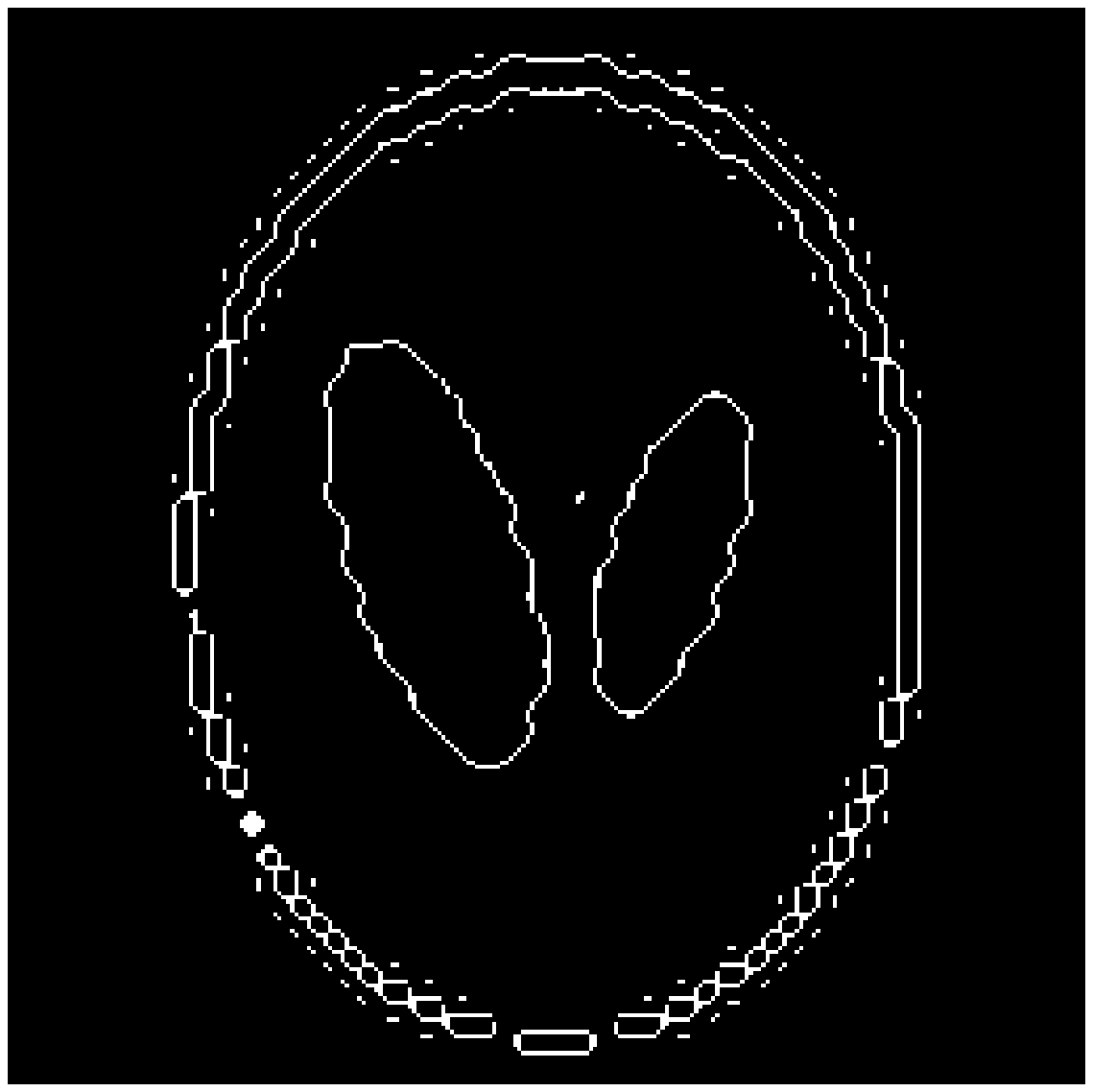}}
\subfigure[]{\label{fig:edgeImg_o}\includegraphics[width=0.24\textwidth]{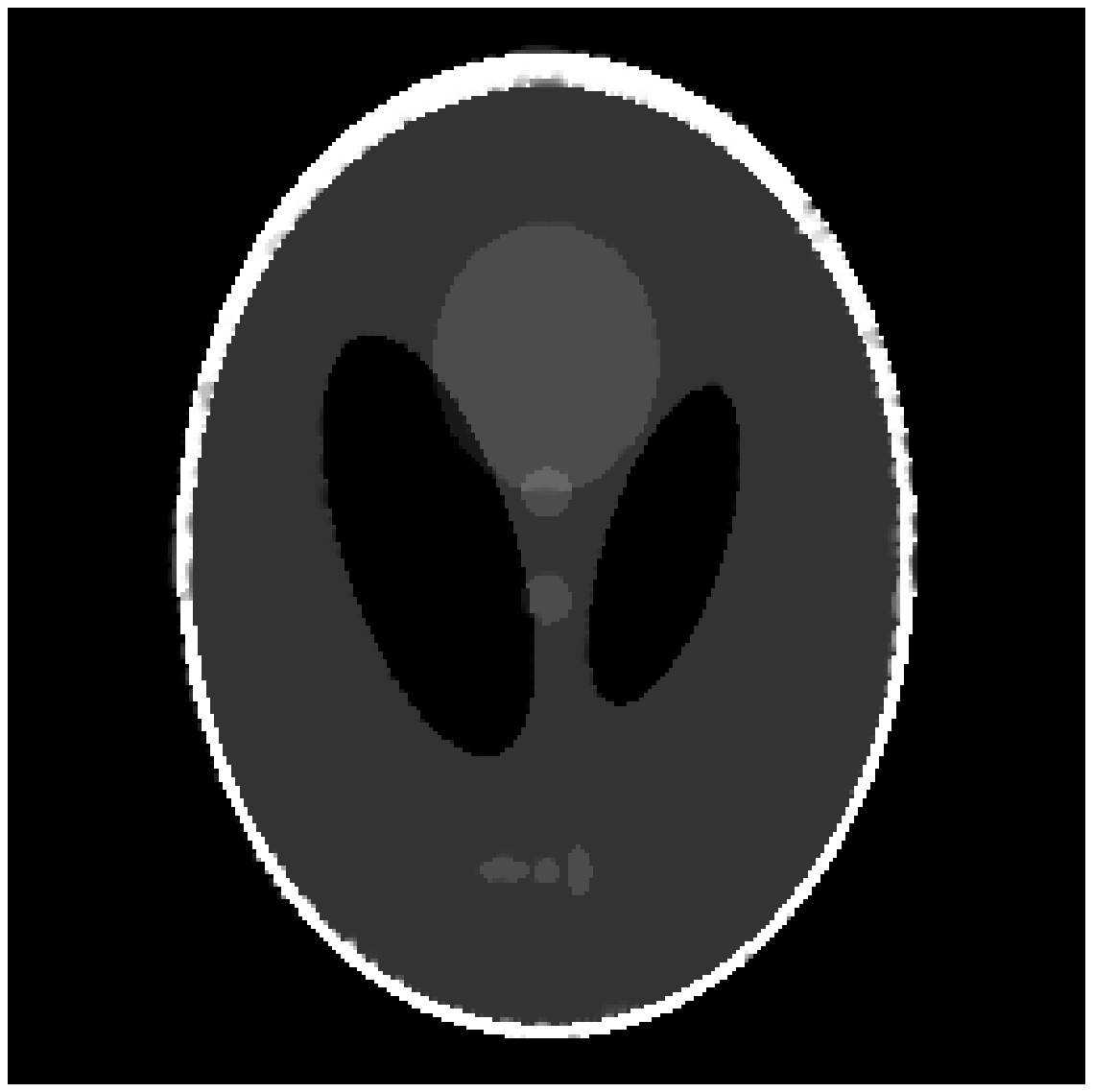}}
\subfigure[]{\label{fig:edgeImg_p}\includegraphics[width=0.24\textwidth]{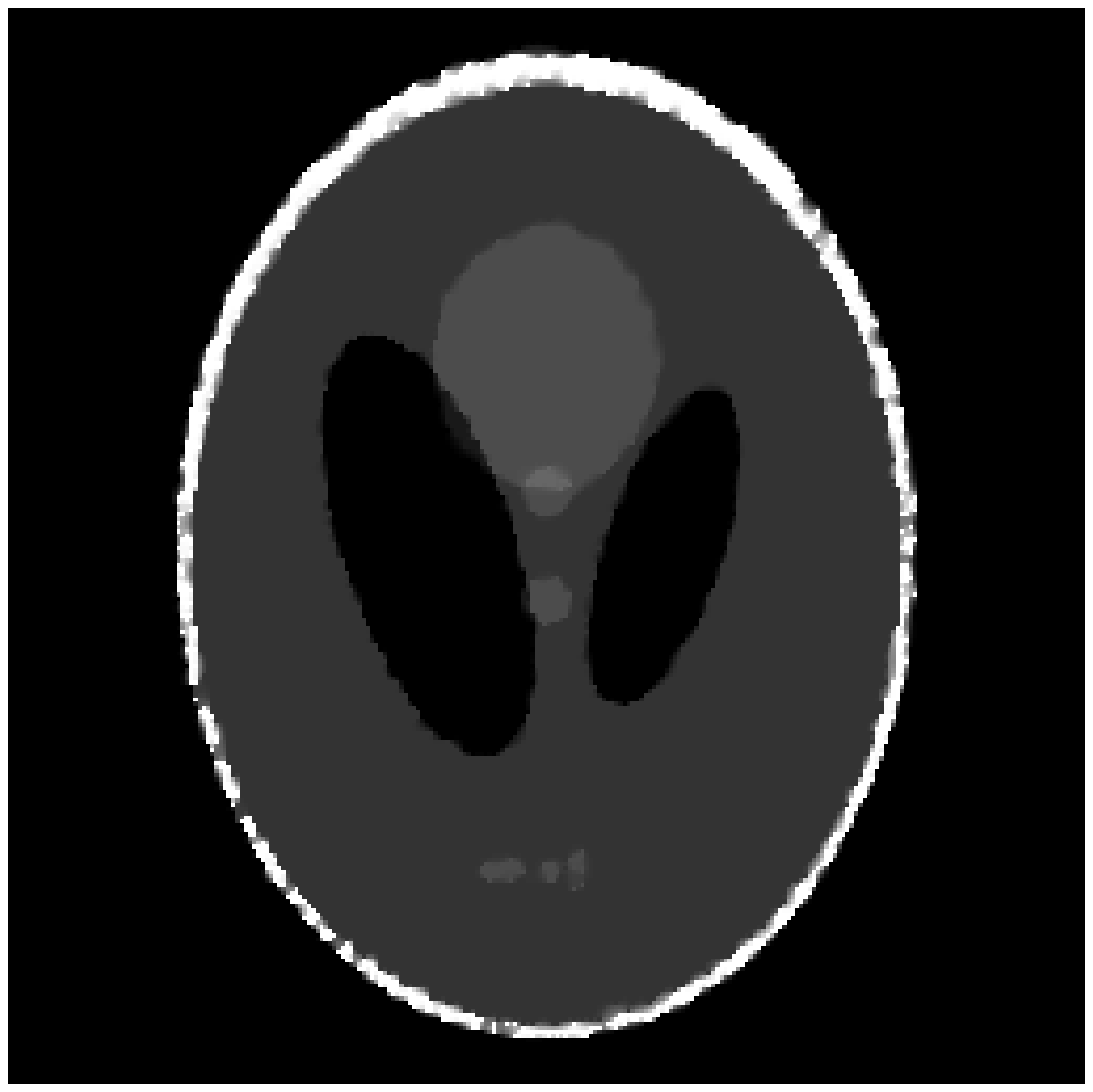}} \\ 

\caption{Reconstructed Phantom images with sensing ratios of $30.24\%$, $29.41\%$, $29.16\%$ $29.85\%$, $29.94\%$, $28.87\%$ and $28.01\%$ respectively for $S_r$, $S_m+M_n$, $S_m+M_n^p$, $S_m+M_d$, $S_m+M_d^p$, $S_m+M_c$ and $S_m+M_c^p$. (a) The original Phantom image; (b) The predicted image $f_p$; (c) Low-resolution sampling pattern $S_l$; (d) $S_r$ sensing: $21.7653$ dB; (e) $M_n$ for edges of (a); (f) $M_n^p$ for edges of (b); (g) $S_m+M_n$ sensing: $30.1195$ dB; (h) $S_m+M_n^p$ sensing: $24.5746$ dB; (i) $M_d$ for edges of (a); (j) $M_d^n$ for edges of (b); (k) $S_m+M_d$ sensing: $65.3853$ dB; (l) $S_m+M_d^p$ sensing: $29.9294$ dB; (m) $M_c$ for edges of (a); (n) $M_c^p$ for edges of (b); (o) $S_m+M_c$ sensing: $31.2032$ dB; (p) $S_m+M_c^p$ sensing: $24.1066$ dB}
\label{fig_edgeImg}
\end{figure}

\begin{figure} [htbp]
\centering
\scriptsize
\subfigure[]{\label{fig:compareImages_a}\includegraphics[width=0.24\textwidth]{fig3_OrigPhantom.eps}}
\subfigure[]{\label{fig:compareImages_b}\includegraphics[width=0.24\textwidth]{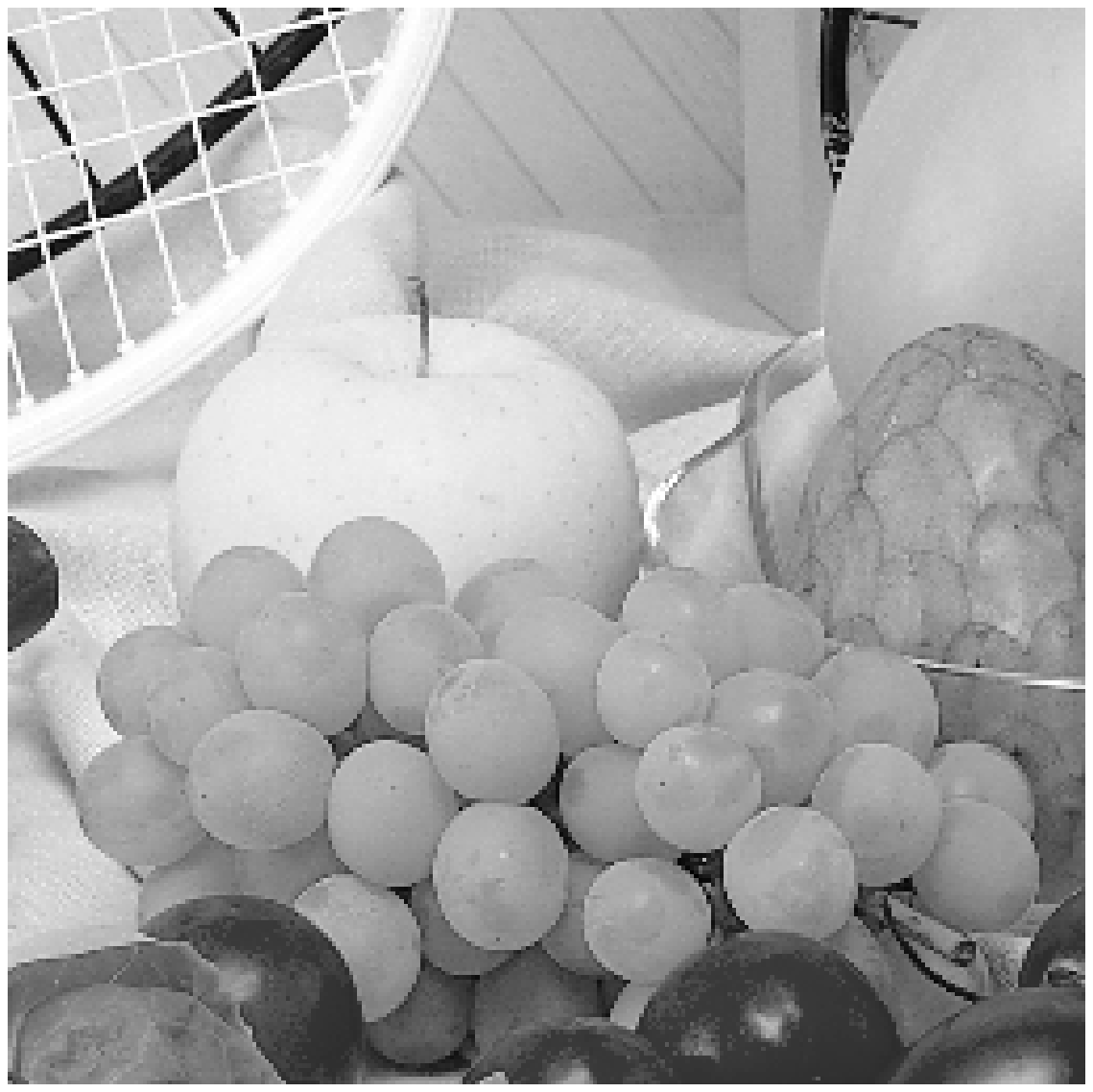}}
\subfigure[]{\label{fig:compareImages_c}\includegraphics[width=0.24\textwidth]{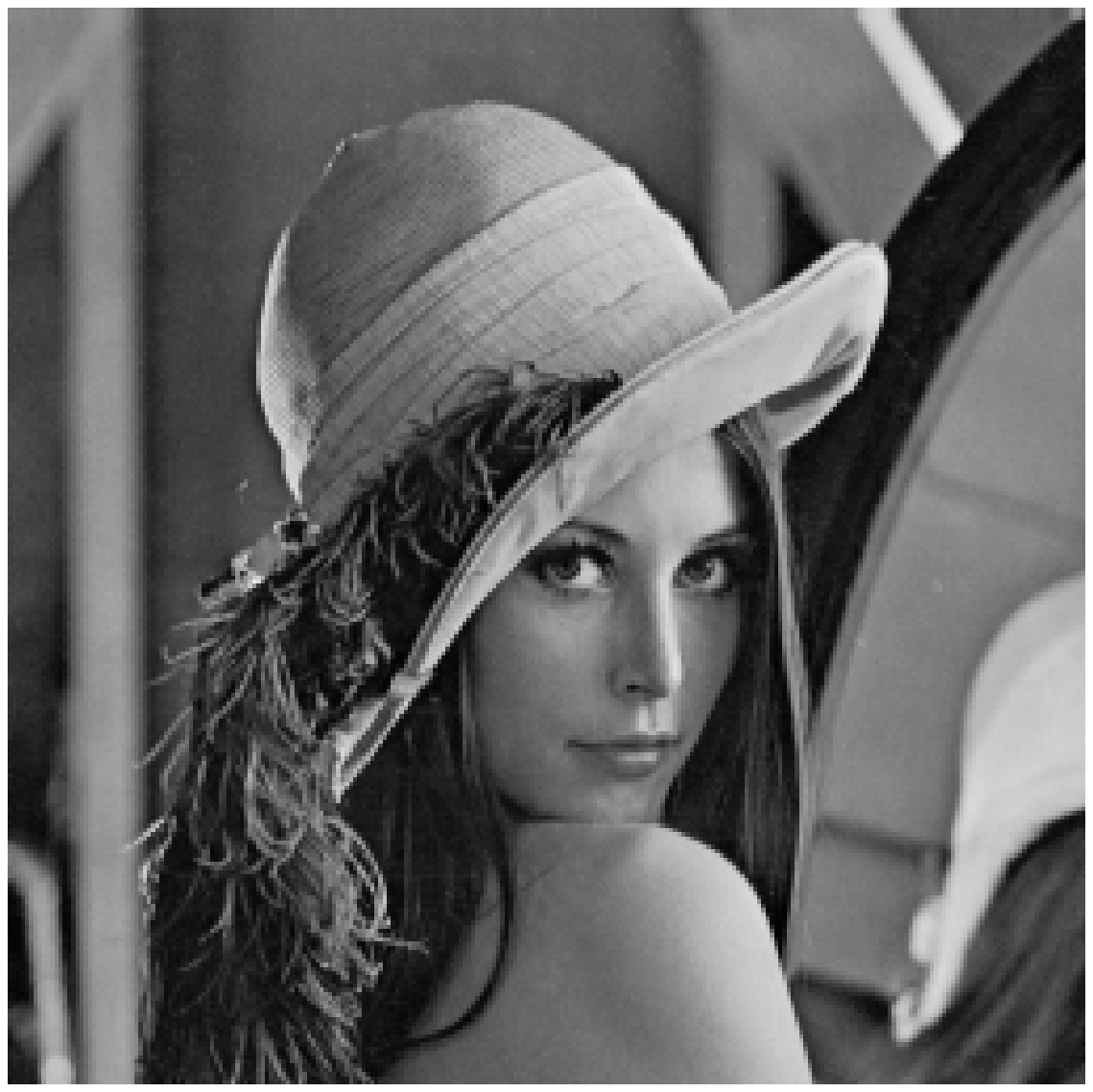}}
\subfigure[]{\label{fig:compareImages_d}\includegraphics[width=0.24\textwidth]{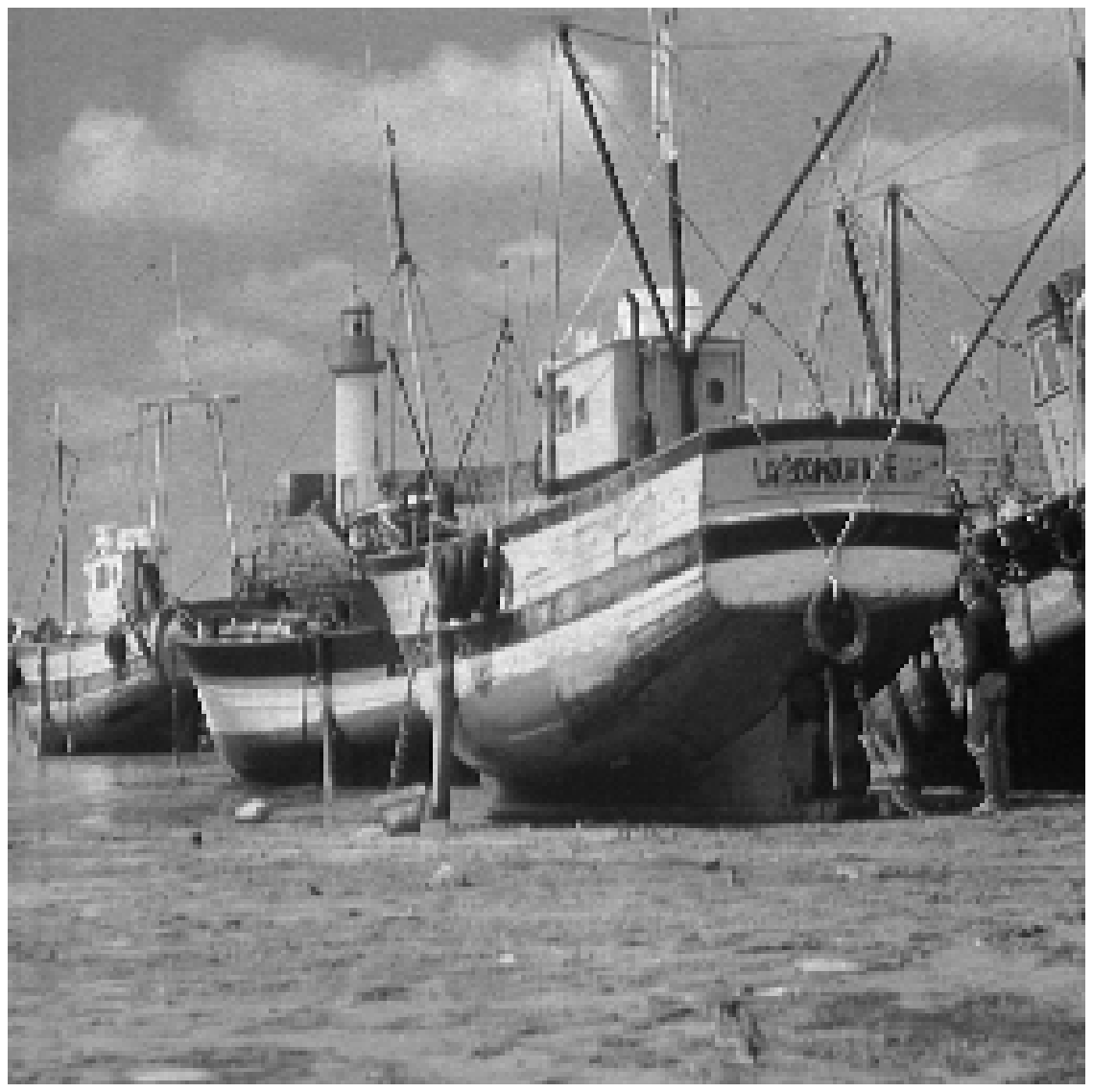}}     \\ 

\subfigure[]{\label{fig:compareImages_e}\includegraphics[width=0.24\textwidth]{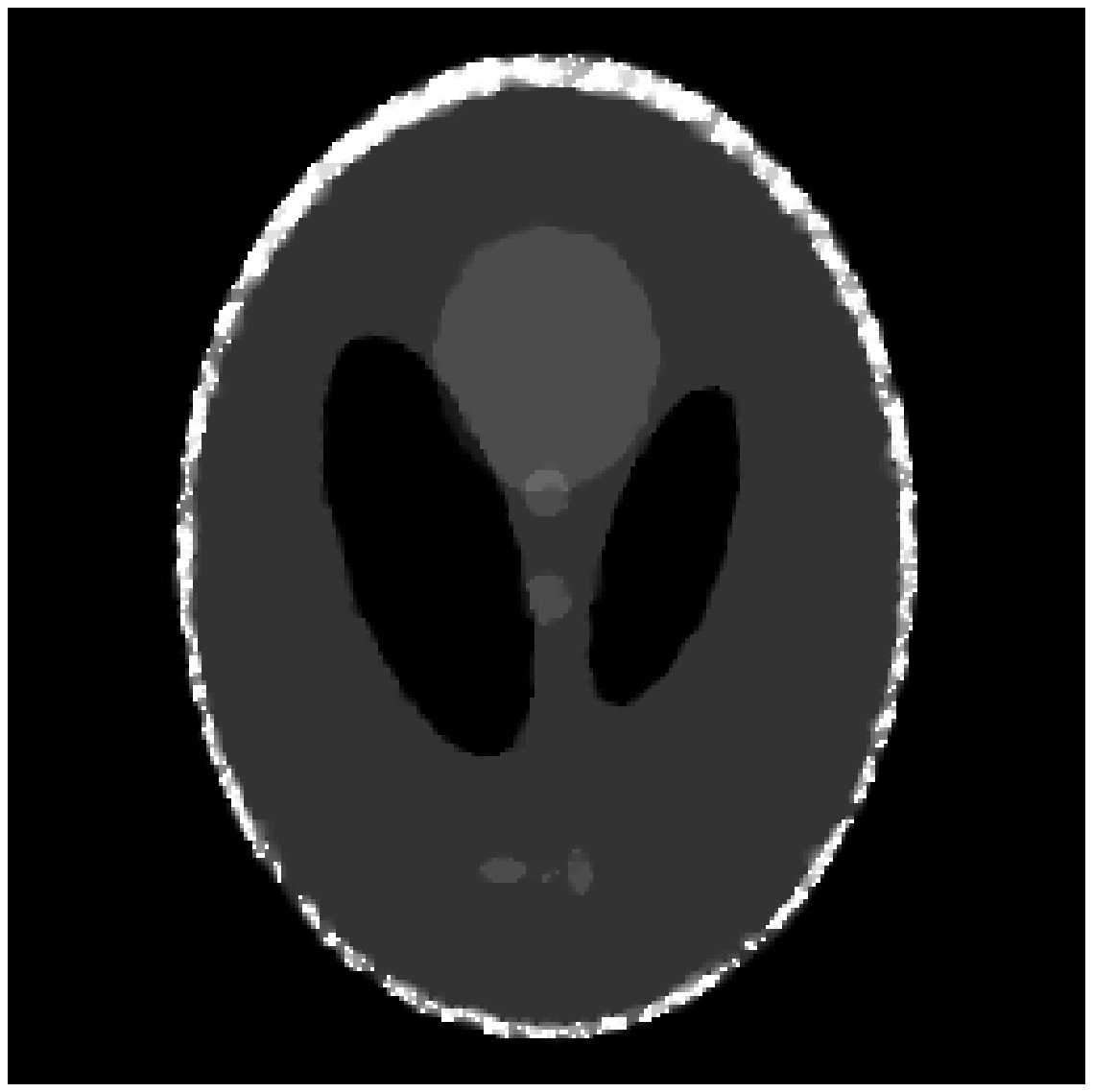}}
\subfigure[]{\label{fig:compareImages_f}\includegraphics[width=0.24\textwidth]{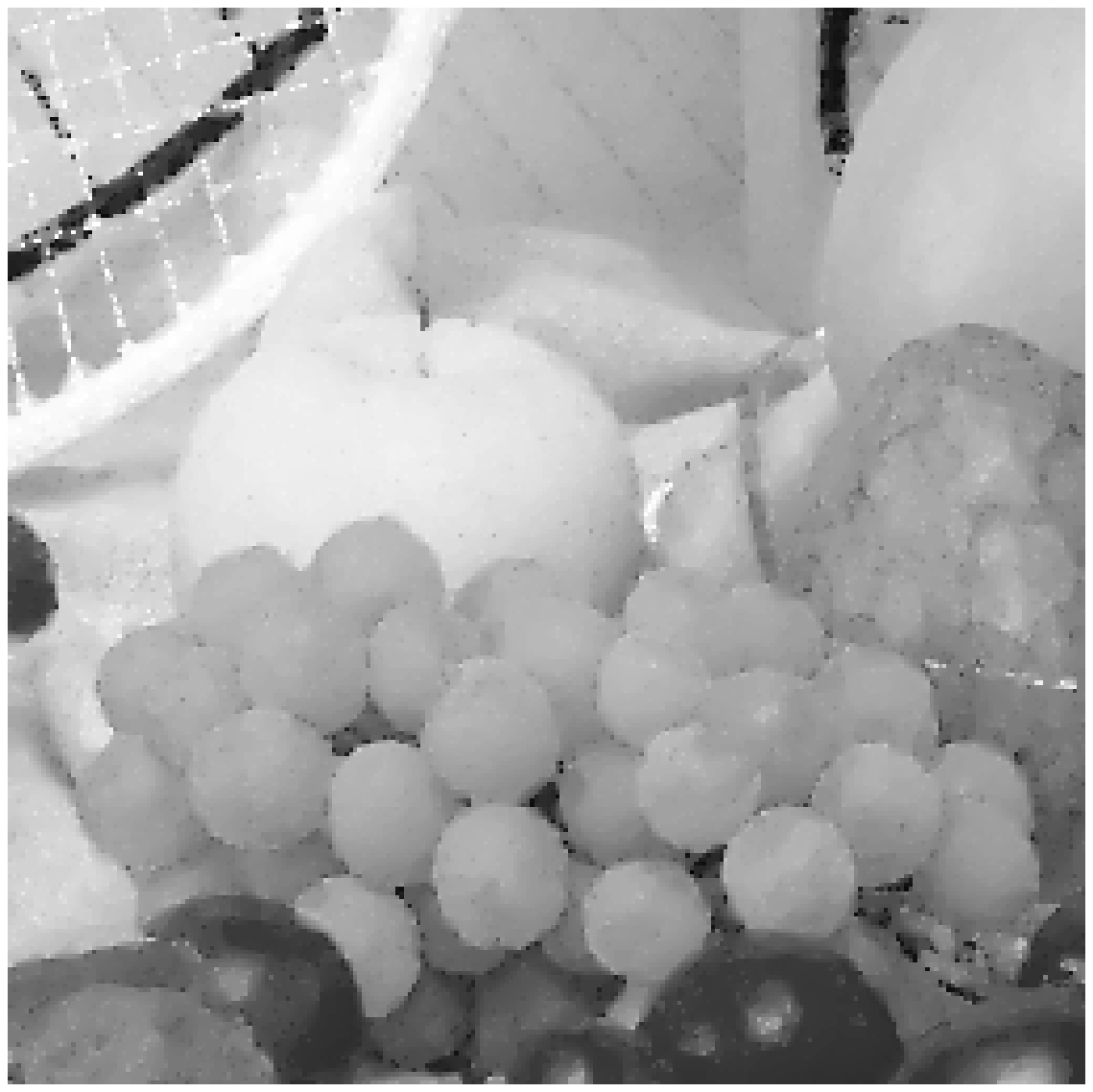}}
\subfigure[]{\label{fig:compareImages_g}\includegraphics[width=0.24\textwidth]{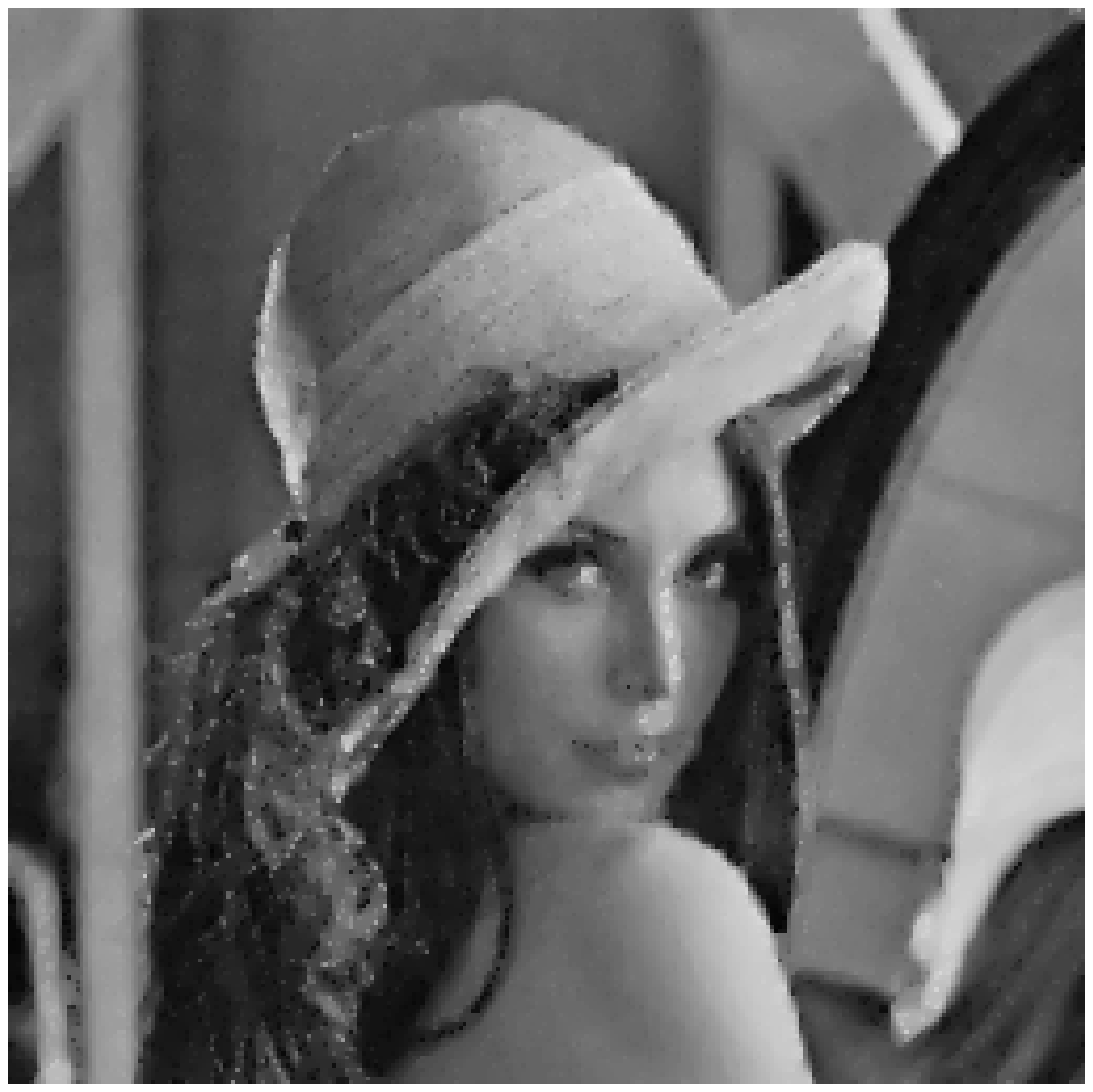}}
\subfigure[]{\label{fig:compareImages_h}\includegraphics[width=0.24\textwidth]{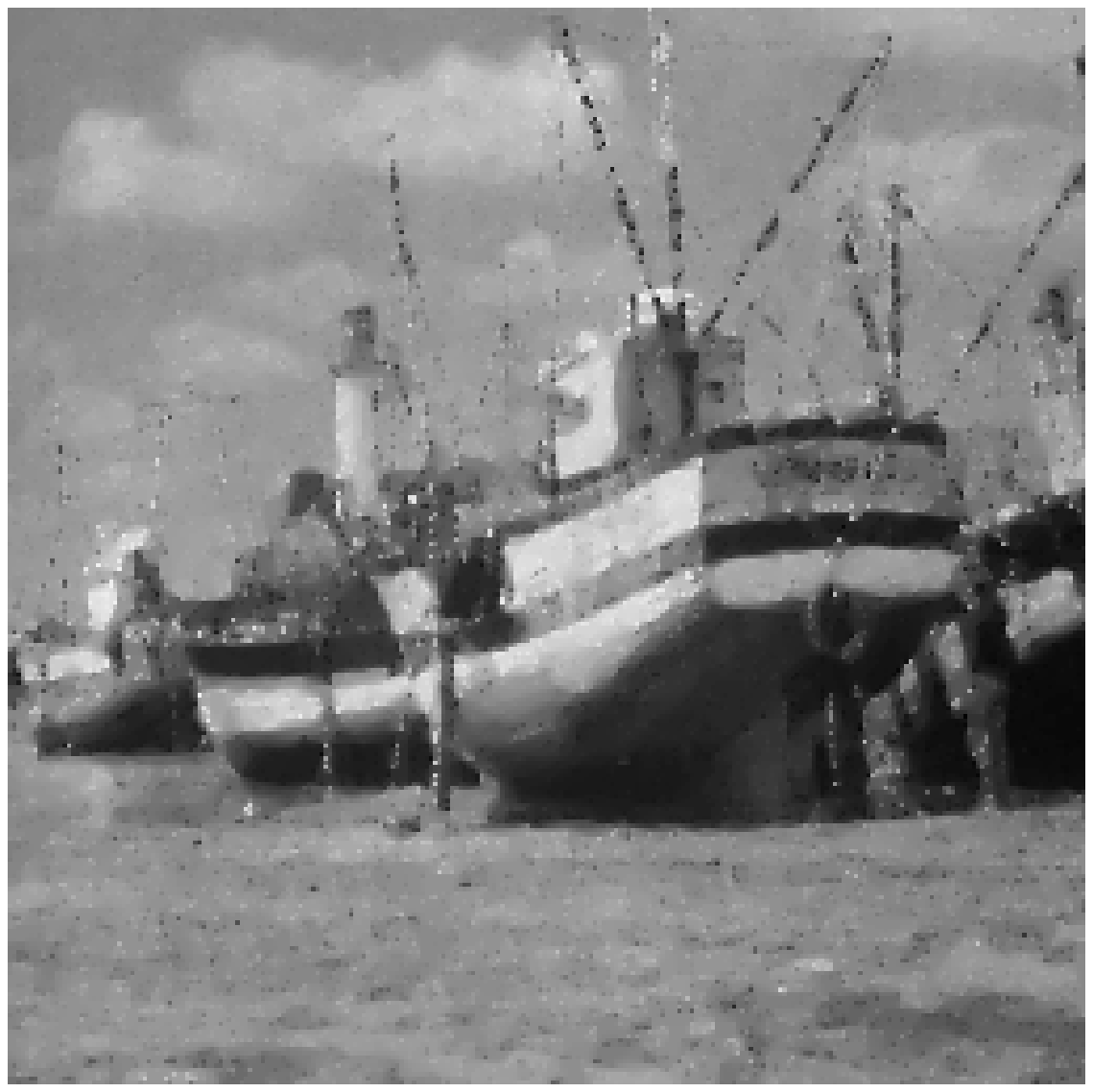}}  \\ 

\subfigure[]{\label{fig:compareImages_i}\includegraphics[width=0.24\textwidth]{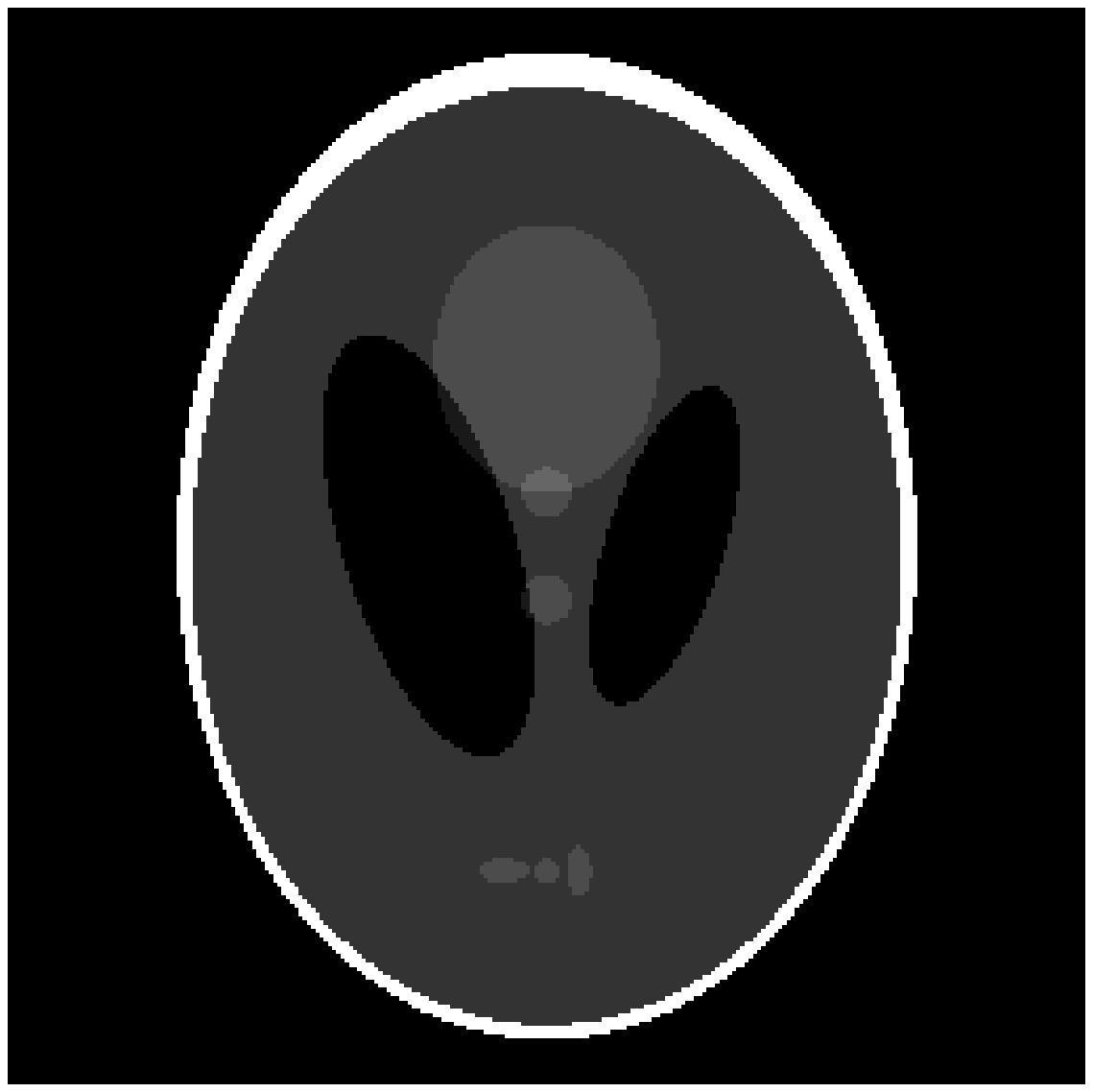}}
\subfigure[]{\label{fig:compareImages_j}\includegraphics[width=0.24\textwidth]{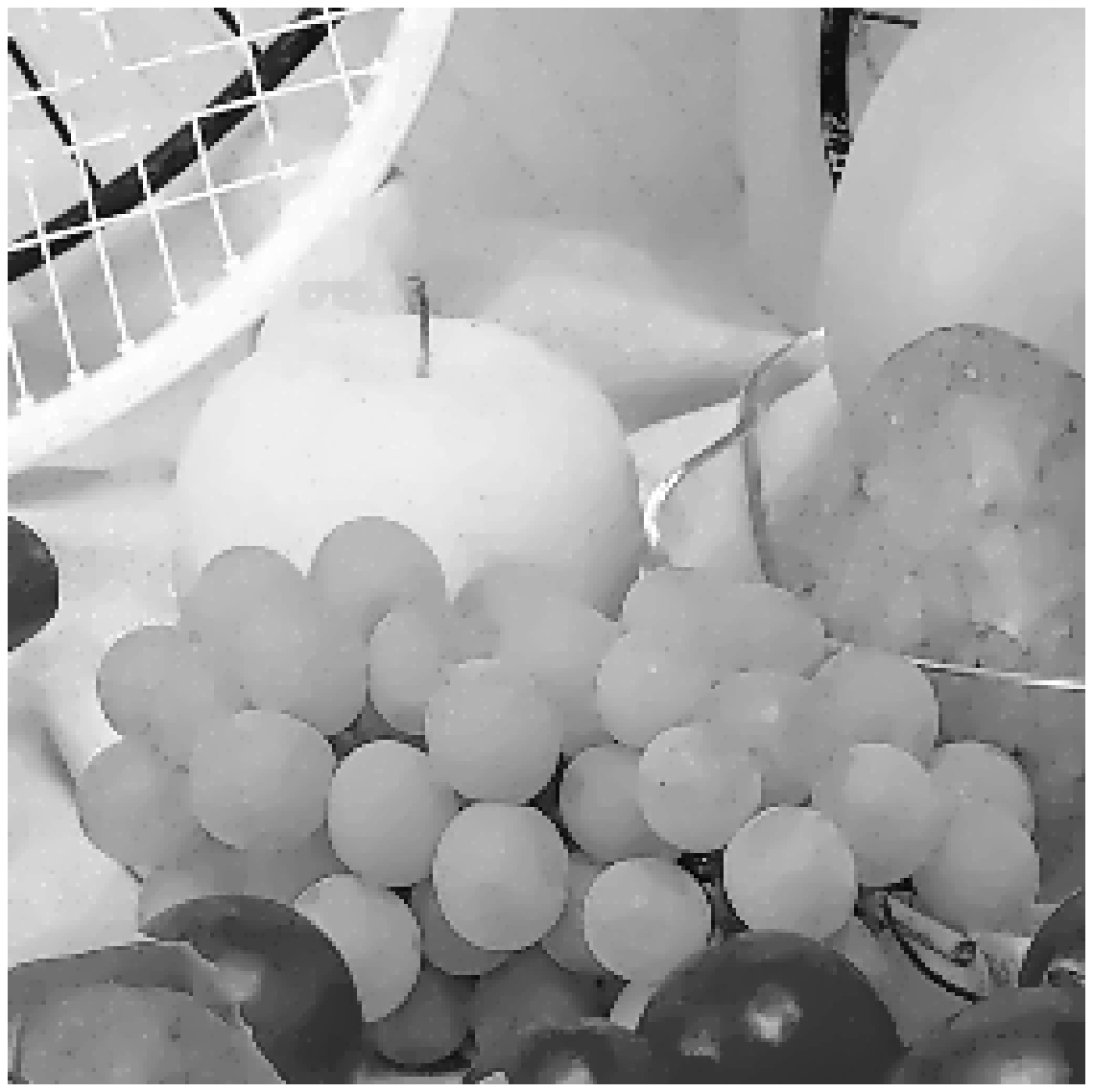}}
\subfigure[]{\label{fig:compareImages_k}\includegraphics[width=0.24\textwidth]{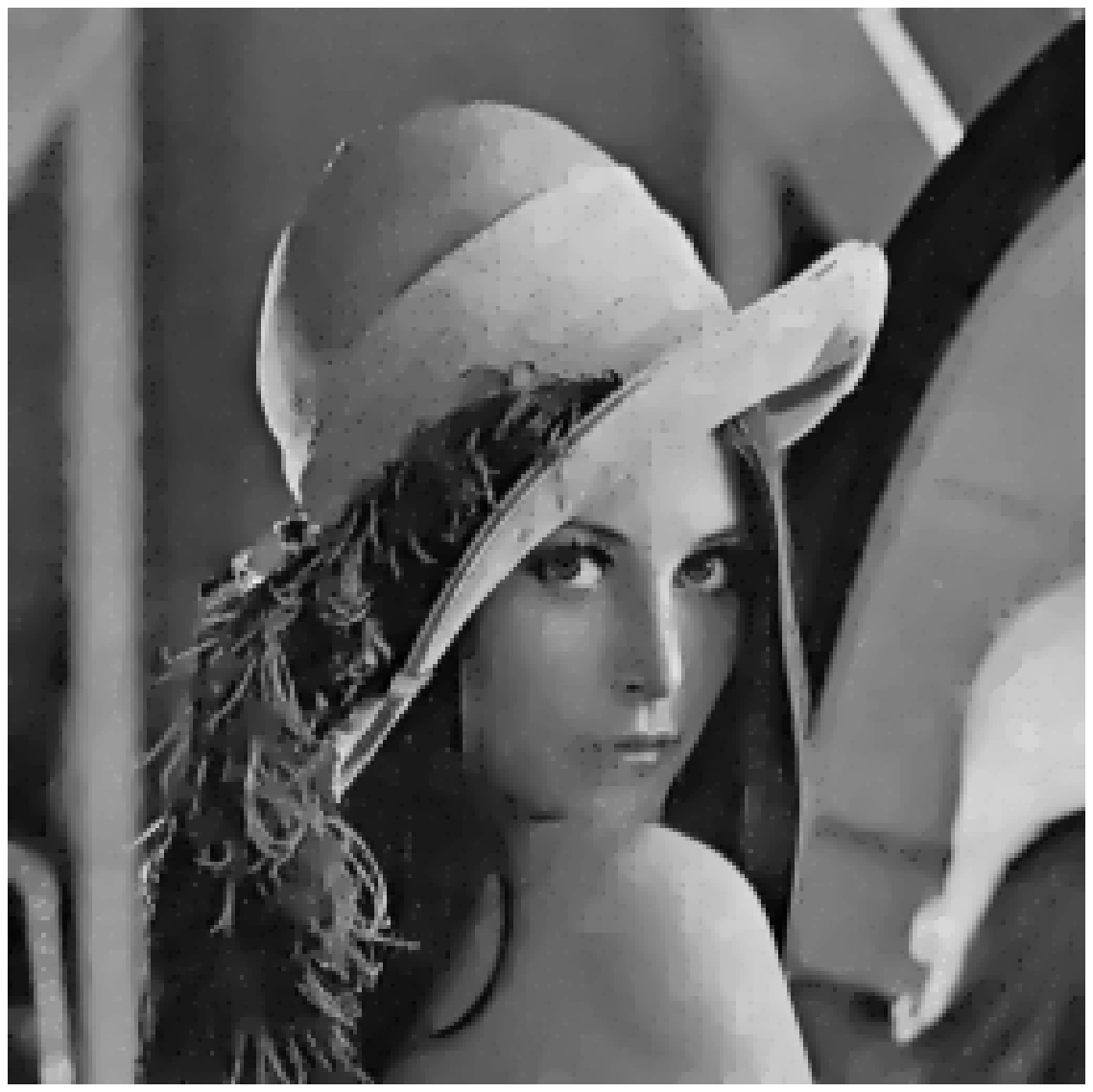}}
\subfigure[]{\label{fig:compareImages_l}\includegraphics[width=0.24\textwidth]{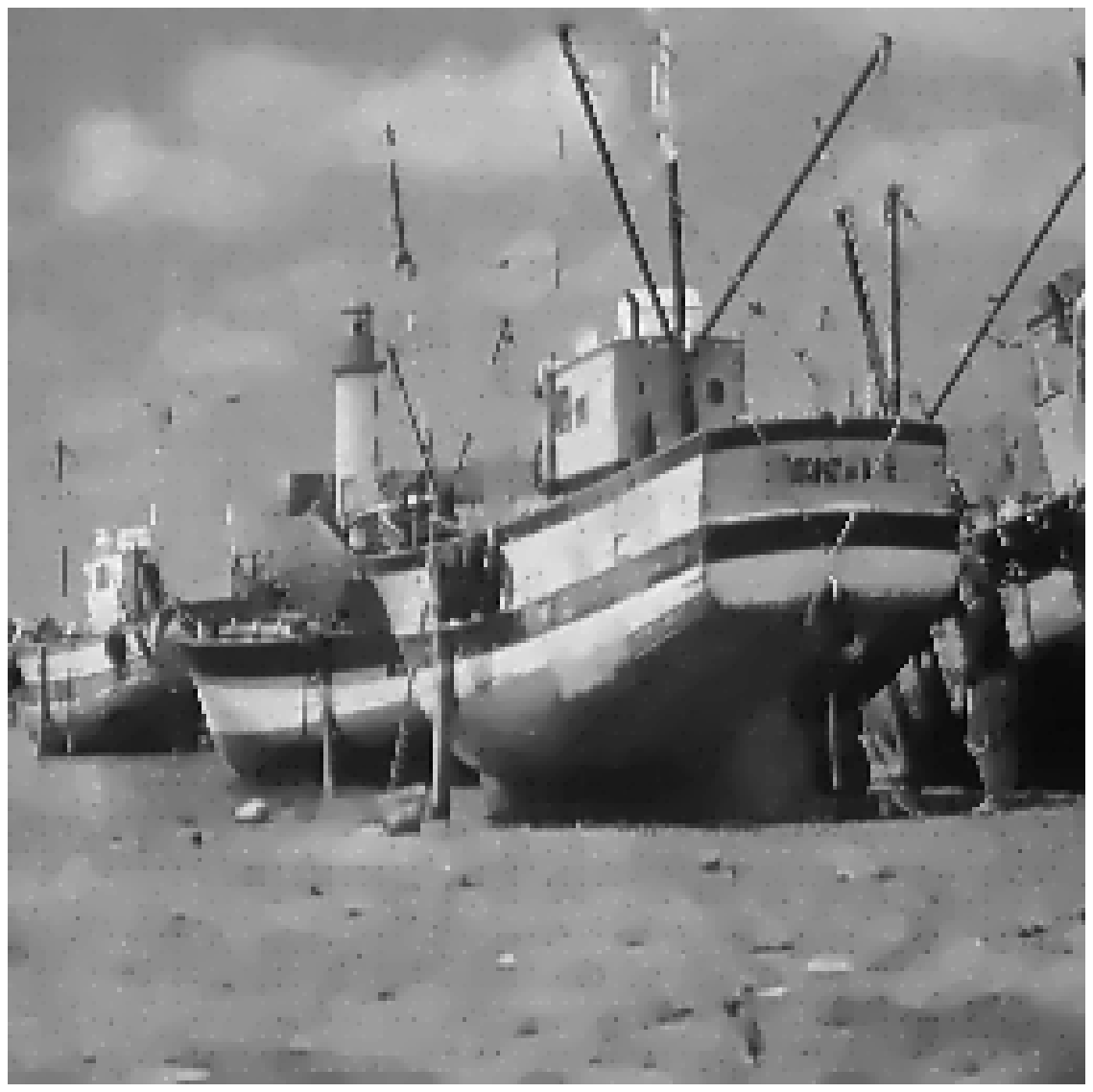}}    \\ 

\subfigure[]{\label{fig:compareImages_m}\includegraphics[width=0.24\textwidth]{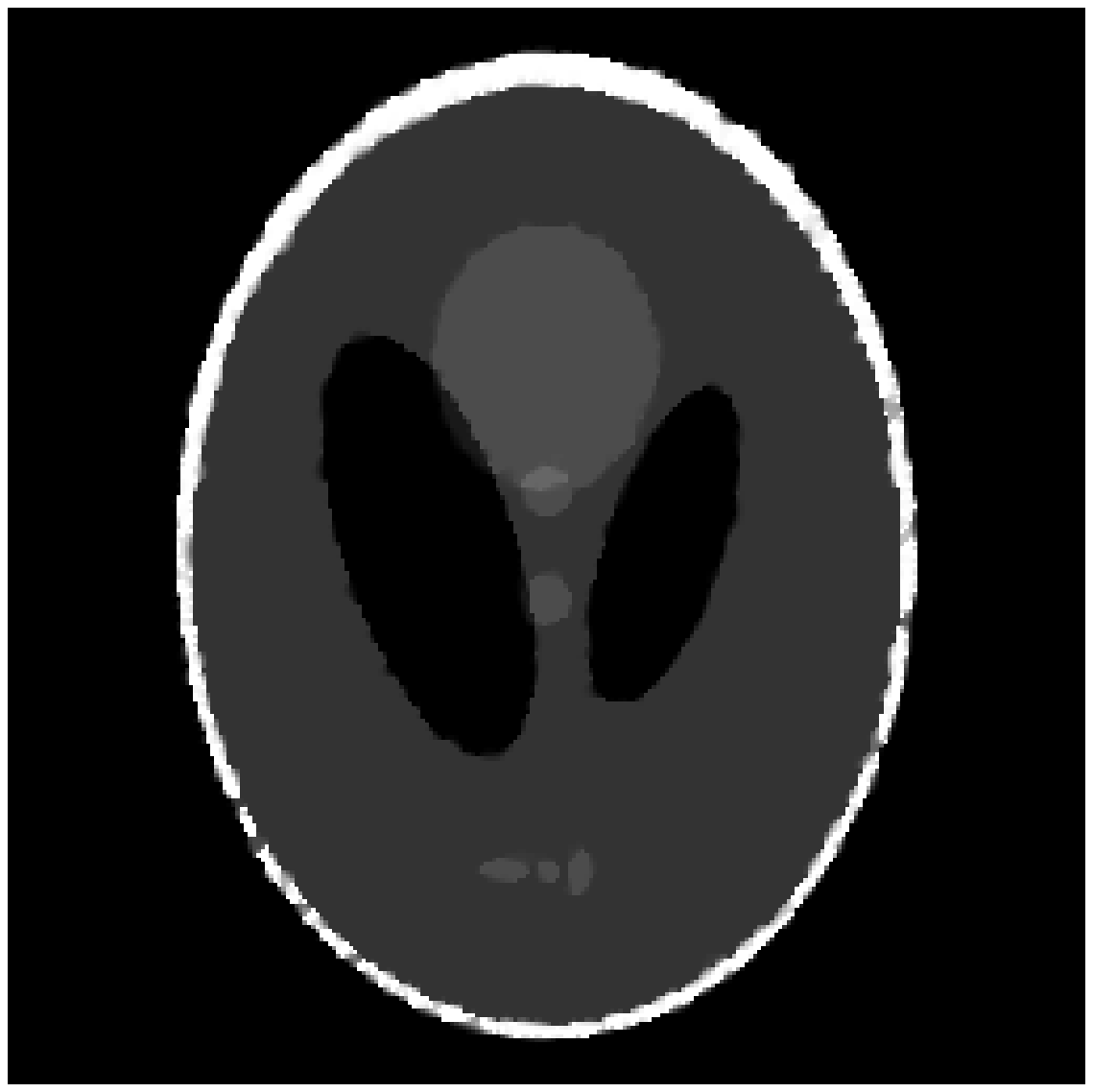}}
\subfigure[]{\label{fig:compareImages_n}\includegraphics[width=0.24\textwidth]{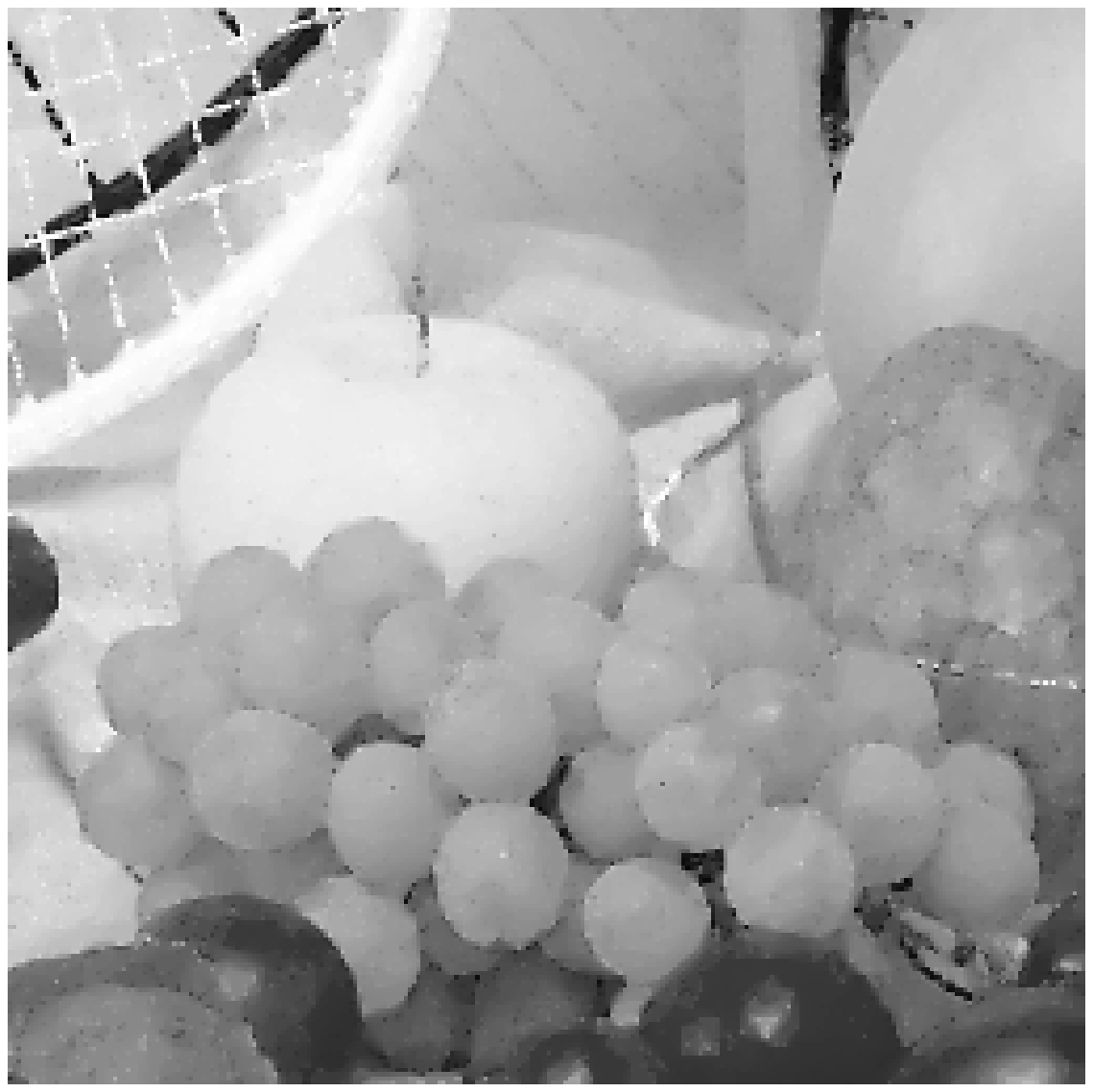}}
\subfigure[]{\label{fig:compareImages_o}\includegraphics[width=0.24\textwidth]{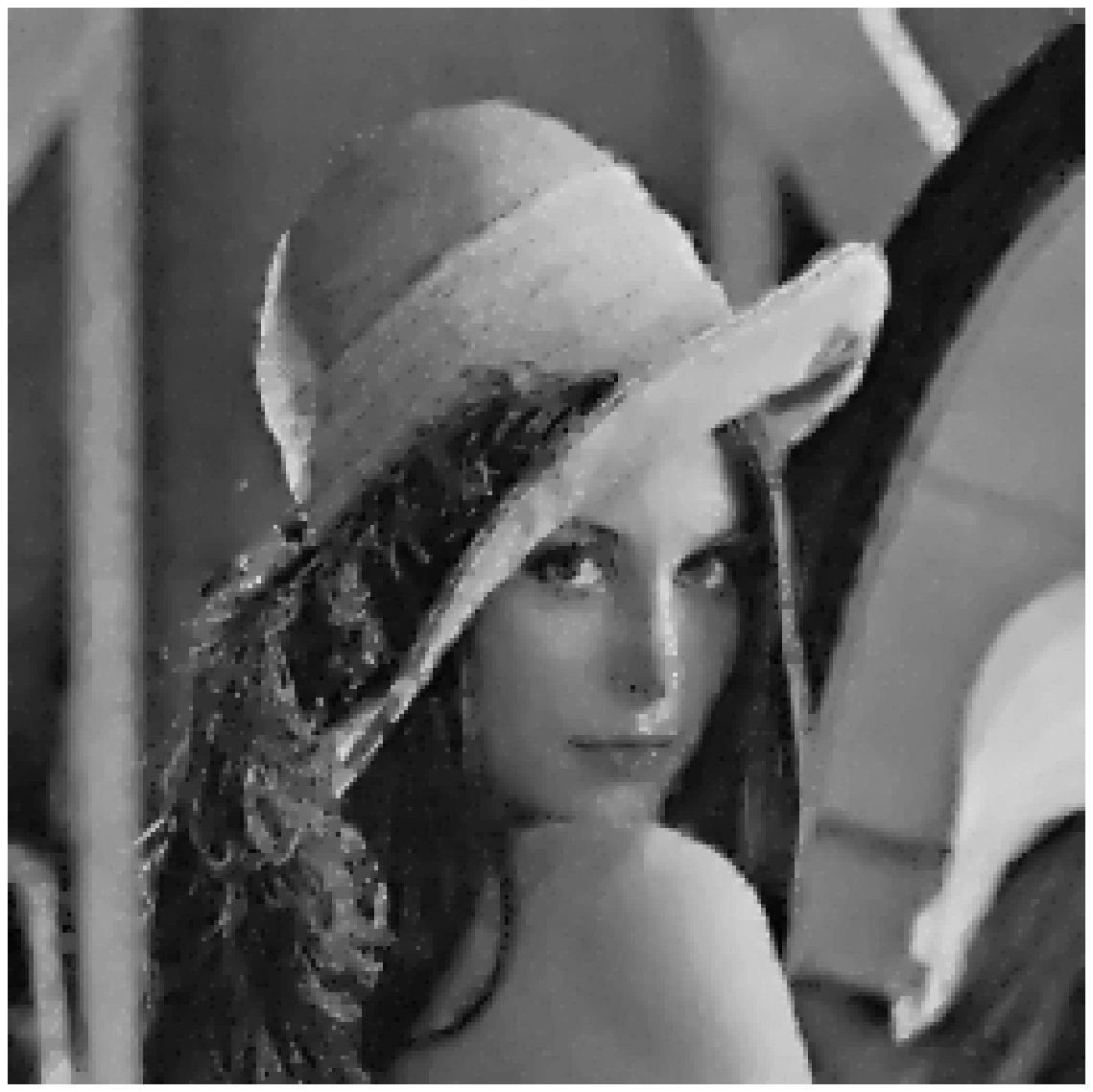}}
\subfigure[]{\label{fig:compareImages_p}\includegraphics[width=0.24\textwidth]{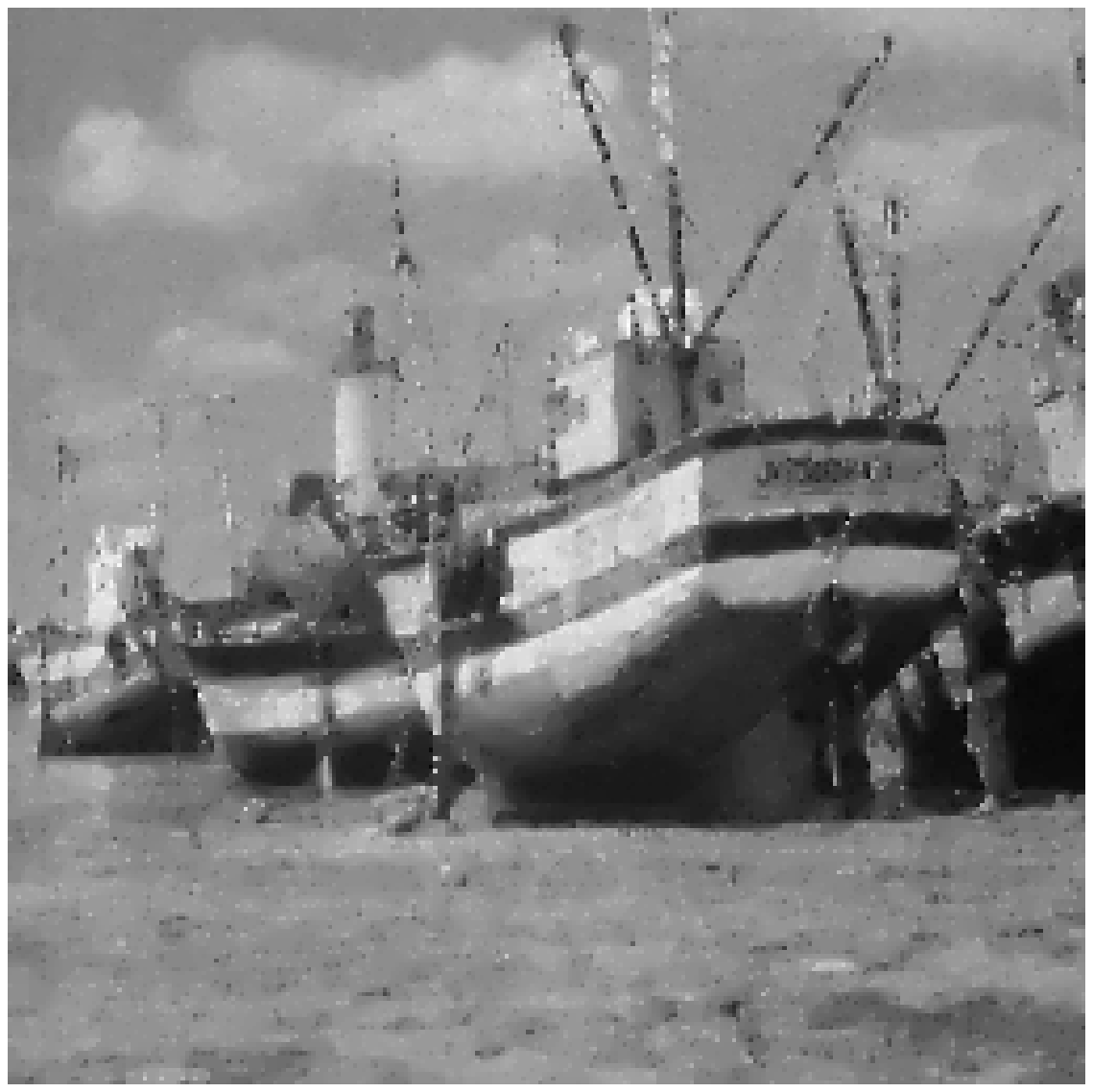}} \\ 

\caption{ Reconstructed Phantom, Fruits, Lena and Boat images using $S_r$, $S_m+M_d^p$ and $S_m+M_c^p$ strategies. (a)-(d): The original images; (e)-(h), (i)-(l) and (m)-(p) are corresponding reconstructed images for $S_r$, $S_m+M_d^p$ and $S_m+M_c^p$, respectively. The corresponding key parameters are listed in Table \ref{table_compare}.}
\label{fig_compareImages}
\end{figure}

 Fig. \ref{fig_edgePSNR} shows the peak signal-to-noise ratio (PSNR) as a function of the sensing ratio $\eta_1$. We observe: (1) the convergence of all the methods are comparable; (2) the performance of $S_m$ sensing is much better than that of $S_r$; (3) the best PSNR is achieved by the $S_m$ sensing involving the dilated edge information. This also suggests the pixels around edges contain very important information of image. Fig. \ref{fig_edgeImg} shows the sensing performance of the Phantom. After comparing Figs. \ref{fig_edgeImg}(g,k,o) to Figs. \ref{fig_edgeImg}(h,l,p), $S_m+M_{n,d,c}$ shows better recovery results than $S_m+M_{n,d,c}^p$. Instead of setting the parameter $\eta_2$ directly, in Fig. 2 and Fig. 3 we select the $1.75\% \times 256 \times 256$ edge pixels combined with the $64\times64$ low resolution samples as the adaptive samples for each $\eta_1$. So the parameter $\eta_2$ in Fig. 2 or Fig. 3 is not fixed for the different $\eta_1$s. However, the MAR sensing matrix incorporating predicted edges by $M_{n,d,c}^p$ operations still achieves high PSNR values (such as $29.93$ dB with the sensing ratio $\eta_1=29.94\%$) in contrast to the completely random sensing matrix ($21.77$ dB with the sensing ratio $\eta_1=30.24\%$). Using standard images, Fig. \ref{fig_compareImages} and Table \ref{table_compare} demonstrate significant advantages of the MAR sensing matrix not only on PSNR but also on structural similarity (SSIM).

\begin{table}[!t]
\scriptsize
\renewcommand{\arraystretch}{1.3}
\caption{Reconstruction performance using the MAR sensing matrix $S_m$ and completely random sensing matirx $S_r$.}
\label{table_compare}
\centering
\begin{tabular}{cccccc}
\hline
\multicolumn{2}{c}{Image}              & PSNR    & SSIM    & $\eta_1$ & $\eta_2$  \\
\hline
\multirow{3}{*}{Phantom} & $S_r$       & 21.7653 & 0.9486  &30.24\%   &         \\
                         & $S_m+M_d^p$ & 72.4857 & 1.0000  &29.81\%   & 43.76\% \\
                         & $S_m+M_c^p$ & 25.5346 & 0.9712  &30.24\%   & 26.64\% \\
\hline

\multirow{3}{*}{Fruits}  & $S_r$       & 25.6011 & 0.8336  &30.46\%   &         \\
                         & $S_m+M_d^p$ & 33.6908 & 0.8767  &30.30\%   & 68.11\%\\
                         & $S_m+M_c^p$ & 27.0509 & 0.8409  &30.28\%   & 23.00\% \\
\hline
\multirow{3}{*}{Lena}    & $S_r$       & 25.5683 & 0.8634  &30.18\%   &         \\
                         & $S_m+M_d^p$ & 31.3728 & 0.8845  &29.82\%   & 68.55\%\\
                         & $S_m+M_c^p$ & 27.5257 & 0.8714  &29.82\%   & 37.38\% \\
\hline
\multirow{3}{*}{Boat}    & $S_r$       & 24.4288 & 0.7433  &29.85\%   &        \\
                         & $S_m+M_d^p$ & 28.7316 & 0.7715  &29.81\%   & 84.69\%\\
                         & $S_m+M_c^p$ & 25.0184 & 0.7513  &29.82\%   & 36.02\% \\
\hline
\end{tabular}
\vspace{-1em}
\end{table}

\begin{figure} [htbp]
\centering
\scriptsize
\subfigure[]{\label{fig:compareImages_ratio2_a}\includegraphics[width=0.32\textwidth]{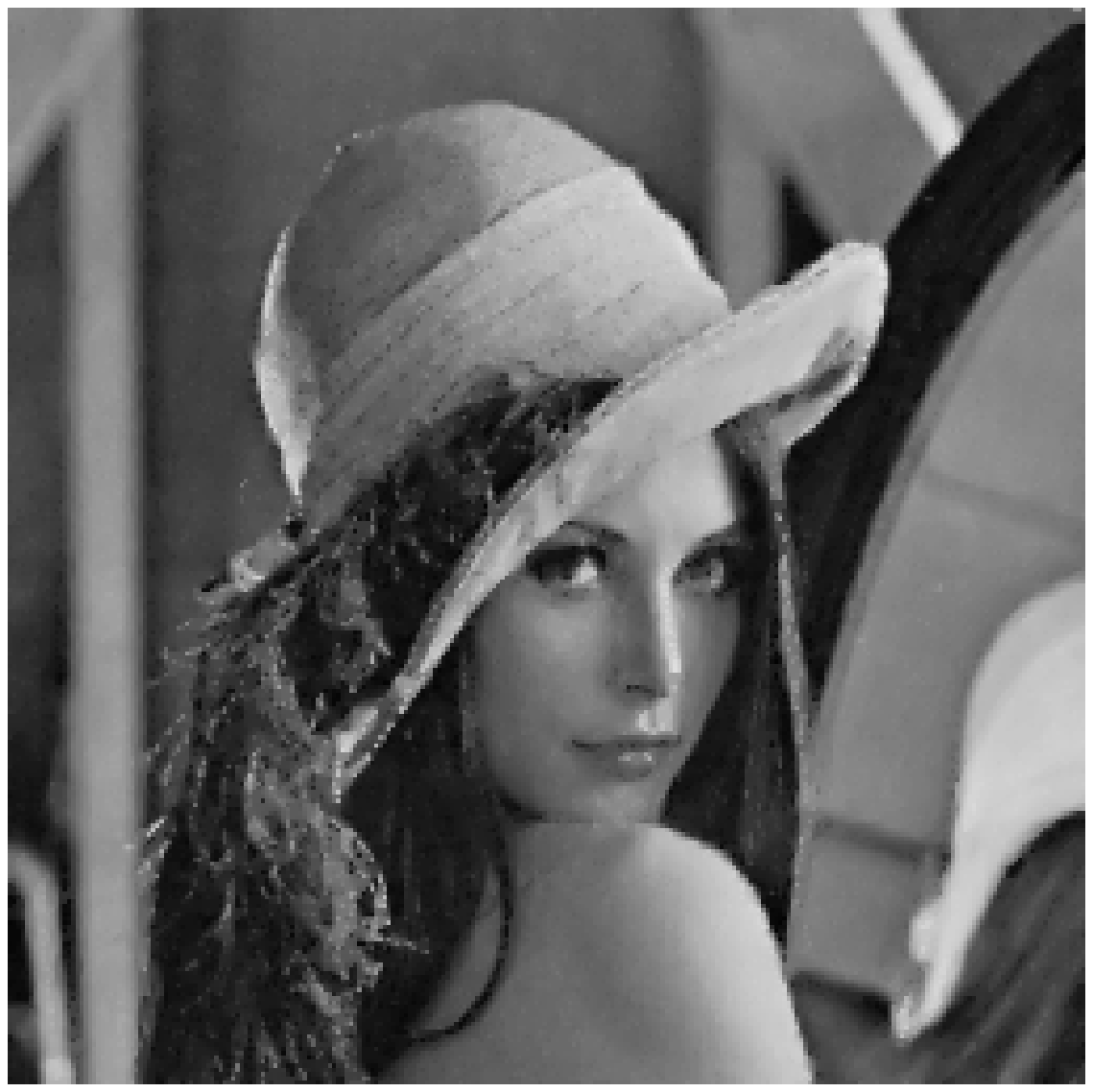}}
\subfigure[]{\label{fig:compareImages_ratio2_b}\includegraphics[width=0.32\textwidth]{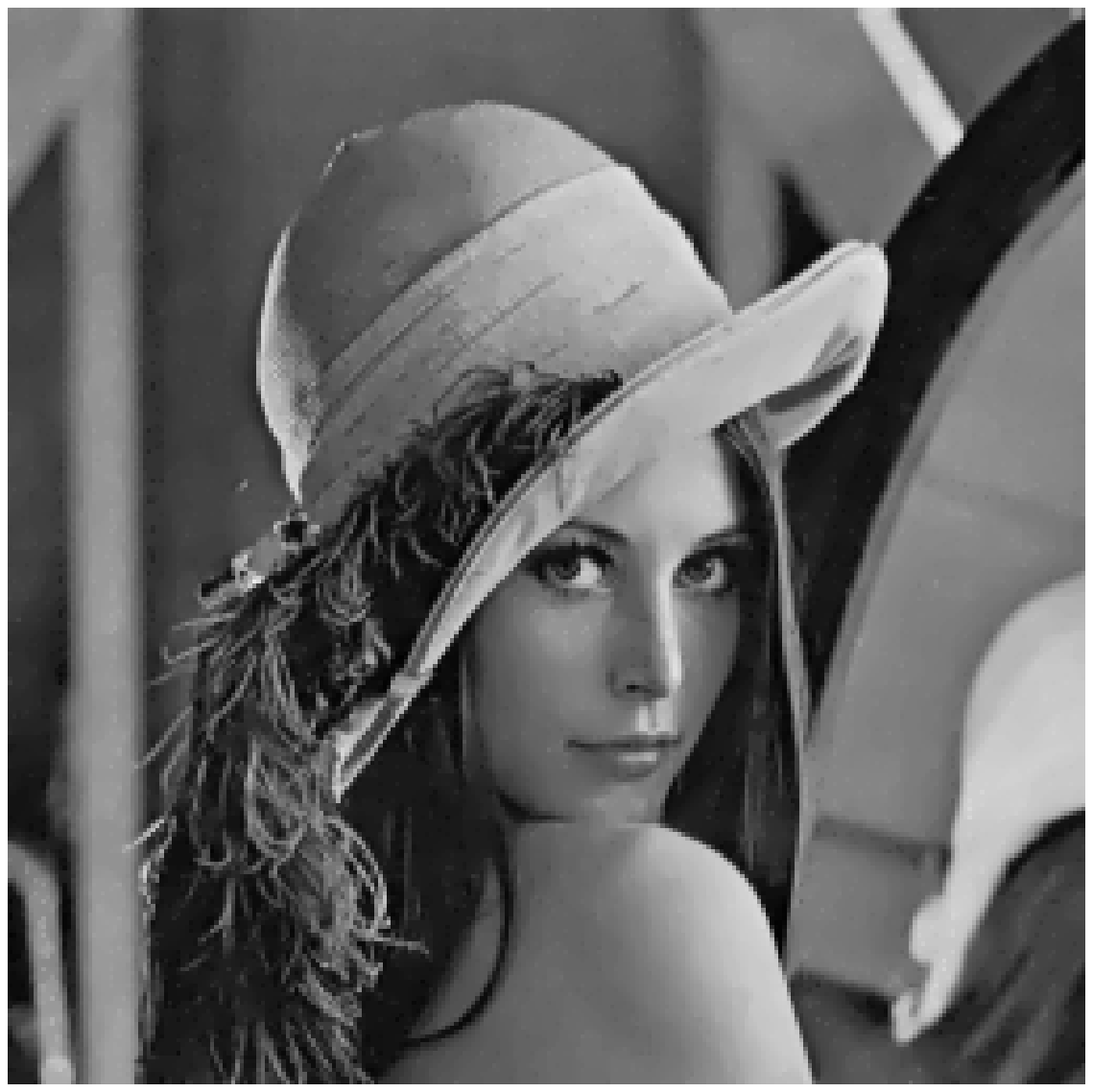}}
\subfigure[]{\label{fig:compareImages_ratio2_c}\includegraphics[width=0.32\textwidth]{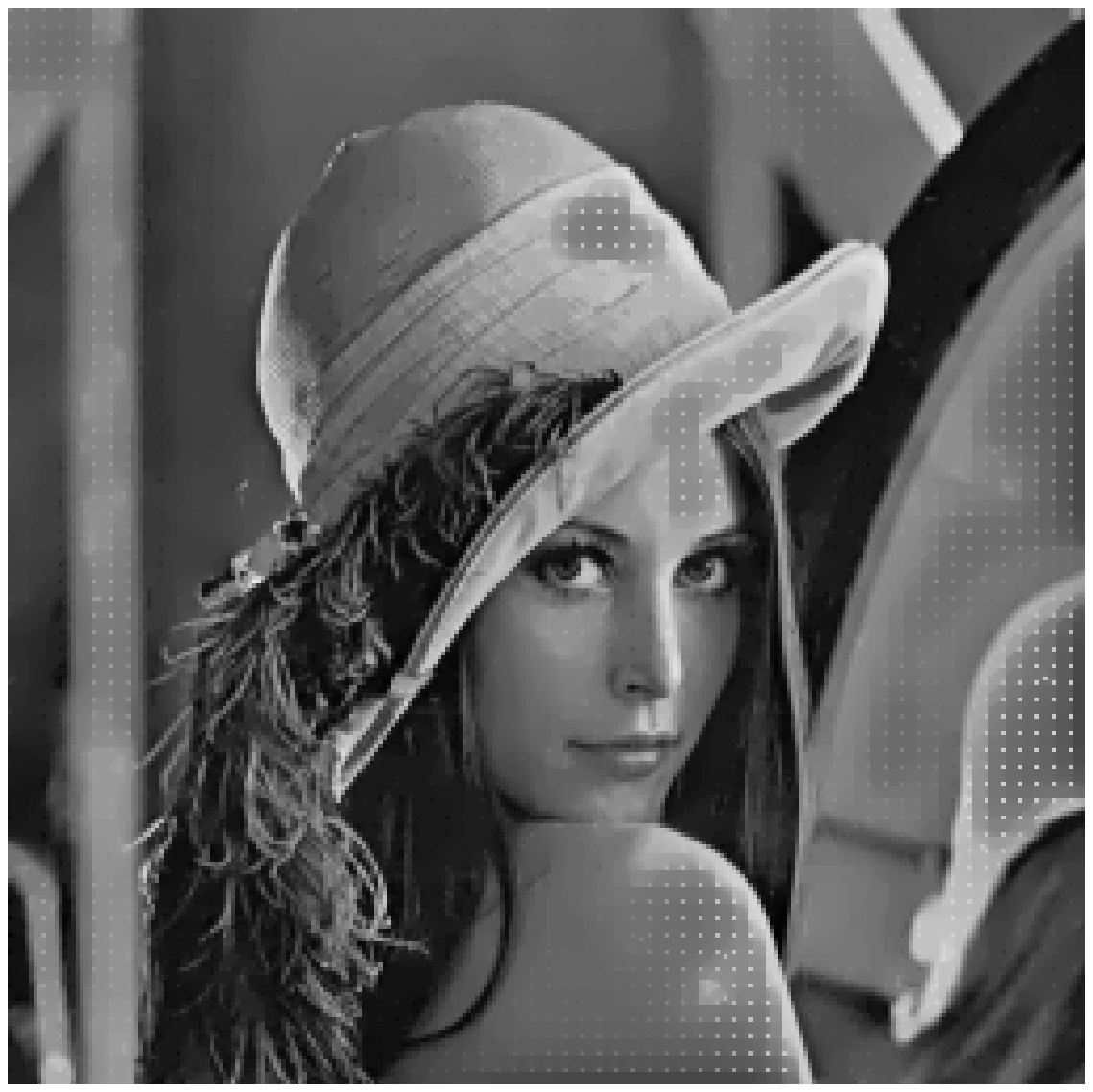}} \\ 

\subfigure[]{\label{fig:compareImages_ratio2_e}\includegraphics[width=0.32\textwidth]{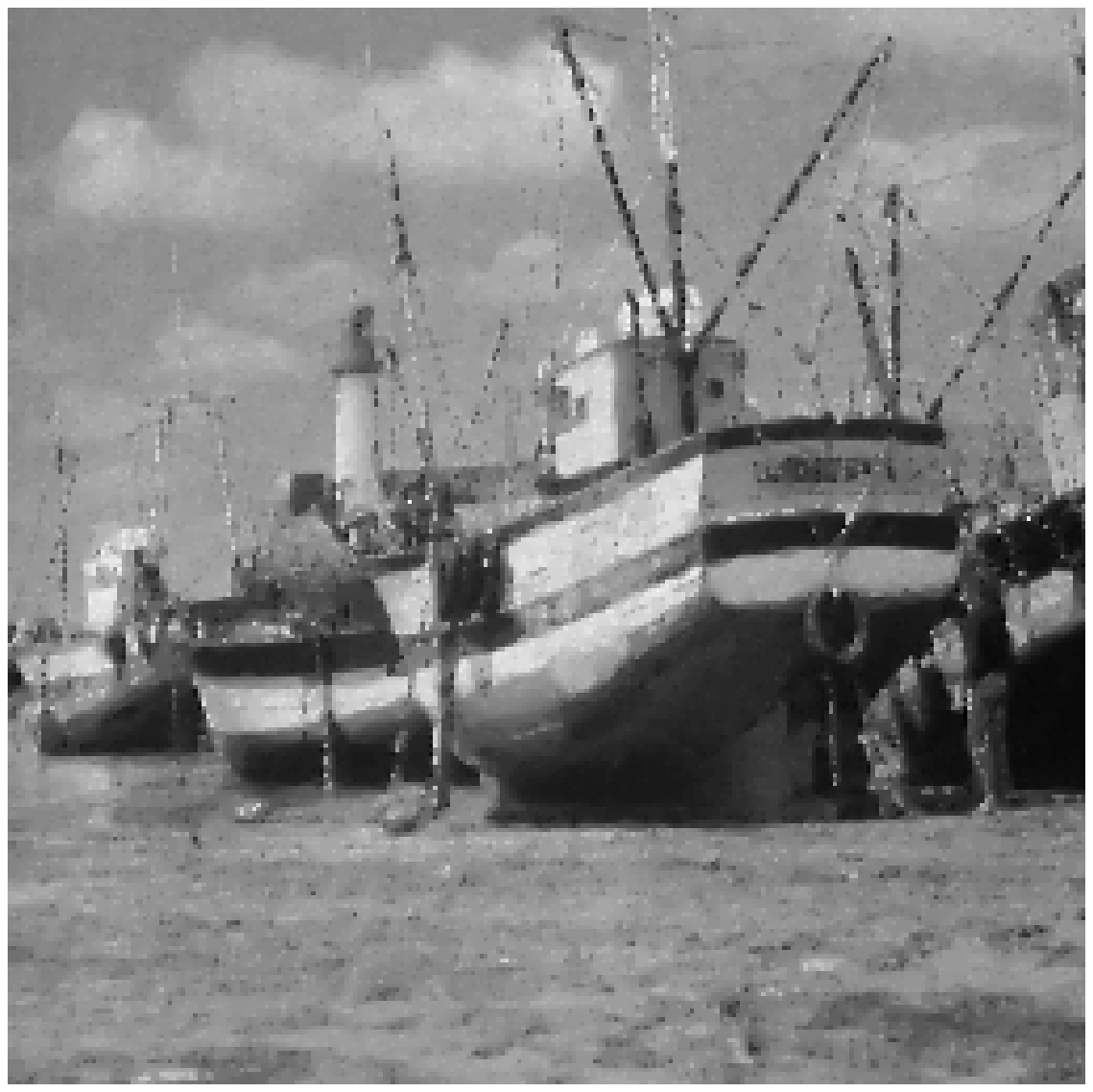}}
\subfigure[]{\label{fig:compareImages_ratio2_f}\includegraphics[width=0.32\textwidth]{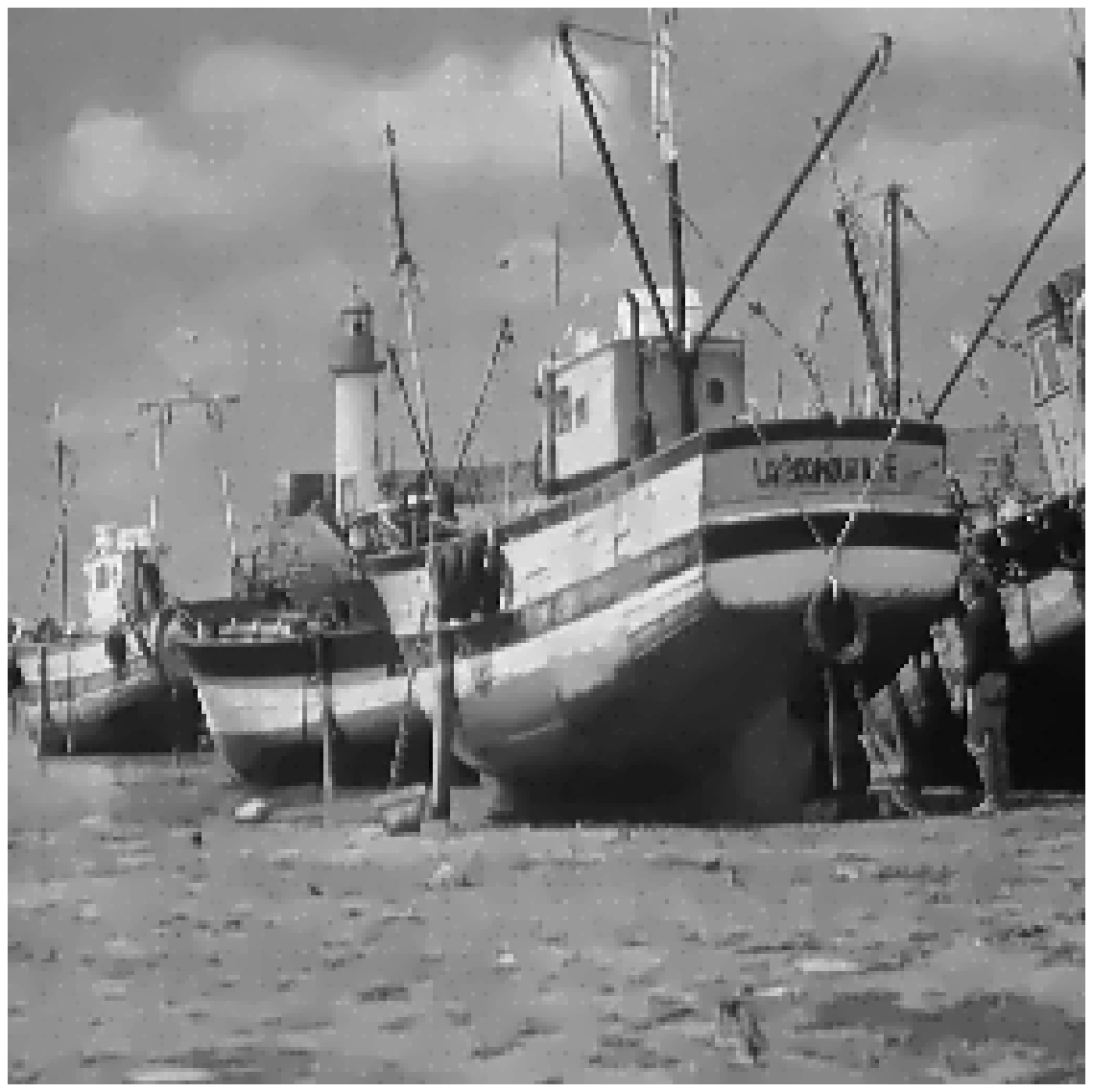}}
\subfigure[]{\label{fig:compareImages_ratio2_g}\includegraphics[width=0.32\textwidth]{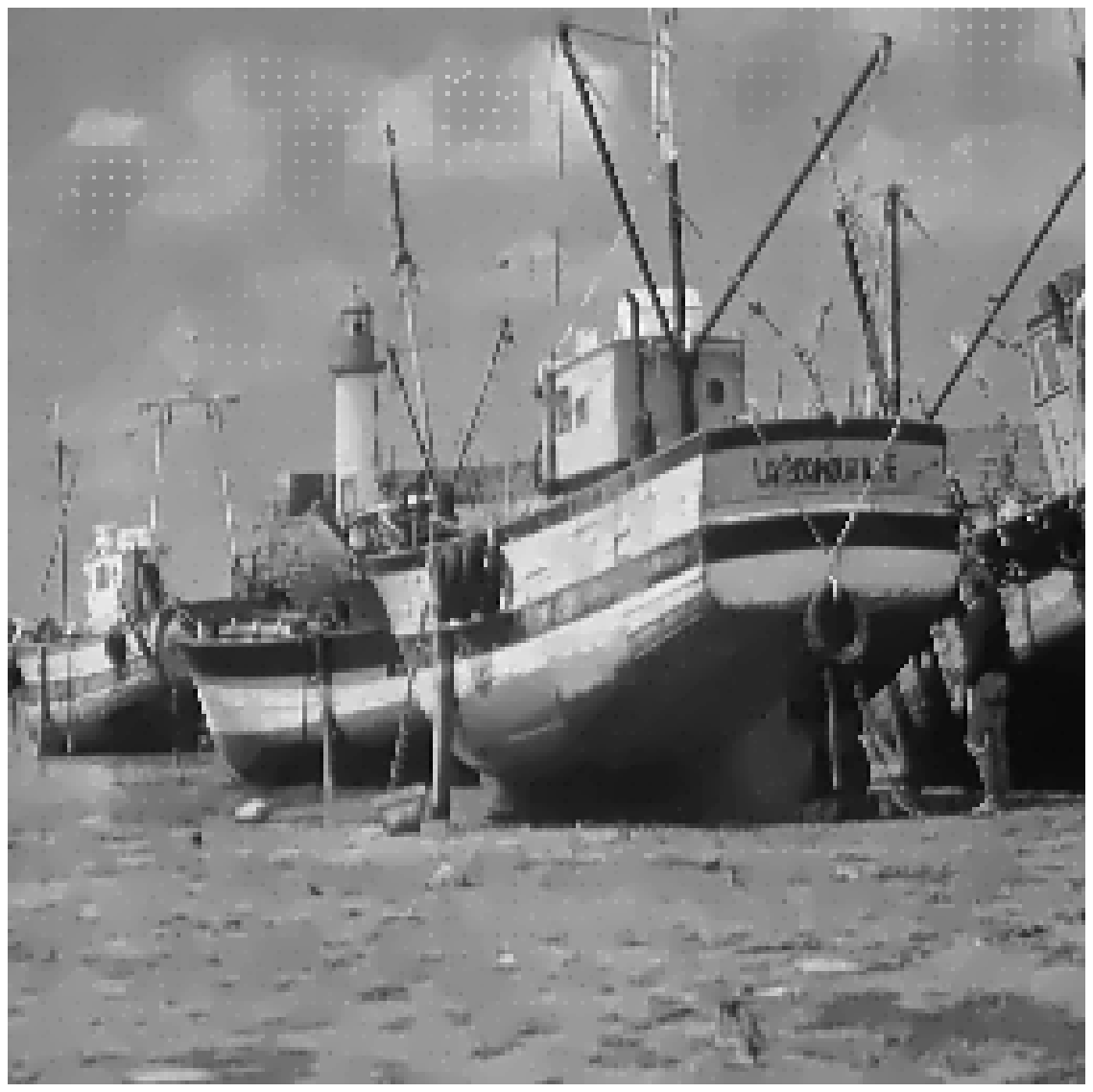}} \\ 
\caption{Reconstructed Lena and Boat images using $S_r$ and $S_m+M_n^p$ strategies. (a) and (d) are reconstructed images for $S_r$. (b) and (e) are reconstructed images for $S_m+M_d^p$ with the optimum $\eta_2$. (c) and (f) are reconstructed images for $S_m+M_d^p$ with the very large $\eta_2$. The corresponding key parameters are listed in Table \ref{table_ratio2}. }
\label{fig_compareImages_ratio2}
\end{figure}

\begin{figure} [htbp]
\centering
\scriptsize
\subfigure[Lena]{\label{fig:fig_eta2_Threshold_a}\includegraphics[width=0.49\textwidth]{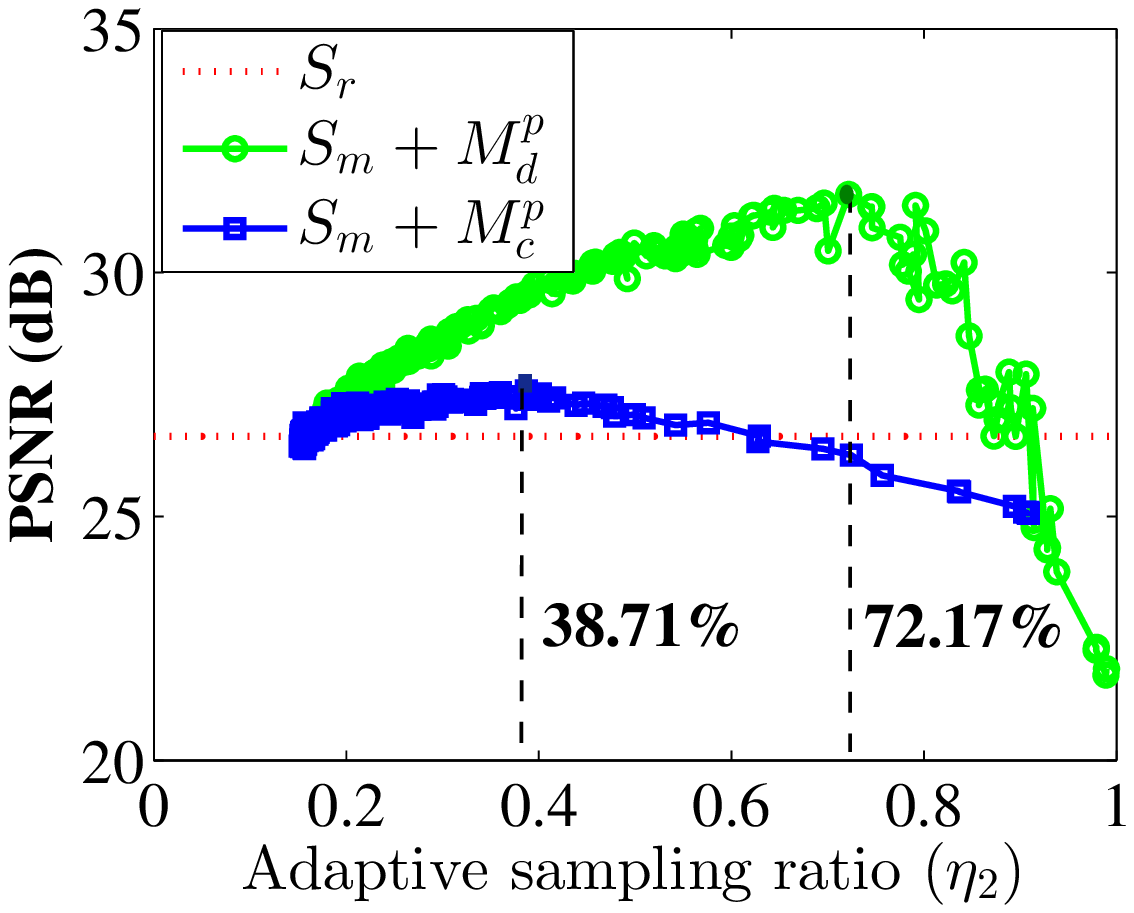}}
\subfigure[Lena]{\label{fig:fig_eta2_Threshold_b}\includegraphics[width=0.49\textwidth]{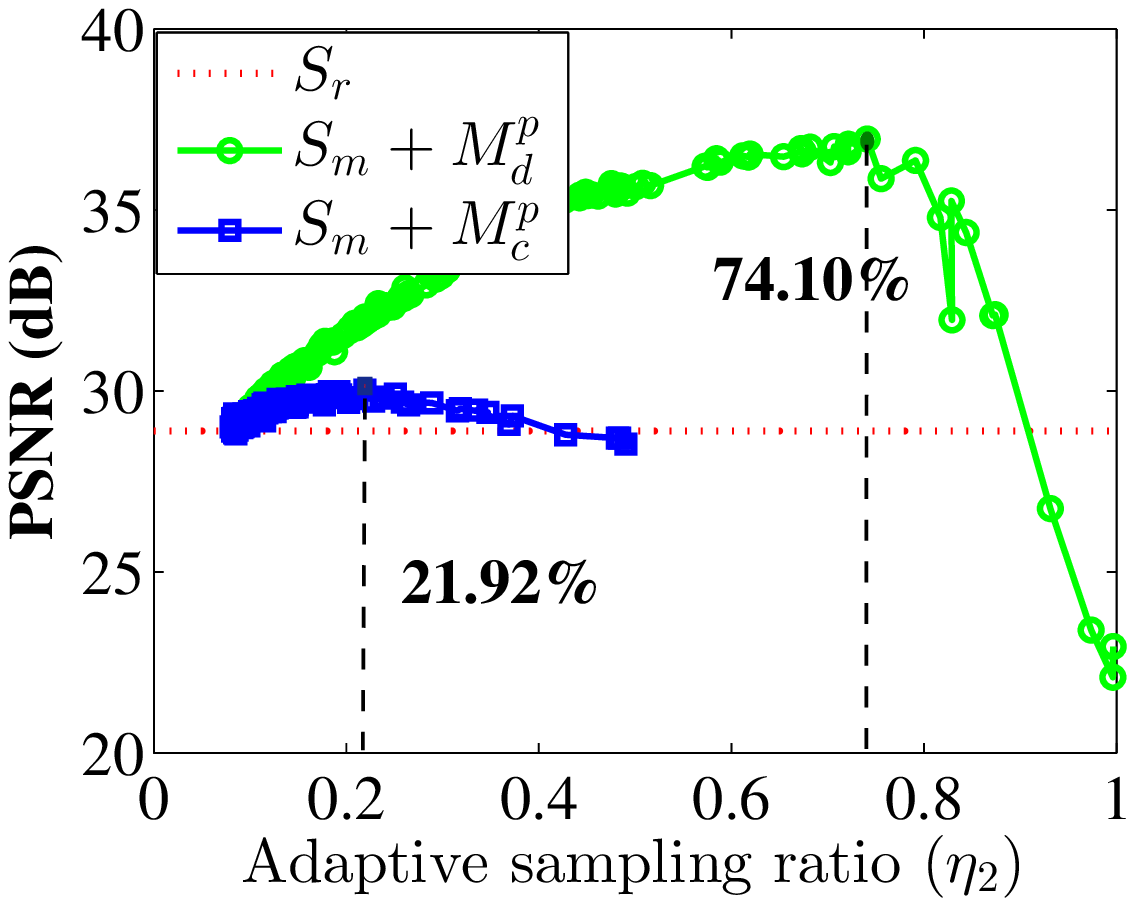}} \\ 
\subfigure[Boat]{\label{fig:fig_eta2_Threshold_c}\includegraphics[width=0.49\textwidth]{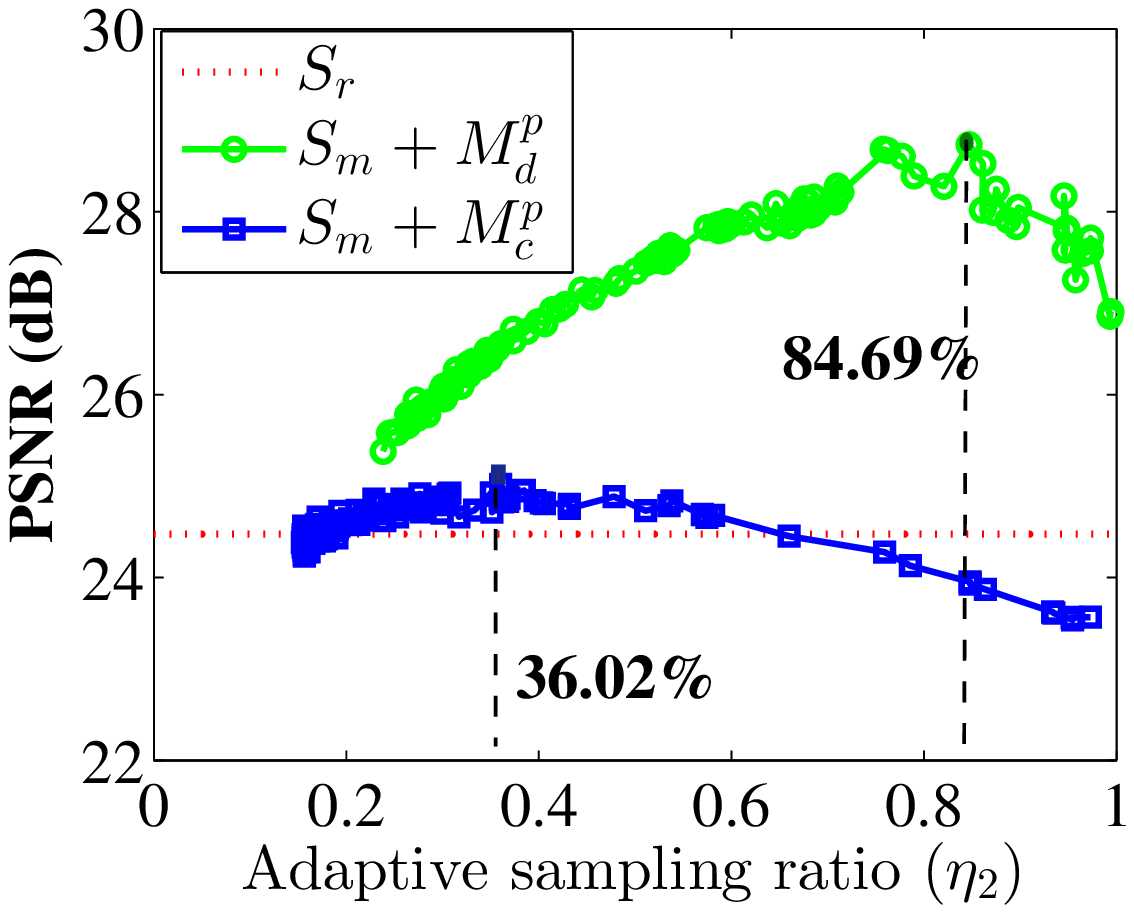}}
\subfigure[Boat]{\label{fig:fig_eta2_Threshold_d}\includegraphics[width=0.49\textwidth]{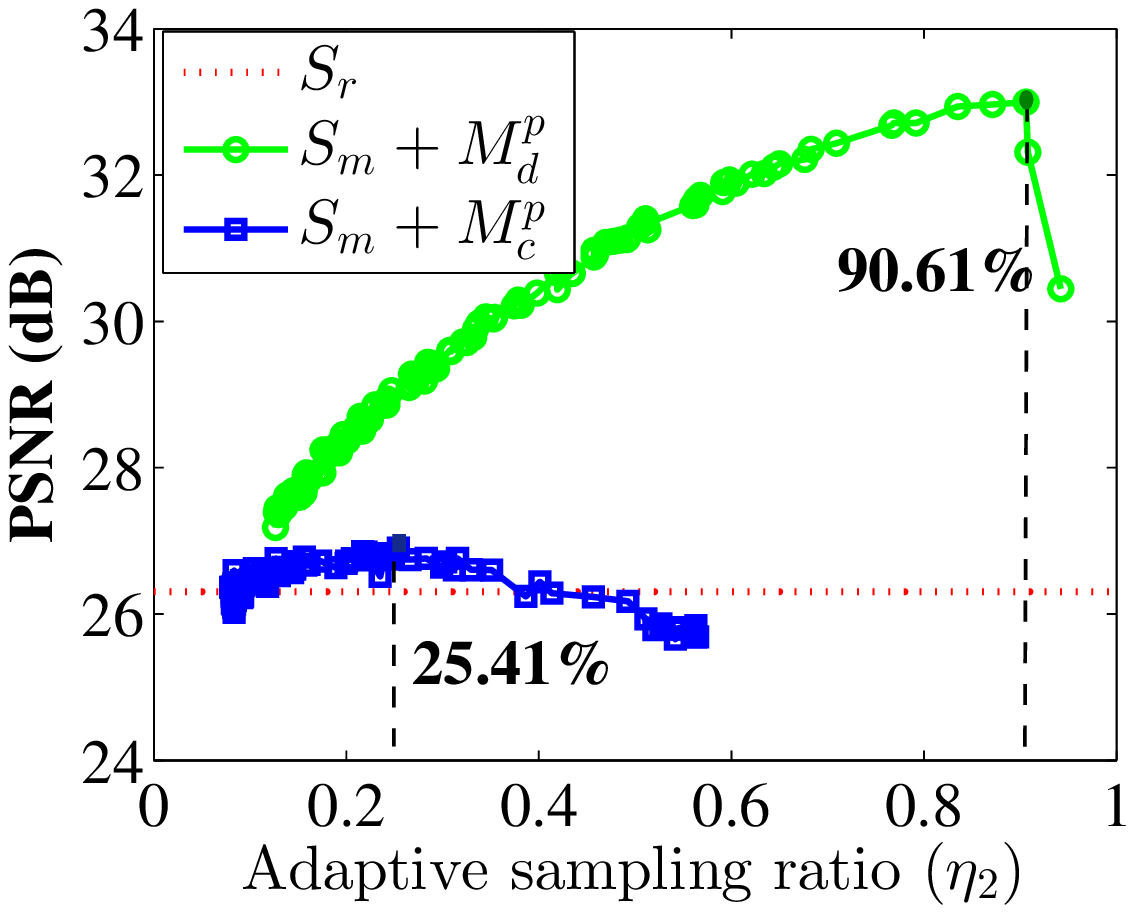}} \\ 
\caption{Evaluating the threshold of $\eta_2$ for $S_m+M_d^p$ and $S_m+M_c^p$. The sensing ratio $\eta_1$ for $S_r$ is set as (a): $30.05\%$, (b): $44.48\%$, (c): $29.81\%$ and (d): $44.69\%$. And the sensing ratio $\eta_1$ for $S_m+M_{d,c}^p$ is almost the same as that of $S_r$ with the error no more than $\pm0.01\%$.}
\label{fig_eta2_Threshold}
\end{figure}

Too much edge information in the MAR sensing matrix will destroy the RIP and incoherence condition of the framework. This situation occurs when an extremely high adaptive sampling ratio $\eta_2$ is set. Generally speaking, there is a tradeoff between the $\eta_2$ and the recovery performance. Fig. \ref{fig_compareImages_ratio2} and Tab. \ref{table_ratio2} show the tradeoff between $\eta_2$ and recovery performance. The bigger $\eta_2$ the better performance. But with very large $\eta_2$ both objective PSNR/SSIM measurements and visually quality will decline. Especially the details in smooth regions of the images are vanished to some extent although the objective measurements are still well for some images when $\eta_2$ is very large. We evaluate the optimum value of $\eta_2$ for $\eta_1$ as illustrated in Fig. \ref{fig_eta2_Threshold}. Using morphology operations, better reconstruction results are achieved by the MAR sensing even if the number of adaptive sampling is larger than that of random sampling (the optimum $\eta_2^{opt}\approx 80\%$ for $S_m+M_d^p$). It is necessary to state that the curve lines in Fig. \ref{fig_eta2_Threshold} (b) and (d) are missing when $\eta_2$ is large. In one hand the $S_m+M_c^p$ curve line will descend when the parameter $\eta_2$ increase and exceed a tradeoff threshold. With the increasing of $\eta_2$ the $S_m+M_c^p$ curve line will lies under the horizontal line of the random sampling pattern $S_r$. In the other hand, we just want to find the optimum $\eta_2$. The optimum $\eta_2$ reasonably locates at the top point of the $S_m+M_c^p$ curve lines and it should lies above the horizontal line of $S_r$. So the $S_m+M_c^p$ curve lines under the horizontal line of $S_r$ can be omitted.

\begin{table}
\scriptsize
\renewcommand{\arraystretch}{1.3}
\caption{The dependence of reconstruction performance on the adaptive sampling ratio $\eta_2$: optimum $\eta_2$ and very large $\eta_2$. The RIP cannot be maintained when $\eta_2$ goes beyond a threshold.}
\label{table_ratio2}
\centering
\begin{tabular}{cccccc}
\hline
\multicolumn{2}{c}{Image}              & PSNR    & SSIM    & $\eta_1$   & $\eta_2$  \\
\hline
\multirow{3}{*}{Lena}    & $S_r$       & 28.8945 & 0.9143  &44.48\%   &         \\
                         & $S_m+M_d^p$ & 36.9584 & 0.9448  &44.48\%   & 74.10\% \\
                         & $S_m+M_d^p$ & 22.9407 & 0.8777  &44.47\%   & 99.64\%\\
\hline
\multirow{3}{*}{Boat}    & $S_r$       & 26.3038 & 0.8337  &44.69\%   &        \\
                         & $S_m+M_d^p$ & 32.9985 & 0.8701  &44.69\%   & 90.61\%\\
                         & $S_m+M_d^p$ & 30.4449 & 0.8524  &44.69\%   & 94.22\%\\
\hline
\end{tabular}
\end{table}

\begin{figure} [htbp]
\centering
\scriptsize
\subfigure[]{\label{fig:fig_ball_a}\includegraphics[width=0.24\textwidth]{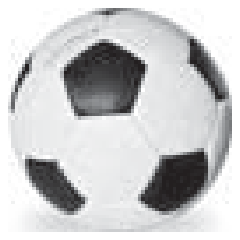}}
\subfigure[]{\label{fig:fig_ball_b}\includegraphics[width=0.24\textwidth]{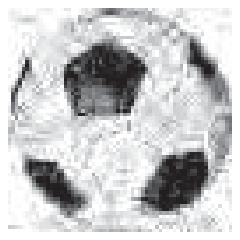}}
\subfigure[]{\label{fig:fig_ball_c}\includegraphics[width=0.24\textwidth]{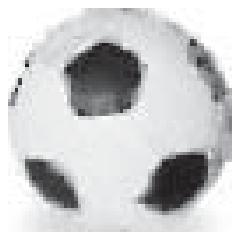}}
\subfigure[]{\label{fig:fig_ball_d}\includegraphics[width=0.24\textwidth]{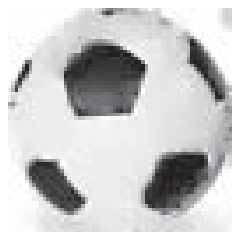}} \\ 
\caption{Experiment results: the visual performance of standard CS and the random partial sampling framework.(a) the original ball image, and the recovered ball by (b): standard CS; (c) completely random partial sampling and (d) mixed adaptive-random sampling: $S_m+M_d^p$}
\label{fig_ball}
\end{figure}

As a sampling method in spatial domain, we also compare it with the projection characterized standard CS with the same sampling rate. In Fig. \ref{fig_ball} we show the original \textit{ball} image with size $64\times64$ in Fig. \ref{fig_ball} (a), and the reconstructed image by the standard CS in Fig. \ref{fig_ball} (b); the completely random partial sampling method in spatial domain in Fig. \ref{fig_ball} (c) and the mixed adaptive-random sampling ($S_m+M_d^p$) method in Fig. \ref{fig_ball} (d). For the three methods above the sample rate is $30\%$ and the PSNR value for the recovered images are 28.9019, 32.2531 and 34.1675 respectively. In the traditional random projection method in CS, the OMP algorithm and the ``sym 8" wavelet transform are employed, and the the \textit{ball} image is denoted as a vector with size $4096\times 1$ to directly sense and recovery. Here there are two main reasons for the poor performance of standard CS. One is the OMP recovery algorithm, which focus on the operation speed rather than the recovery efficiency, and the other is the limited sparse representation ability of wavelet transformation. Other good recovery algorithm and better sparse representation will achieve better performance, which may surpass ours. However this paper do not aim at developing a reconstruction algorithm but proposing a new concept for measurement learning.

We mainly focus on the importance of edges as the prior information exploited in the sampling processing. Our sampling strategy is the random partial sampling in spatial domain of signals. The paper did not aim at developing a reconstruction algorithm but proposing a new concept for measurement learning. In our method the sampling strategy includes three successive sampling processes. The first one is to get the low-resolution image and then the predicted edges are employed to guide for the second adaptive sampling process. By combining the third random sampling process the total samplings can achieve much better performance than completely random sampling with the same sampling rate and recovery algorithm as demonstrated above. Some improved recovery algorithm can also achieve much better performance in the partial random sampling methods \cite{zhang2011highquality,zhang2014group}.

In the end of this section we compare our method with the partial random sampling method, which we note it as the random sampling and GSR recovery (R-GSR), in the literature \cite{zhang2011highquality,zhang2014group}. R-GSR employs the nonlocal adaptive 3-D sparse representation prior to developed the Split-Bregman based iterative recovery algorithm. Our method and R-GSR use the different strategies to achieve better performance. Our method focus on the adaptive sampling strategy with TV recovery algorithm while the later pays close attention to the improved recovery algorithm with completely random sampling strategy. MAR-GSR denotes the method with MAR sampling strategy and the GSR recovery algorithm. As MAR-GSR, TRPS-GSR denotes the method with TRPS sampling strategy and the GSR recovery algorithm.

In practice we can directly use the random partial sampling $\bar{S}_r$ to get the low-resolution image and then the predicted edges. The edges are employed to guide for the adaptive sampling $\bar{S}_a$. We name this sampling strategy, $\bar{S}_m= \bar{S}_r\bigcup \bar{S}_a$, as the two random partial samplings (TRPS). The $\bar{\eta}_2$ is defined as $\bar{\eta}_2=1-\frac{\sum_{i,j}\bar{S}_r(i,j)}{\sum_{i,j}\bar{S}_m(i,j)}$. In this way we can get rich edges by adjusting the ratio between the fist random sampling $\bar{S}_r$ and the second adaptive sampling $\bar{S}_a$.

For fair comparison, we use four standard test gray images with the size of $256\times256$ including House, Barbara, Lean and Boat. The iterative number in the literature \cite{zhang2011highquality,zhang2014group} are set to 60, 50 and 40 for the sample ratio 0.3, 0.5 and 0.8 respectively. We set $\eta_2 = 0.6$ and $\bar{\eta}_2=0.4$ for MAR and TRPS respectively. Tab. \ref{table_compare_GSR} shows the PSNR comparisons among different algorithms.

\begin{table}[!t]
\scriptsize
\renewcommand{\arraystretch}{1.3}
\caption{The PSNR (dB) comparisons among different algorithms.}
\label{table_compare_GSR}
\centering
\begin{tabular}{ccccccc}
\hline
Image  & Data Ratio       & MAR    & R-GSR \cite{zhang2011highquality,zhang2014group}  & TRPS   & MAR-GSR     & TRPS-GSR\\
\hline
\multirow{3}{*}{House}   & $30\%$     & 32.7723    & 36.7562                &33.6687      &\textbf{37.1779}     &37.0893    \\
                         & $50\%$     & 35.6070    & 40.3797                &38.7431      &40.6712     &\textbf{41.7113}    \\
                         & $80\%$     & 38.7992    & 46.3753                &47.7785      &45.6682     &\textbf{49.8807}    \\
\hline

\multirow{3}{*}{Barbara} & $30\%$     & 24.5645    & \textbf{34.9229}                &24.7261      &33.5012     &34.3876     \\
                         & $50\%$     & 27.4300    & 39.2226                &29.3758      &39.6593     &\textbf{39.8125}     \\
                         & $80\%$     & 29.8954    & 45.7269                &42.1341      &42.3528     &\textbf{47.7081}     \\
\hline
\multirow{3}{*}{Lena}    & $30\%$     & 28.1117    & 30.3159                &28.7301      &31.2212     &\textbf{31.7998}     \\
                         & $50\%$     & 32.3837    & 33.9816                &35.1140      &36.1792     &\textbf{37.7648}     \\
                         & $80\%$     & 35.5468    & 40.2195                &43.2416      &40.2195     &\textbf{46.1186}     \\
\hline
\multirow{3}{*}{Boat}    & $30\%$     & 25.4256    & 25.7810                &26.3588      &26.7768     &\textbf{27.1711}     \\
                         & $50\%$     & 28.5856    & 28.7616                &30.8805      &30.3538     &\textbf{32.0989}     \\
                         & $80\%$     & 31.7591    & 34.4550                &39.3448      &34.1296     &\textbf{40.0811}     \\
\hline
\end{tabular}
\vspace{-1em}
\end{table}

\begin{figure} [htbp]
\centering
\scriptsize
\subfigure[]{\label{fig:fig_compareRPS_a}\includegraphics[width=0.32\textwidth]{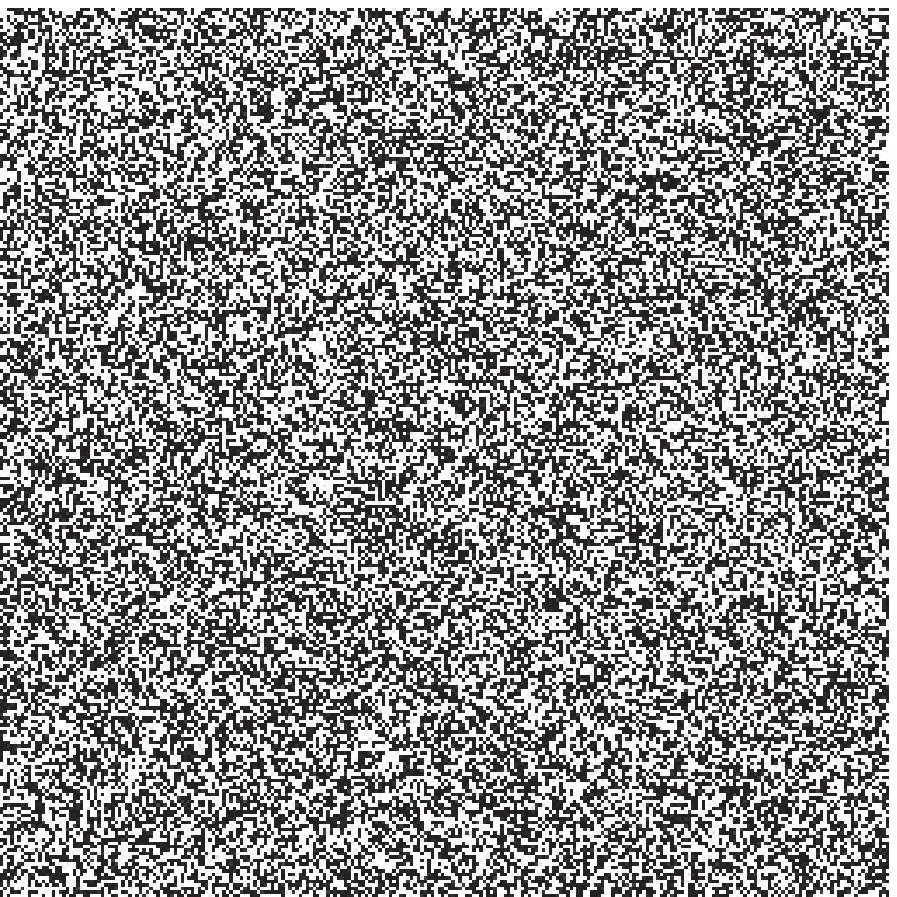}}
\subfigure[]{\label{fig:fig_compareRPS_b}\includegraphics[width=0.32\textwidth]{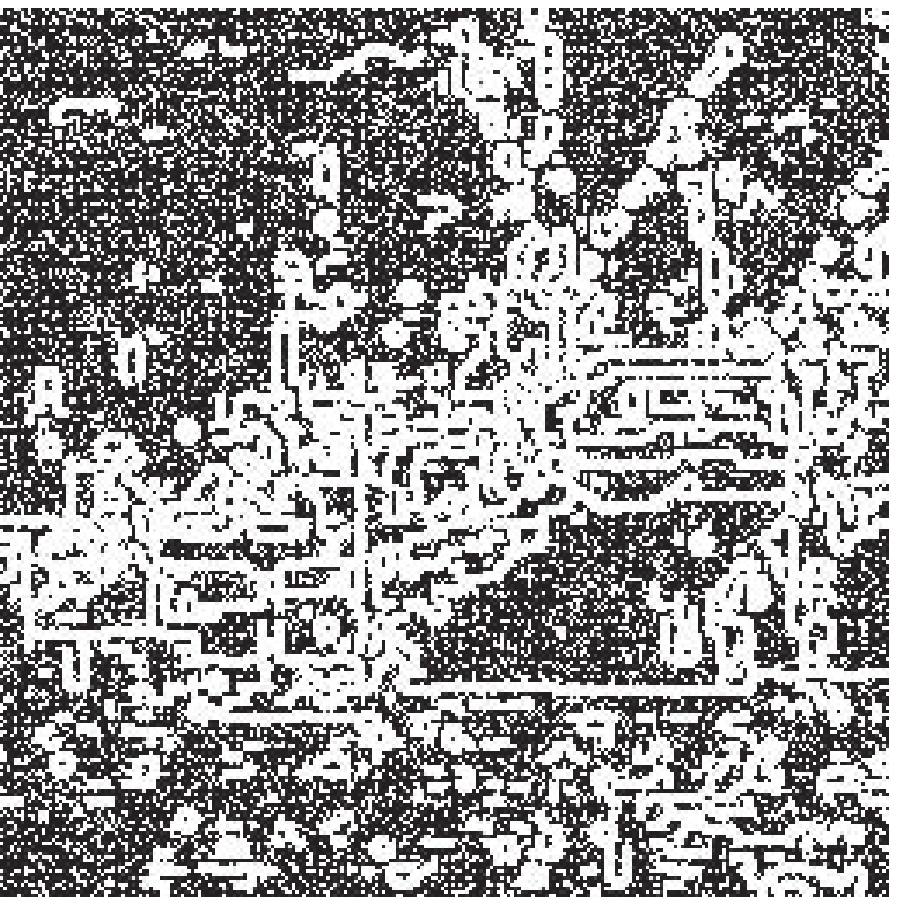}}
\subfigure[]{\label{fig:fig_compareRPS_c}\includegraphics[width=0.32\textwidth]{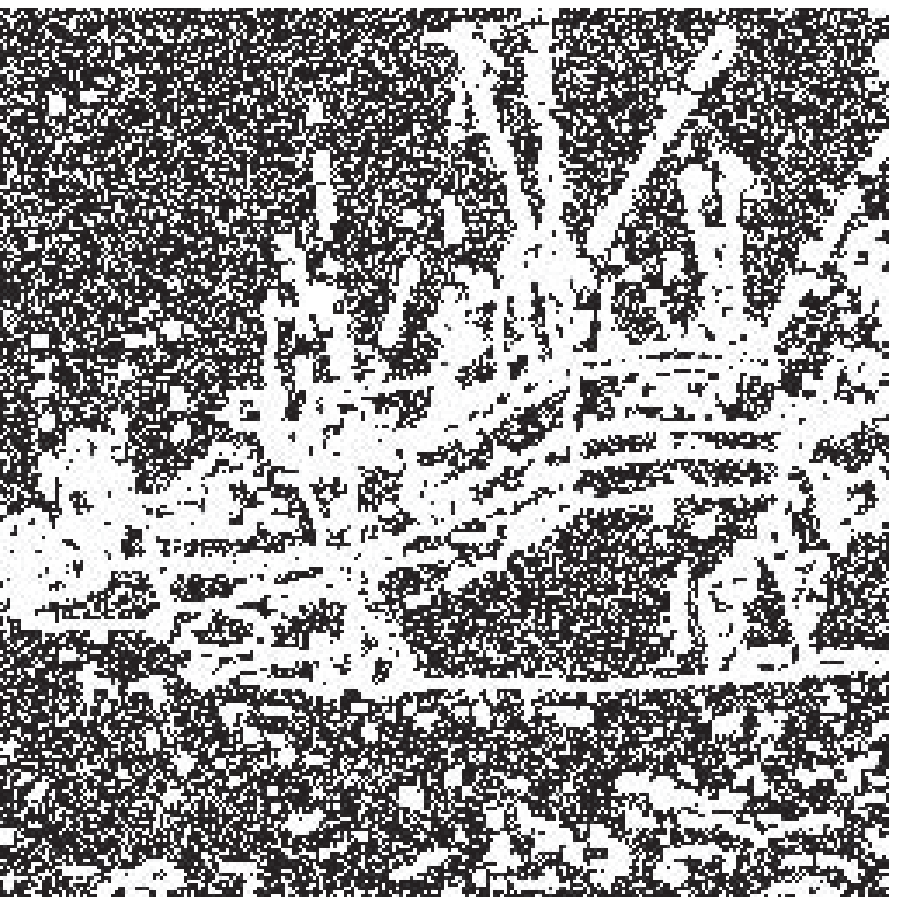}} \\ \vspace{-0.12in}
\subfigure[]{\label{fig:fig_compareRPS_d}\includegraphics[width=0.32\textwidth]{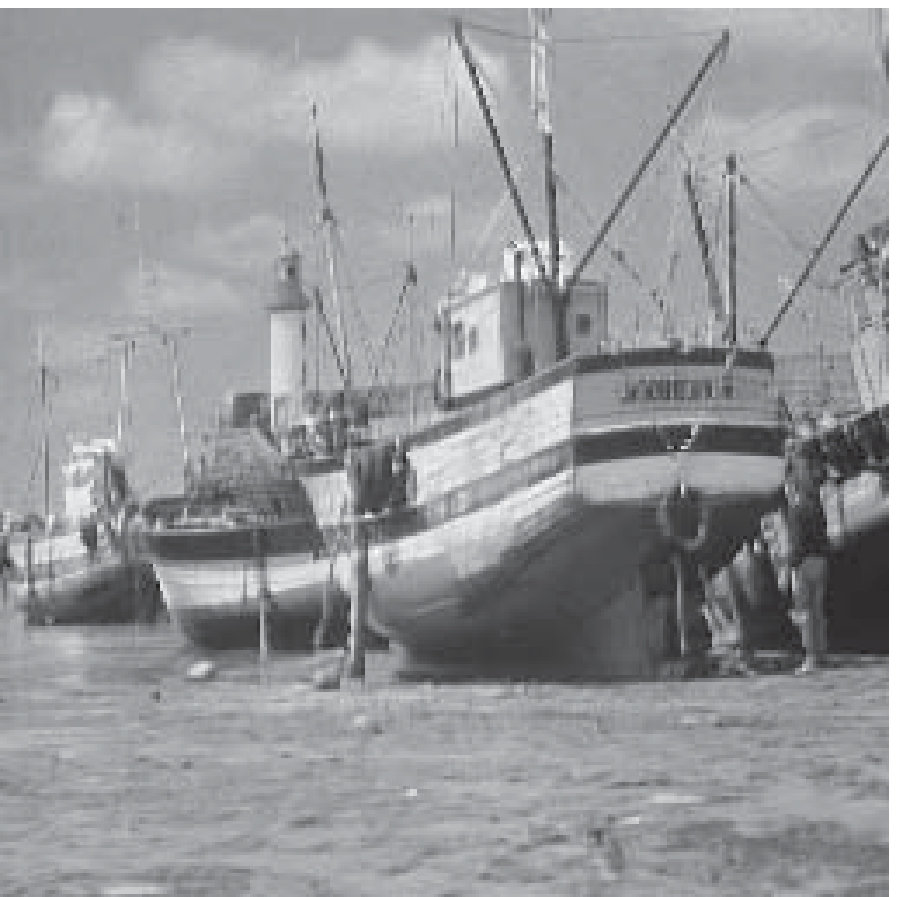}}
\subfigure[]{\label{fig:fig_compareRPS_e}\includegraphics[width=0.32\textwidth]{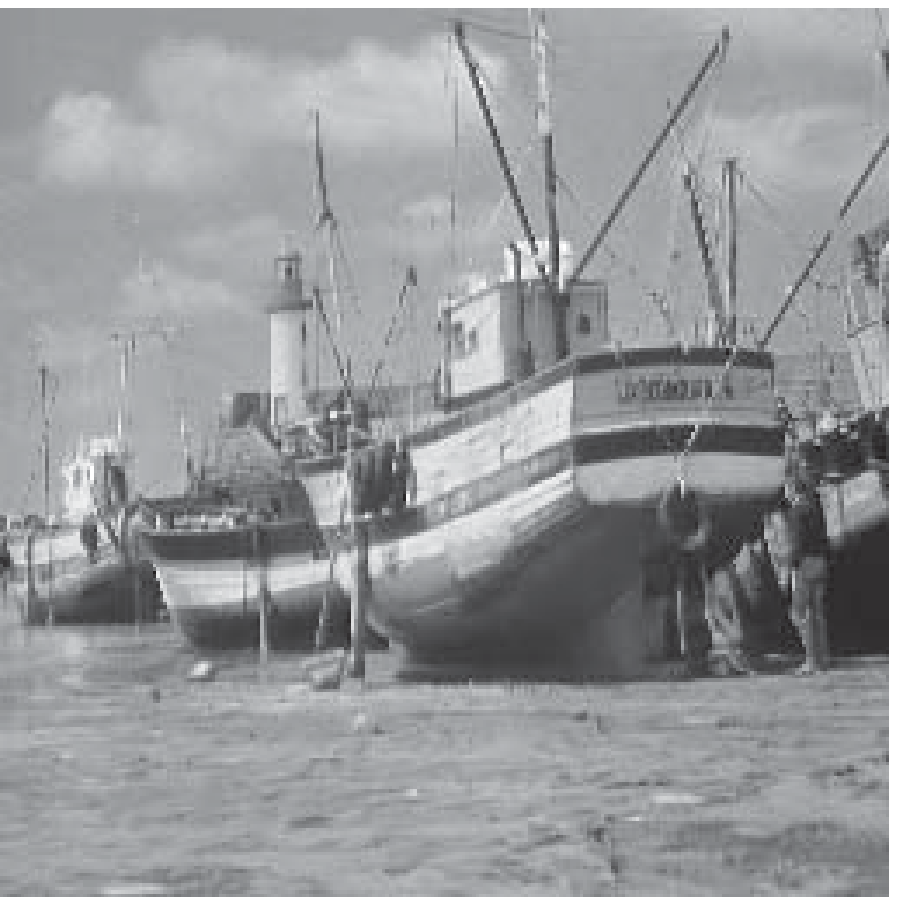}}
\subfigure[]{\label{fig:fig_compareRPS_f}\includegraphics[width=0.32\textwidth]{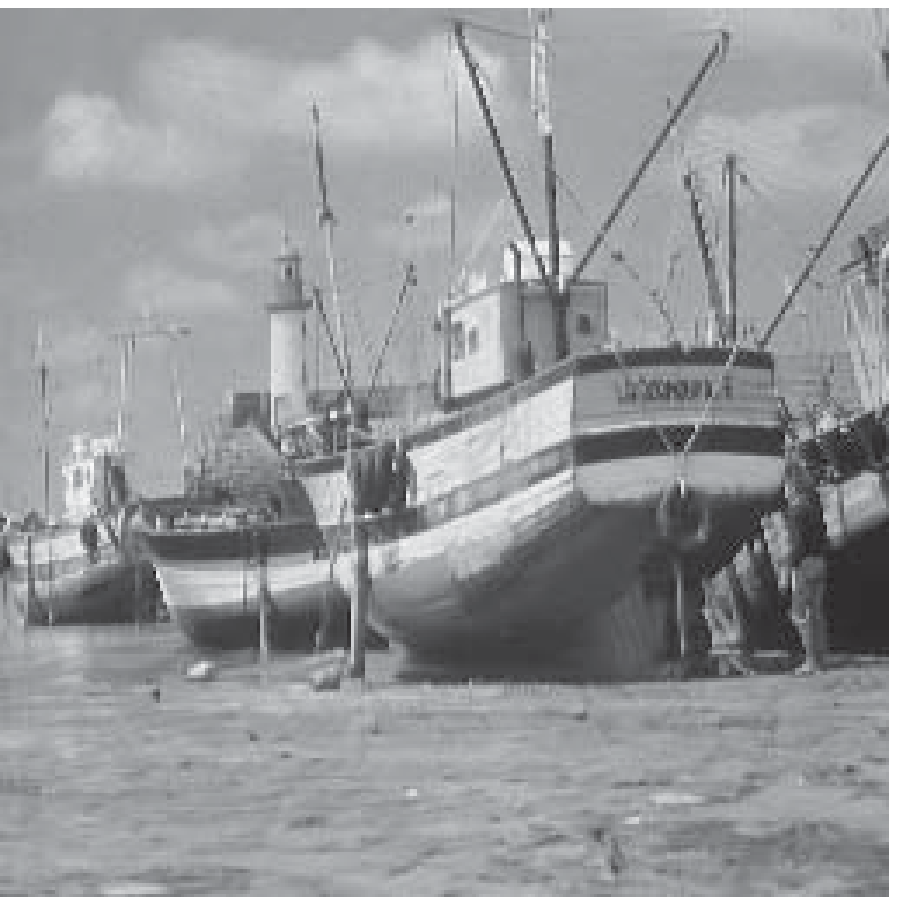}} \\ 
\caption{Experiment results: the sampling masks with sample ratio $50\%$ and the visual performance of the random partial sampling frameworks.(a),(b) and (c) are the partial sampling masks with the same ratio $50\%$ for random sampling, MAR and TRPS respectively. (d), (e) and (f) are the corresponding recovered images for R-GSR, MAR-GSR and TRPS-GSR with the GSR recovery algorithm.}
\label{fig_compareRPS}
\end{figure}

We can see that our sampling strategy can achieve better or comparable recovery performance when we use the same recovery algorithm GSR. Fig. \ref{fig_compareRPS} shows the random partial sampling masks and the recovered images for R-GSR, MAR-GSR and TRPS-GSR respectively with the same ratio ($50\%$) and the same recovery algorithm (GSR).

\section{Conclusion}
In this paper, a novel MAR sensing protocol is proposed to acquire a compressed image representation in space domain. Incorporating adaptive edge information that can be trivially extracted from a low-resolution sampling, the MAR measurements show much better reconstruction results in comparison with the completely random measurements. The RIP and incoherence condition of the MAR sensing matrix can be satisfied by balancing the number of adaptive sampling with that of completely random sampling. The mixed sensing concept opens up a bright and unexplored way for high-resolution and lost-cost data acquisition.


\bibliographystyle{spmpsci}      
\bibliography{myreferences}   

%
%

\end{document}